\documentclass[aps,10pt,onecolumn,rmp,showpacs,amsmath,amsfonts,amssymb,longbibliography]{revtex4-1} 
\usepackage{graphicx}
\usepackage{algorithm}
\usepackage{algpseudocode}
\usepackage{amsmath}
\usepackage{amssymb}
\usepackage{wrapfig,caption,sidecap}
\usepackage{shadow}
\usepackage{bm}
\usepackage{bbm}
\usepackage{mathrsfs}
\usepackage{soul}

\newcommand{\be}{\begin{eqnarray}}
\newcommand{\ee}{\end{eqnarray}}
\newcommand{\la}{\langle}
\newcommand{\ra}{\rangle}
\newcommand{\bh}{{\rm BH}}

\newcommand{\out}{{\rm out}}

\newcommand{\gp}{g^\prime}

\newcommand{\Tr}{{\rm Tr}}
\newcommand{\inv}{{\rm in}}
\newcommand{\one}{\mathbbm 1}

\def\dg{\dagger}

\def\om{\omega}

\def\om{\omega}

\newcommand{\ket}[1]{\mathop{|#1\rangle}\nolimits}

\def\df{\overset{\rm df}{=}}

\begin{document}
\title{Stimulated Emission of Radiation and the Black Hole Information Problem}

\author{Christoph Adami}
\email[]{adami@msu.edu}


\affiliation{Department of Physics and Astronomy, Michigan State University, East Lansing, Michigan 48824, USA}

\date{\today} 

\begin{abstract}The quantum theory of black holes has opend up a window to study the intersection of general relativity and quantum field theory, but perceived paradoxes concerning the fate of classical information directed at a black hole horizon, as well as concerning the unitarity of the evaporation process, have led researchers to question the very foundations of physics. In this pedagogical review I clarify the ramifications of the fact that black holes not only emit radiation spontaneously, but also respond to infalling matter and radiation by emitting approximate clones of those fields in a {\em stimulated} manner. I review early purely statistical arguments based on Einstein's treatment of black bodies, and then show that the Holevo capacity of the black hole (the capacity to transmit classical information through a quantum channel) is always positive. I then show how stimulated emission turns the black hole into an almost optimal quantum cloning machine, and furthermore discuss the capacity of black holes to transmit {\em quantum} information. Taking advantage of an analogy between black hole physics and non-linear optics I show that a calculation of the evolution of a black hole over time, using a discretization of the black hole $S$-matrix path integral, yields well-behaved Page curves suggesting that black hole evaporation is unitary. Finally, I speculate about possible observable consequences of stimulated emission of radiation in black holes. 
\end{abstract}
\maketitle


\numberwithin{equation}{section}
\renewcommand{\thefigure}{\arabic{section}.\arabic{figure}}
\section{Introduction}
Black-hole quantum physics has been an exciting but frustrating area of research for almost fifty years---ever since Hawking discovered that black holes described in quantum field theory become less black than their name suggests~\cite{Hawking1975}. In classical physics black holes are completely black~\footnote{This holds true only for incident radiation with vanishing impact parameter, as modes with angular momentum can be scattered.} and do not emit any particles because all classical trajectories must end up in the black hole singularity. According to a semi-classical calculation in curved-space quantum field theory, however, black holes emit radiation through the spontaneous emission of particles near the event-horizon, while the black hole gives up mass in the process. Ultimately, if no new mass is accreted, black holes may even evaporate completely, leaving behind only Hawking's eponymous radiation.

The mere existence of Hawking radiation revealed a daunting problem almost immediately~\cite{Hawking1976b}. Particles that accrete onto a black hole can be viewed as carriers of information. For example, a particle's identity (whether it is a photon, electron, or proton), its angular momentum, mass, etc., can encode information that appears to be lost completely and irretrievably behind the event horizon. While it is expected that the Hawking radiation theoretically consists out of all kinds of particles (all those that can be created in pairs via vacuum fluctuations), the {\em thermal} nature of the radiation seems to imply that its quantum numbers are completely uncorrelated to those that enter the horizon of an already formed black hole, or even of the matter and radiation that created it. If Hawking radiation was the only energy left over after the evaporation of the black hole, all information about the state of matter that initially created the black hole, as well as the information carried by particles that accreted on to it later, would forever be lost. 

This is a more serious problem than is perhaps obvious from the start. Such a loss of information is not merely inconvenient. It would signal the breakdown of some of the most fundamental and cherished laws of nature that we have been able to establish, namely the conservation of probability. Probability conservation is built into quantum mechanics by describing the time-dependence of a wavefunction by unitary operators $U=e^{-i/\hbar H}$, where $H$ is a Hermitian Hamiltonian operator. In quantum field theory, unitarity is ensured by the unitarity of the $S$-matrix, which is in itself a consequence of a Hermitian interaction Hamiltonian (see, e.g., \cite{Sakurai1967}). On the one hand, both quantum mechanics and in particular quantum field theory have been exceedingly accurate in their prediction of the microscopic properties of matter and light, to such an extent that it would be shocking if unitarity would be violated by such macroscopic objects as black holes. 

On the other hand, we do not have a consistent theory of quantum gravity, in particular not one that is expected to accurately describe the late stage of black-hole evaporation, where gravitational fields are expected to be very strong. One could therefore not dismiss out of hand that our current theories simply break down under such extreme circumstances, and that a consistent theory of quantum gravity would remedy the dilemma. But the following reasoning suggests that a full-fledged theory of quantum gravity cannot be necessary to solve the problem of probability conservation in black hole evaporation. Classical information, as mentioned above, is carried by ordinary degrees of freedom such as spin, polarization, momentum etc., all of which are adequately described in a semiclassical theory of gravity. The approximations involved in the semiclassical theory concern the treatment of the space-time metric: it is left unquantized, meaning that it is treated as a background field. While a consistent theory of quantum gravity should treat the space-time metric as a quantum mechanical variable that can be entangled with other degrees of freedom, it is not plausible that the uncertainty associated with the decoherence of the metric will have a significant impact on black hole dynamics until perhaps the black hole is of Planck mass~\cite{Wald1994}. Yet, during the period where quantum effects of the metric are small (large, massive black holes), predictability would {\em already} be lost because the incident quantum numbers are inaccessible behind the horizon. Indeed, the principle of microscopic time reversibility relies on predictability at {\em all} times.
Thus, it is not necessary to wait for the complete evaporation of the black hole in order to see a problem with the standard description of black holes. Moreover, at the point where quantum effects on the metric are expected to be significant, it is implausible that information can be recovered from the evaporation of a Planck-size black hole because it is unclear how it could store that much entropy. Thus, we are encouraged to look for a consistent treatment of black hole dynamics that allows for predictability {\em at all times}, and  explains the apparent lack of coherence in a completely unitary manner.

Before discussing possible solutions to the "black hole information problem", I'd like to emphasize that there are really {\em two} such problems. The one that gets most of the attention is the problem of whether black holes can turn pure states into mixed states, as Hawking claimed they do~\cite{Hawking1976b}. Even though this question is often couched in language that suggests that it is a problem of information conservation ("what happens to the information about the initial state of the matter and radiation that formed the black hole"?) it is strictly speaking not a problem of information transmission, but rather a problem in showing that the quantum state after the evaporation of the black holes has returned to the pure state it started out as, before the formation of the black hole. Quite generally, the question should be: "Is black hole formation and evaporation described by unitary dynamics?"

The literature usually distinguishes four standard alternatives to deal with the apparent "information loss". I briefly summarize them here, but refer to~\cite{Preskill1992,FabbriNavarro-Salas2005,UnruhWald2017} for a more thorough exposition. The first alternative is the most conservative: it claims that information is released with the Hawking radiation after all, or to put it more concisely, that the state after evaporation of the black hole somehow returns to purity. This type of scenario has been advocated by Page and by Bekenstein~\cite{Page1993, Bekenstein1993}, and also has more modern support from theories in which quantum gravity is coupled to dilatons~\cite{Callanetal1992,Almheirietal2020}. One of the most common objections to this scenario is that the thermal nature of the Hawking radiation precludes it from carrying any information. We will see below that there is in fact no basis to this objection, as information can be encrypted in maximum entropy states.

The second alternative posits that, conceivably, the information remains locked inside of the black hole, but the black hole does not disappear but rather becomes a stable remnant. This explanation suffers from several problems, such as the necessity of requiring almost infinitely-degenerate Planck-sized black holes that should be observable via pair formation. Moreover, it does not address how those remnants would constitute a pure state.
The third general category of explanations claims that it is possible that all the accumulated information ``comes out at the very end", that is, after most (or all) of the mass of the black hole was radiated away. This suggestion runs into the problem that an arbitrary large amount of information cannot be radiated away in a finite amount of time. Finally, a fourth alternative  suggested that the information is not lost, but rather is sealed away into one or more nucleating baby universes. This type of explanation has now even fewer defenders than it had at its inception. 

The second "information problem" concerns the fate of information that interacts with an already-formed black hole. While sometimes discussed at the same time as the problem of the unitarity of the black-hole evaporation process, these two problems are actually quite separate. While Hawking radiation (according to the first alternative to information loss discussed above) might carry the imprint of the formation of the black hole, it is very unlikely that it carries information about the identity of particles that interact with the black hole horizon at late times. This problem turns out to be a problem in quantum channel theory: can information that is absorbed at the black hole event horizon be reconstituted by an inertial observer that can only observe radiation emanating from the black hole? Because the black hole is treated as a static quantity here (the "communication channel"), statements about the capacity of this channel do not address the unitarity of the evaporation process.

In this pedagocical review, I will address both problems using modern methods of quantum information theory. The problem of communicating via a black hole is solved by realizing that information is not lost within a black hole because a perfect copy of the information is always maintained outside of the event horizon, thanks to a well-known physical mechanism: the stimulated emission of particles at the event horizon that must accompany the spontaneous emission (i.e., Hawking radiation) in any consistent theory of black body radiation. It turns out that particles emitted via stimulated emission provide the "quantum hair"~\cite{Colemanetal1992} necessary for a conceptual understanding of macroscopic black holes. 

I will first present a simple statistical argument that appeared in the same year as Hawking's paper that points out that the process of stimulated emission was missing from the latter's discussion. I then discuss the curved-space quantum field theory derivation of stimulated emission in response to early- and late-time modes, and go on to show how to calculate the classical and quantum information transmission capacity of black holes. We will see that the capacity to transmit classical information with arbitrary accuracy over a quantum channel (the so-called "Holevo capacity") is always positive, meaning that information is never lost in black holes. The capacity to transmit quantum information turns out to vanish in come cases, but we'll see that this ensures that the laws of physics are {\em not} broken.

I then present an approach that allows us to move beyond the semi-classical description of black holes to illustrate how black holes might evaporate in a unitary manner, giving rise to the well-known Page curves. This approach is speculative in the sense that I have to assume a particular form for the interaction Hamiltonian of black holes with radiation modes, but is arguably less speculative than approaches that rely coupling gravity to dilaton fields or conformal field theories. Moreover, because the coupling between black-hole and Hawking/partner modes is exactly equivalent to the coupling of modes that governs optical parametric amplifiers, there is a chance that laboratory experiments can shed light on black-hole evaporation dynamics in a manner similar to black-hole analogue experiments. I close with speculations about observable consequences of stimulated emission in black holes. 

\section{Statistical Black Hole Thermodynamics}
\setcounter{figure}{0}
To set the stage (and my notation), I will repeat a simple "maximum entropy" argument due to Bekenstein~\cite{Bekenstein1975} to obtain the probability distribution $p(n)$ of quanta in any of $n$ outgoing modes emitted by a black hole via spontaneous emission (the distribution of Hawking radiation), using only Hawking's result that the mean number of outgoing quanta is~\cite{Hawking1975}
\be
\la n\ra=\frac{\Gamma}{e^{x}-1}\;. \label{hawking}
\ee
Here, $\Gamma$ is the absorption coefficient of the black hole (the ``gray-body factor", so that $1-\Gamma$ is the black hole's reflectivity), and $x=\hbar\omega/T_\bh$\footnote{In general, for a charged and rotating black hole, $x=\hbar\omega/T_\bh-\hbar m\Omega-\epsilon\Phi$, with $m$ the azimuthal quantum number, $\epsilon$ the electric charge, $\Omega$ the rotational frequency, and $\Phi$ the electrical potential. To keep matters simple, I will only treat uncharged non-rotating black holes here.}, where $\omega$ is the mode's frequency, and $T_\bh$ the black hole temperature. Bekenstein derived the result (for a single massless scalar bosonic mode $n$)
\be
p(n)=(1-e^{-\lambda})e^{-n\lambda}  \label{distrib}
\ee
simply by demanding that the entropy of the outgoing radiation
\be
S=-\sum_n p(n)\log p(n) \label{shannon}
\ee
is maximal, with the constraints $\sum_n p(n)=1$ and $\sum_n np(n)=\frac {\Gamma}{e^{x}-1}$ implemented via Lagrange multipliers. In Eq.~(\ref{distrib}), the Lagrange multiplier $\lambda$ is related to black hole parameters via
\be
e^{-\lambda}=\frac{\Gamma}{e^x-1+\Gamma}\;.
\ee
It should be noted that distribution (\ref{distrib}) (which was not given by Hawking) was also derived independently using full-fledged curved-space quantum field theory by Wald~\cite{Wald1975}.

\begin{figure}[htbp] 
\centering
   \includegraphics[width=0.75\textwidth]{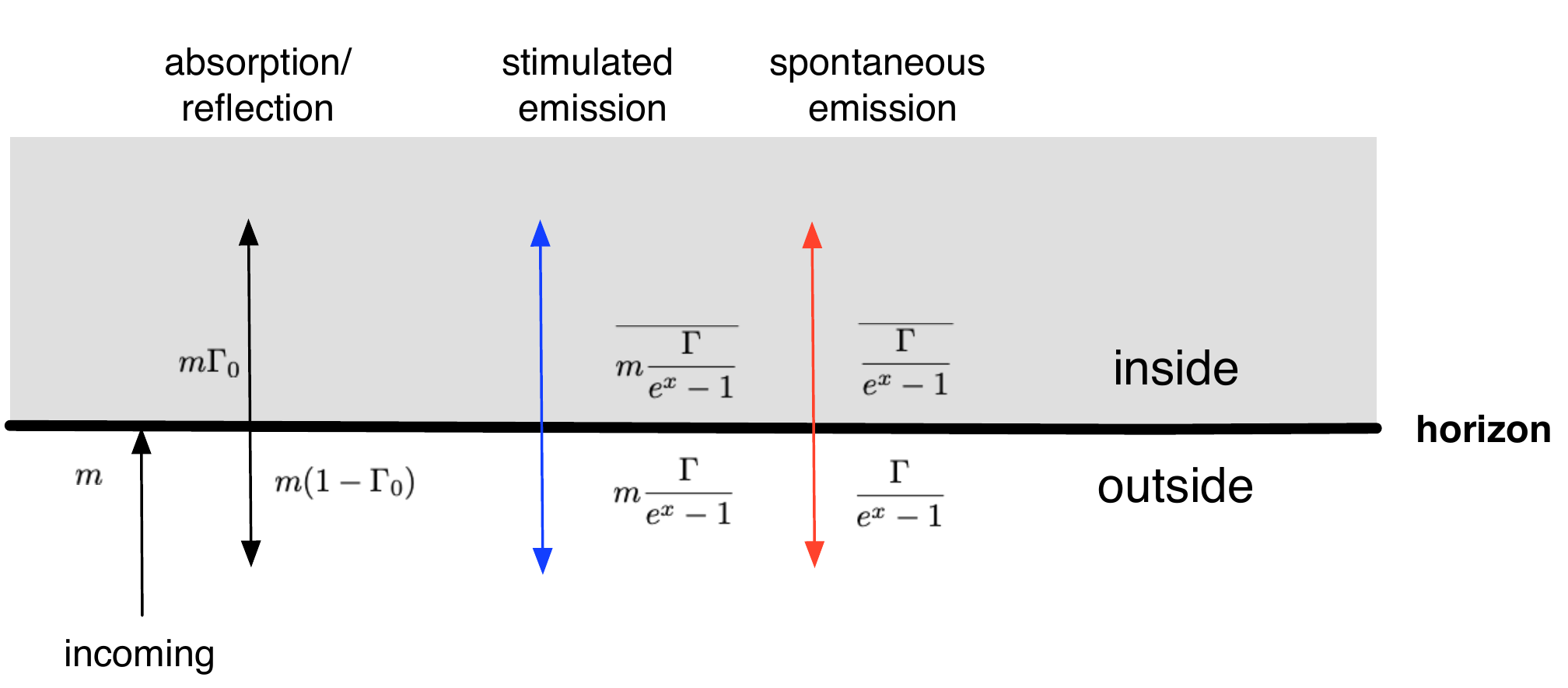} 
   \caption{Outgoing particles according to Eq.~(\ref{out2}) (arrows pointing down) for $m$ incident particles on the horizon. The terms $ \overline{m\frac{\Gamma}{e^x-1}}$ and $\overline{\frac{\Gamma}{e^x-1}}$ refer to anti-particles generated behind the horizon to conserve particle number, and will be discussed further below.}
   \label{fig1}
\end{figure}

Bekenstein now considered what would happen if a black hole that produces outgoing radiation with distribution $p(n)$ is immersed in a heat bath with temperature\footnote{This radiation has the distribution $p_\star(n)=(1-e^{-y})e^{-ny}$ where $y=\hbar\omega/T$ and $T$ the temperature of the heat bath.}  $T$. Together with his graduate student Amnon Meisels, Bekenstein found that the distribution for the ensuing outgoing radiation $p_o(n)$ was {\em not} compatible with the assumption that this radiation was composed only of spontaneous emission and radiation scattered from the surface of the black hole with a reflectivity $1-\Gamma_0$ (so that $\Gamma_0$ is the probability that a single quantum is absorbed by the black hole). Instead, they found that 
\be
p_o(n)=\sum_{m=0}^\infty p(n|m)p_\star(m)\;, \label{outdist}
\ee
where $p(n|m)$ is the probability distribution to observe $n$ outgoing particles {\em given} that $m$ such particles were incident on the black hole. This distribution is independent of the environment that the black hole finds itself in, and respects the detailed balance condition
\be
e^{-xm}p(n|m)=e^{-xn}p(m|n) \;.  \label{detbal}
\ee
This condition implies microscopic reversibility, and allows us to see that the mean number of outgoing particles $\la n\ra$ is given by the sum of Hawking's spontaneous emission term (\ref{hawking}) {\em and} an average fraction $m(1-\Gamma)$ 
returned outward
\be 
\la n\ra=\frac{\Gamma}{e^{x}-1}+m(1-\Gamma)\;. \label{out1}
\ee
By using the result that the fraction that is returned by pure scattering has to be $m(1-\Gamma_0)$ (where $1-\Gamma_0$ is the pure scattering reflectivity introduced earlier), Bekenstein and Meisels could show that
\be
\Gamma=\Gamma_0(1-e^{-x})\;.
\ee
This result shows that $1-\Gamma$ is in fact the {\em effective} reflectivity of the black hole, so that the black hole absorptivity $\Gamma$ is the sum of $\Gamma_0$ and a {\em negative} contribution due to stimulated emission $-\Gamma_0e^{-x}$, which is present in {\em all} modes.
Specifically, (\ref{out1}) can be rewritten to read
\be
\la n\ra=(1-\Gamma_0)m+(m+1)\frac\Gamma {e^x-1}\;, \label{out2}
\ee
where the $m$ in the second term on the right hand side of Eq.~(\ref{out2}) refers to the stimulated response to $m$ incoming particles, and the ``1" to the spontaneously emitted particles (Hawking radiation).  These different terms are indicated in Fig.~\ref{fig1}, which also shows the number of anti-particles generated inside of the black hole via spontaneous and stimulated emission (in order to conserve particle numbers, both spontaneous and stimulated emission occur via pair formation, as we will see later).

The exact form of the distribution $p(n|m)$ is complicated, and had to be inferred from Eq.~(\ref{outdist}) by a power series expansion. However, it was later shown to be {\em exactly equal} to the result obtained with a full treatment in curved-space quantum field theory by Panangaden and Wald~\cite{PanangadenWald1977}. The result was later confirmed by Audretsch and M\"uller~\cite{AudretschMueller1992}, who redid the calculation using wave packets, taking into account the red shift, and studying the effect of both incoming particles and anti-particles.

\section{Curved-space Semi-classical Quantum Field Theory}
\setcounter{figure}{0}
The results of Bekenstein and Meisels, Panangaden and Wald, and Audretsch and M\"uller did not convince everybody that information in black holes was preserved even though that work demonstrated microscopic reversibility. Schiffer, for example, argued that thermal radiation still overpowers stimulated emission for the vast majority of modes~\cite{Schiffer1993}. Furthermore, taking the red shift into account revealed that outgoing modes at observed frequencies $\sim1/8\pi M_\bh$ (where $M_\bh$ is the mass of the black hole in units where $G=c=k=\hbar=1$ as usual) ought to be due to incoming modes that were present just as the black hole was forming, and therefore must have been enormously blue-shifted with respect to the outgoing late-time radiation~\cite{Jacobsen1991}. Thus, it was not clear how those particular calculations confront the question of what happens to late-time particles absorbed by an already-formed black hole. I will address this last point in this section by introducing Sorkin's treatment of early- and late incoming modes, and then answer the first question (does stimulated emission really conserve information?) by explicitly calculating the capacity of the black hole to transmit information encoded in both early-time and late-time modes in the section that follows, using the standard methods of quantum information theory. 

To establish notation, I will first treat the simplest case: a perfectly absorbing black hole ($\Gamma=1$) and only early-time complex massless modes of energy $\omega_k$ ($\omega_{-k}$ for anti-particles). Even though we will later see that such a choice of absorptivity is inconsistent (and we will understand why), it is instructive to do this calculation first as it is simpler than the general case.

To introduce complex fields (which allow us to describe both particles and anti-particles) might at first glance seem like an unnecessary complication. Indeed, it is simpler (and still instructive) to study Hawking radiation using scalar fields only. However, because of the crucial role that negative-frequency modes play in this discussion, ignoring anti-particles (which are equivalent to particles traveling backwards in time) obscures some fundamental aspects of black hole physics, in which time-reversal invariance is key.

After the initial exposition of known (and therefore canonical, if not classical) results, I will introduce particles to the in-vacuum to understand how they stimulate the emission of particles in the out-vacuum outside of the horizon (and the emission of anti-particles beyond the horizon),  
and then introduce late-time modes using Sorkin's trick, allowing me to recover the gray-body absorptivity (and alleviate any worries about transplanckian frequencies at past infinity). 

\begin{figure}[htbp] 
   \centering
   \includegraphics[width=2in]{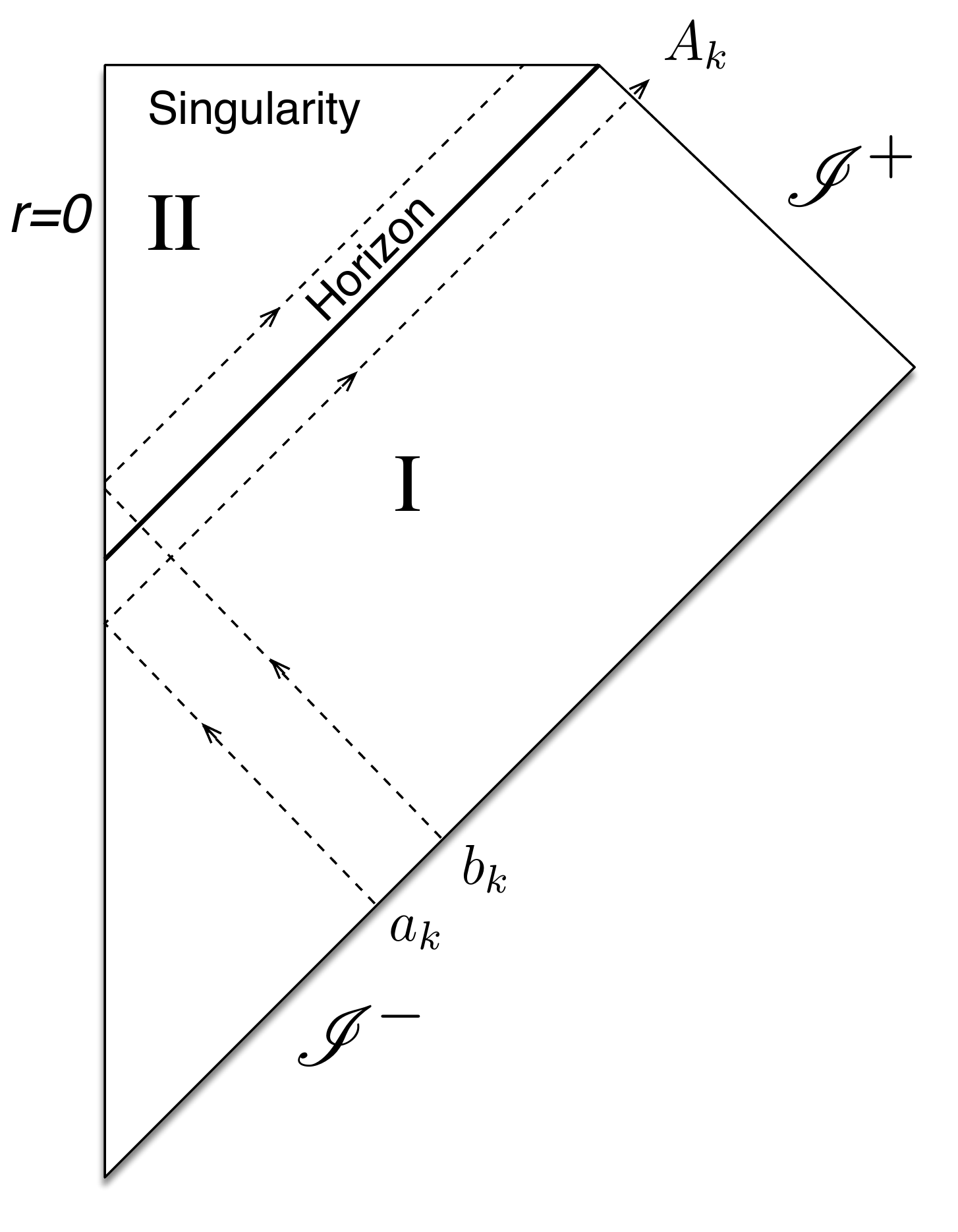} 
   \caption{Penrose diagram showing the early-time modes $a_k$ and $b_k$, created at past infinity ${\mathscr I}^-$ (just as the black hole formed) and traveling just outside and just inside of the event horizon towards future infinity ${\mathscr I}^+$. The outgoing mode at future infinity is annihilated by $A_k$.}
   \label{penrose}
\end{figure}

Consider the Penrose diagram in Fig.~\ref{penrose}.
The relation between the operators annihilating the incoming modes $a_k$ and $b_k$ and the outgoing mode $A_k$ is given by a Bogoliubov transformation (the annihilation and creation operators satisfy the commutation relations $[a_k,a_{k'}^\dagger]=[b_k,b_{k'}^\dagger]=\delta_{k,k'}$ and $[a_k,a_{k'}]=[b_k,b_{k'}]=[a_k,b^\dagger_{k'}]=0$)
\be
A_k=e^{-iH}a_ke^{iH}=\alpha_ka_k-\beta_kb^{\dagger}_{-k}\; \label{bol}
\ee
so that the unitary operator $U=e^{-iH}$ (I set $\hbar=1$) maps the in-vacuum to the out-vacuum:
\be
|0\ra_{\rm out}=e^{-iH}|0\ra_{\rm in}\;.  \label{out}
\ee
It is not an accident that I chose the letter $H$ for the Hermitian operator. In general, we can write the time-dependent mapping from in-states to out-states in terms of the time evolution operator
\be
U(t_2,t_1)=\mathsf{T}e^{-i \int_{t_1}^{t_2} H(t') dt'}\;,  \label{smatrix}
\ee
where $\mathsf{T}$ stands for Dyson's time-ordering operator and $H(t)$ is the Hamiltonian describing the unitary evolution of the quantum state. This time evolution operator can be approximated using $N$ small time slices $\Delta t$ so that with $t=N\Delta t$ 
\begin{equation}
U(t)=\mathsf{T}e^{-i\int_0^tH(t')dt'}\approx \prod_{i=1}^N e^{-i\Delta t H_i}, \label{unitary}
\end{equation}
where $H_i$ is the $i$-th time-slice Hamiltonian. In the static path approximation (one time slice) and absorbing $\Delta t$ into the interaction strength, the operator (\ref{unitary}) becomes the one implementing
(\ref{out}) with the Hamiltonian
\be
H=i\sum_{k=-\infty}^\infty g_k\big(a_k^\dg b_{-k}^\dg -a_kb_{-k}\big)\;. \label{ham1}
\ee 
where $g_k$ is an "interaction strength" that, as we will see, sets the black hole temperature. Using the Baker-Campbell-Hausdorff theorem we can relate the coefficients $\alpha_k$ and $\beta_k$ in (\ref{bol}) to $g_k$ via
\be
\alpha_k^2=\cosh^2 g_k\;, \beta_k^2=\sinh^2 g_k
\ee
and we have $\alpha_k^2-\beta_k^2=1$. The standard arguments of Hawking~\cite{Hawking1975} enforcing analyticity on the solutions to the free field equations\footnote{Despite the appearance of an interaction strength $g_k$, $H$ is a free-field Hamiltonian.} allow us to deduce that
\be
\alpha_k^2=e^{\omega_k/T_\bh}\beta_k^2\;, \label{ratio}
\ee
and relate $g_k$ to $T_\bh$ since $\frac{\omega_k}{T_\bh}\approx \log(g_k)+{\cal O}(g_k^2)$.
With these definitions out of the way, we can write down the out-vacuum state as
\be
|0\ra_{\rm out}=\prod_{k=-\infty}^\infty e^{g_k\big(a_k^\dg b_{-k}^\dg -a_kb_{-k}\big)}|0\ra_{\rm in}\;.
\label{out3}
\ee
Using the disentangling theorem for SU(1,1), we can evaluate (\ref{out3}) to become (writing the in-vaccum as the product state $|0\ra_{\rm in}=|0\ra_a|0\ra_b$)
\be
|0\ra_{\rm out}=\prod_{k=-\infty}^\infty \frac1{\cosh^2 g_k} \sum_{n_k,n_k'}e^{-(n_k+n'_{-k})\omega_k/2T_\bh}|n_k,n'_{-k}\ra_a|n_k',n_{-k}\ra_b\;. \label{outwf}
\ee
This is enough for us to recover the probability distribution of outgoing particles (\ref{distrib}) (albeit for the case $\Gamma=1$ as we do not treat reflection here), by calculating the density matrix of outgoing radiation (the radiation in "region I", see Fig~\ref{penrose}) via tracing out the interior of the black hole (region II)
\be
\rho_I=\Tr_{\rm II}|0\ra_{\rm out}\la 0|=\prod_k\rho_k\otimes\rho_{-k}\;.
\ee
As expected, the density matrix factorizes into a particle term and an anti-particle term with
\be
\rho_k=\frac1{1+\beta_k^2}\sum_{n=0}^\infty\biggl(\frac{\beta_k^2}{1+\beta_k^2}\biggr)^n|n_k\ra\la n_k|=(1-e^{-\omega_k/T_\bh})\sum_{n_k=0}^\infty e^{-n_k\omega_k/T_\bh} |n_k\ra\la n_k|\;. \label{mat}
\ee
This expression implies Bekenstein's (and therefore Hawking's) result for the single mode spontaneous emission probability Eq.~(\ref{distrib}), as $p(n)=\la n|\rho_k|n\ra$ (note that $e^{-\lambda}=e^{-x}$ for total absoption). The mean number of outgoing particles becomes
\be
\sum_{k=-\infty}^\infty  {}_{\rm out}\la0| a_k^\dagger a_k|0\rangle_{\rm out}=\sum_{k=-\infty}^\infty \beta_k^2\;,
\ee
which is the celebrated Planck distribution of Hawking radiation since
\be
\beta_k^2=\frac{e^{-\omega_k/T_\bh}}{1-e^{-\omega_k/t_\bh}}\;.
\ee  
Before considering the impact of particles entering the in-vacuum, let us take a closer look at the Hamiltonian (\ref{ham1}) that we used to map past-infinity states to future-infinity states. This Hamiltonian is, as a matter of fact, very common in quantum optics, where it is known as the "squeezing" Hamiltonian that describes optical parametric amplification (generally, all quantum amplification processes can be described by Bogoliubov transformations~\cite{Leonhardt2010}). 
Quantum amplification is inherently a nonlinear process. In the simplest system, a {\em pump} photon with frequency $\omega_p$ is converted into two photons, called the {\em signal} and {\em idler}, with frequencies $\omega_s$ and $\omega_i$, where $\omega_p=\omega_s+\omega_i$. This process, called parametric downconversion, creates entangled photon pairs via the Hamiltonian (for a single mode)
\be
H_{\rm OPA}=i\eta(a^\dagger_sb^\dagger_i-a_sb_i)\;, \label{ham_opa}
\ee
where $\eta$ is the coupling strength (which depends on the pump amplitude) and $a^\dagger_s$ and $b^\dagger_i$ are the creation operators for the signal and idler modes, respectively ("OPA" stands for "optical parametric amplification"). If we compare Hamiltonian (\ref{ham_opa}) to Eq.~(\ref{ham1}), we see that the role of positive and negative frequency modes of Hawking radiation are here played by the signal and idler modes, and indeed the wavefunction of the signal-idler pair is simply~\cite{Nationetal2012}
\be
|\Psi(t)\ra=\frac1{\cosh \eta t}\sum_{n=o}^\infty e^{-\eta t n/2}|n_i\ra|n_s\ra\;, \label{outwf1}
\ee
which we can compare to Eq.~(\ref{outwf}). In contrast to (\ref{outwf}) that has both particles and anti-particles, the wave function for the signal/idler mode is that of a real (rather than a complex) scalar field and is explicitly time-dependent.
Other quantum amplification processes that can be described by Bogoliubov transformations are the Unruh effect~\cite{Unruh1976} (particle creation by an accelerated observer) and the dynamical Casimir effect~\cite{Moore1970} (particle creation by oscillating mirrors), but only in the Unruh effect and Hawking radiation are the "signal" and "idler" modes causally disconnected. So, while the mathematics of mapping "in" to "out" states is similar, the physics of the processes can be quite different.

Bekenstein and Meisels found that (just as Einstein had derived~\cite{Einstein1917}) not only does the vaccum emit radiation spontaneously, it can also be stimulated to do so. To see this effect in curved-space quantum field theory, we need to consider $m$ particles in the initial state 
\be
|\psi\ra_{\rm out}=e^{-iH} |m\ra_a|0\ra_b=e^{-iH} \frac1{\sqrt{m!}}(a_k^\dagger)^{m}|0\ra_{\rm in}= \frac1{\sqrt{m!}}(A_k^\dagger)^{m}e^{-iH}
|0\ra_{\rm in}\;.
\ee
with $H$ from Eq.~(\ref{ham1}) and using the Bogoliubov transformation (\ref{bol}).
Carrying out this calculation for $m_k$ particles in a single mode $k$ incident on the black hole leads to an outgoing density matrix in region I (outside the black hole horizon)~\cite{AdamiVersteeg2014}
\be
\rho_{\rm I}=\Tr_{\rm II} |\psi\ra_\out \la \psi |=\rho_{k|m}\otimes \rho_{-k|0}\;. \label{dens1}
\ee
Here, the density matrix $\rho_{k|m}$ of outgoing particles given that $m$ particles were incident in mode $k$ is  (the density matrix of outgoing anti-particles $\rho_{-k|0}$ looks just like the particle matrix with no incoming particles, shown in (\ref{mat}))
\be
\rho_{k|m}=\frac1{(1+\beta_k^2)^{m+1}}
\sum_{n=0}^\infty\left(\frac{\beta_k^2}{1+\beta_k^2}\right)^{n}{{m+n}\choose{n}}|m+n\ra\la m+n|\;,\ee
which can be rewritten as 
\be
\rho_{k|m}=\sum_{n=0}^\infty p(n|m)|n\ra\la n| \label{denmat}
\ee
with the conditional probability (here $x=\frac{\omega_k}{T_\bh}$ as before)
\be
p(n|m)=(1-e^{-x})^{m+1}{{m+n}\choose{n}}e^{-nx}\;. \label{condprob}
\ee
If there are no particles entering the black hole ($m=0$), Eq.~(\ref{condprob}) reduces to 
$p(n)=(1-e^{-x})e^{-nx}$, which is the $\Gamma\to 1$ limit of  (\ref{distrib}).
However, it is easily checked that the detailed balance condition (\ref{detbal}) does not hold for the conditional probability (\ref{condprob}). We will now see that this is due to our setting $\Gamma=1$, which will turn out to be inconsistent: if there is stimulated emission, then the black hole cannot be fully absorbing: it {\em must} return something to the outside world.

\begin{figure}[htbp] %
   \centering
   \includegraphics[width=2in]{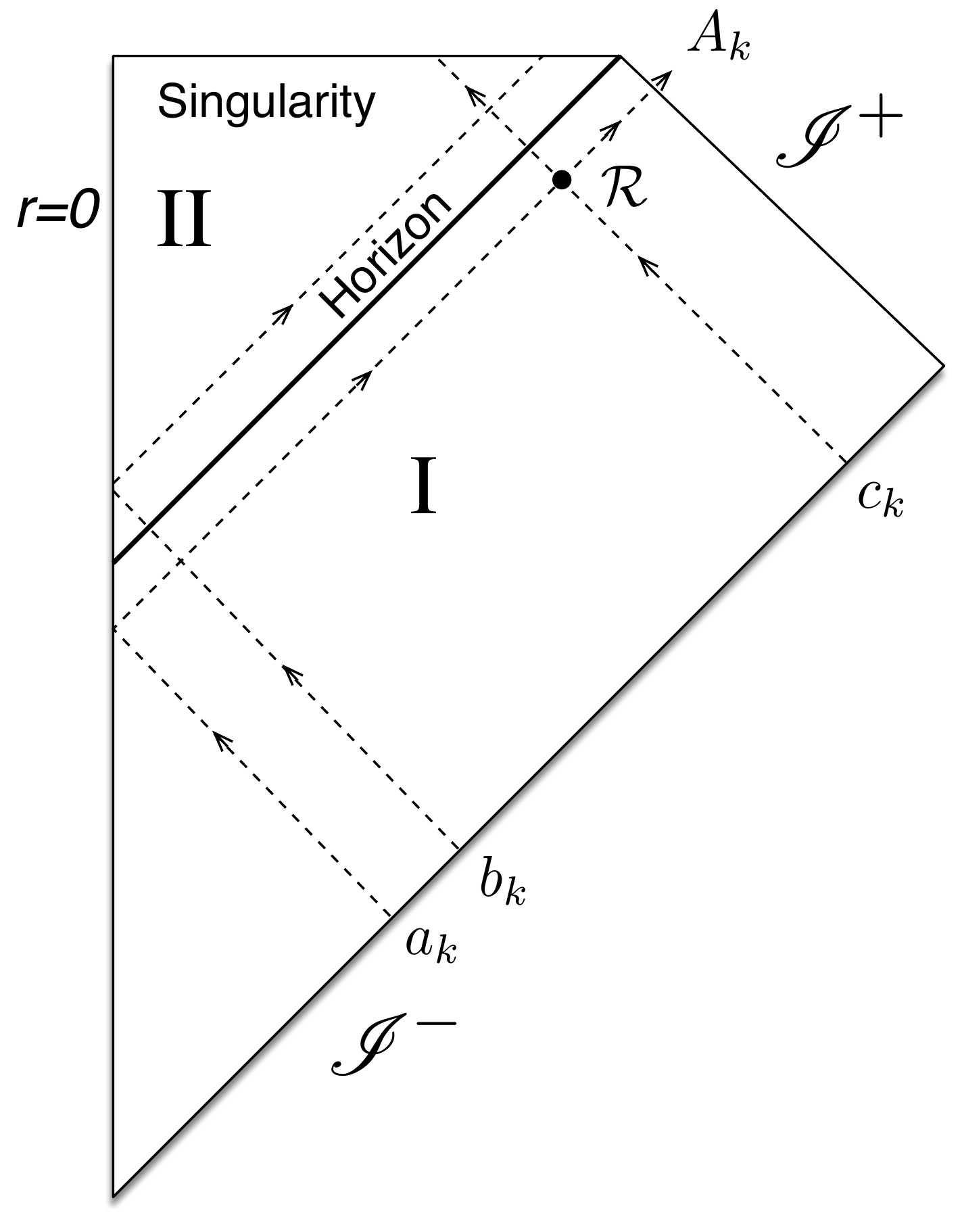} 
   \caption{Penrose diagram showing early-time modes ($a_k$ and $b_{-k}$) and late-time modes $c_k$. Late-time modes are scattered at the horizon with probability ${\cal R}$ (the black hole reflectivity). A perfectly absorbing black hole has ${\cal R}=0$.}
   \label{penrose1}
\end{figure}

In order to correctly describe scattering off of the black hole horizon in curved-space quantum field theory, I will use a trick due to Sorkin~\cite{Sorkin1987}, which will ultimately allow us to recover Bekenstein and Meisels' $p(n|m)$ that observes detailed balance. Sorkin's insight comes from the observation that when we discuss the capacity of a black hole to transmit information, we are not really interested in the information that was encoded in the particles that were present during the formation of the black hole (the modes $a_k$ and $b_{-k}$). After all, whether a collapsing star is going to form a black hole in the future is uncertain, and choosing the timing of informational modes in such a way that they travel just outside of the black hole would be rather difficult. Besides, we know that such modes will be exponentially redshifted. Sorkin instead introduces {\em late-time} modes $c_k$ that at future infinity are exponentially blue-shifted\footnote{Note that in order to keep with the previous definition of $a_k$ and $b_{-k}$ modes, I have changed the nomenclature of ~\cite{Sorkin1987}.}  with respect to the early-time modes $a_k$ and $b_{-k}$, and therefore commute with them. Sorkin's late-time mode along with the early-time modes are shown in Fig.~\ref{penrose1}.

The Bogoliubov transformation that connects the outgoing mode $A_k$ to $a_k$ and the late-time mode $c_k$ can now be written as~\cite{Sorkin1987}: 
\be A_k &=& e^{-iH}a_k e^{iH} = \alpha_k
a_k - \beta_k b^\dagger_{-k}+ \gamma_k c_k\;,\label{bogol} \ee
with $\alpha_k^2-\beta_k^2+\gamma_k^2=1$ to ensure unitarity. What is the Hamiltonian $H$ that gives rise to this transformation? It turns out that it is given by the sum of the term we already had (which turned out to be formally equivalent to the Hamiltonian of an active optical element) and the Hamiltonian of a {\em passive} optical element: a beam splitter~\cite{Leonhardt2010}:
 \be
 H = \sum_{k=-\infty}^\infty ig_k(a_k^\dagger b_{-k}^\dagger - a_k b_{-k} )+ig^\prime_k (a_k^\dagger c_k- a_k c_k^\dagger)  \;.\ \ \ \  \ \ \label{ham2}
 \ee
The beam-splitter term in (\ref{ham2}) describes the interaction of late-time modes with the black hole horizon using an interaction strength $g_k^\prime$, which we will be able to relate to the black hole's reflectivity ${\cal R}$.

The Bogoliubov coefficients
$\alpha_k$, $\beta_k$, and $\gamma_k$ can be written in terms of $g_k$
and $g^\prime_k$ by using (\ref{ham2}) in (\ref{bogol}),  
\be
\alpha_k&=& \cos(\gp_k w) \label{alpha}\;,\\
\beta_k&=&\frac{g_k}{\gp_k}\frac{\sin(\gp_k w)}{w}\; \label{beta}\;,\\
\gamma_k&=&-\frac{\sin({\gp_k w})}{w} \label{gamma}\;,
\ee 
where
$w=\sqrt{1-(g_k/\gp_k)^2}$. 

We can now proceed as before and construct the out-state using this amended Hamiltonian 
\be
|0\ra_\out=e^{-iH}|0\ra_a|0\ra_b|0\ra_c\;,  \label{unitary-sorkin}
\ee
now acting on a product state of early-and late-time modes (as we have assumed these modes commute). We can calculate the density matrix of radiation outside the black hole horizon by tracing the full density matrix over region II (the inside of the black hole that contains both modes $b_{-k}$ and $c_k$)
\be
\rho_{\rm I}=\Tr_{\rm II}|0\ra_\out\la 0| \;.
\ee
Just as before, the anti-particle density matrix $\rho_{-k|0}$ factorizes, and we find 
\be 
\rho_{k|0}=\frac1{1+\beta_k^2}\sum_{n_k=0}^\infty
\left(\frac{\beta_k^2}{1+\beta_k^2}\right)^{n} |n_k\ra\la n_k|\;. \label{rho0}
 \ee
This expression is formally identical to expression 
(\ref{mat}) when written in terms of $\beta_k^2$, except that due to the term $\gamma_k^2$ in the unitarity relation $\alpha_k^2-\beta_k^2+\gamma_k^2=1$ we now have
\be
\beta_k^2= \frac{\Gamma}{e^{\omega_k/T}-1}\;, \label{beta2}
\ee
where $\Gamma$ is the absorptivity of the black hole (for this particular mode), with $\Gamma=1-\gamma_k^2$.
Thus, (\ref{rho0}) is just the standard density matrix of spontaneous emission (a.k.a. Hawking radiation) including gray-body factors.  It reduces to (\ref{mat}) in the unphysical limit $\Gamma\rightarrow1$. 

Now that we have seen how we can use Sorkin's trick along with a simple beam-splitter term to recover the gray-body factor in Hawking's radiation (note that Hawking obtained this factor in a very different manner, by following particles from inside the black hole backwards in time into region I), we can study how the black hole reacts to particles that fall into the black hole at late times, in mode $c_k$. Note that because these modes will {\em not} be significantly red-shifted by the black hole, the corresponding Hawking radiation will be fully commensurate with the size of the black hole and a "transplanckian" problem is avoided. 

I will now construct the outgoing state $|\psi\ra_\out$ when $m_k$ late-time particles are directed at the horizon (all in the same mode $k$), 
 \be
|\psi\ra_\out = e^{-iH}|m_k\ra_\inv\;, \label{state}
 \ee 
with $H$ from (\ref{ham2}) and where $|m_k\ra_\inv=|0\ra_a|0\ra_b|m_k \ra_c$, i.e., the state with $m_k$ incoming particles in mode $c_k$ on ${\mathscr I}^-$.
Using Eq.~(\ref{bogol}), we can immediately calculate the number of
particles emitted into mode $A_k$ if $m_k$ were incident in mode $c_k$, since
 \be {_\inv}\la \psi|A_k^\dagger A_k|\psi\ra_\inv=\beta_k^2 +m_k\gamma_k^2 \;. \ \ \ \ \ \ \label{partnum}
\ee 
Using $\gamma_k^2=1-\alpha_k^2 +\beta_k^2$, we can see that we have recovered Bekenstein and Meisels' result (\ref{out1}), using (\ref{beta2}) and identifying $\gamma_k^2=1-\Gamma$. 
Furthermore, we can also recover 
\be
\Gamma=\Gamma_0(1-e^{-\omega_k/T_\bh})\;,
\ee
by noting that $\alpha_k^2=\Gamma_0$ actually sets the reflectivity of the black hole, that is, ${\cal R}$ in Fig.~\ref{penrose1} is just $1-\alpha_k^2$. We can also see that on account of Eqs.~(\ref{alpha}-\ref{gamma}), we have
\be
\biggl(\frac{g_k^\prime}{g_k}\biggr)^2=\frac{\gamma_k^2}{\beta_k^2}=1+\frac{1-\alpha_k^2}{\alpha_k^2}e^{\omega_k/T_\bh}\;,
\ee
which implies that because the reflectivity ${\cal R}=1-\alpha_k^2=\Gamma_0\leq1$, the parameter  $g_k^\prime$ that sets the scattering rate for modes of energy $\omega_k$ is bounded from below by $g_k$, which is itself bounded from above by 1 since $g_k\sim e^{-\omega_k/T_\bh}$. 

This explains why setting $\Gamma=1$ could never be consistent. As previously discovered by Bekenstein and Meisels using maximum entropy arguments only~\cite{BekensteinMeisels1977}, the total absorptivity of the black hole $\Gamma$ is the product of the "bare" absorptivity $\Gamma_0$ times the factor $1-e^{-\omega_k/T_\bh}$ due to stimulated emission, so that $\Gamma<1$ always. A "classical" black hole has $\Gamma_0=1$ (for incoming $s$-waves) but a quantum black hole can never be totally black, and as a consequence information is never lost in black hole dynamics. To make this statement quantitative, we should proceed to calculate the capacity of quantum black holes to transmit classical information, using the tools of quantum information theory.

\section{Classical Information Capacity of Quantum Black Holes}
\setcounter{figure}{0}
The primary application of the classical theory of information due to Shannon~\cite{Shannon1948} was to quantify the capacity of channels to transmit information. One of the surprising results of that theory is that it is possible to send information through a channel with perfect accuracy even when there is substantial noise that affects the transmitted message. For example, we might imagine the problem of communicating by throwing copies of two different books into a fire repeatedly: books that have exactly the same number of words, and weigh the same\footnote{I choose books that differ only in their words but not in their mass to connect with a paradox where, throwing these books into a black hole, the identity of the books would be lost because the increase in the mass of the black hole is the same for either book.}. The two different books allow us to encode information into a sequence of zeros and ones (denoting which of the books we choose to incinerate). But how could an individual that can only observe the flames extract the information encoded in the series of books? The answer is that the information is not lost in the flames: the two different books when burning give rise to slightly different ways in which the flames and smoke behave when turning the pages to ashes (due to the different ways in which words are arranged on the pages). While we certainly do not have the technology to detect these differences among the much more pronounced variation they are embedded in, this information is in principle accessible. And as a consequence, error correction techniques will allow us to retrieve this information with perfect accuracy. One way to do this is to coat the pages of the books with different pyrotechnic colorants for each, making information retrieval trivial. 

Classical black holes, however, are very different from fire. If no marker carrying the information is available to the observer (as it is in the case of communication through fire), no amount of error correction can recover the information. Because Schwarzschild black holes (classical, non-spinning, neutral black holes) are only characterized by their mass, absorbing two books with different writing but equal mass would give rise to the same exact final state. Such dynamics would dictate that two separate phase-space trajectories merge into one, which is a direct violation of microscopic reversibility: an abomination.

In 1973, Bekenstein painted the picture of information loss that is still being discussed today~\cite{Bekenstein1973}: 
\begin{quote}``We imagine a particle goes down a (...) black hole. As it disappears some information is lost with it".
\end{quote}
In fact, Bekenstein estimated that the information loss must be at least one bit, which is the amount of uncertainty created by not knowing whether the particle still existed behind the horizon or not. It is now clear that this line of thinking is fundamentally rooted in a misunderstanding of the concept of information in physics, namely, that information is necessarily attached to the object that encodes it. Information, rather, is a relative state between an observer and a system~\cite{Adami2016}, and more importantly, is {\em not} tied to the encoding body. In the case of the fiery communication described earlier, the information was first encoded in ones and zeros, then translated to two kinds of books, and then retrieved into a list of ones and zeros after the color of the flames was decoded. In communication through black holes, information is first encoded in particles (for example, the polarization of photons, or particle/anti-particle identity). The particles are absorbed at the event horizon, which stimulates the emission of {\em exact copies} of those particles outside (and inside) the horizon (I will discuss how such a process complies with the no-cloning theorem in section {\ref{cloning}). The process of stimulated emission
copies the information from the absorbed particle and {\em transfers} it to other carriers outside the black hole, where they are accessible to an observer so that separate phase-space trajectories remain separate.

\subsection{Classical late-time capacity}
We can follow the fate of information
interacting with a black hole by using a {\it preparer} to encode
information into late-time quantum states that are then sent into the event
horizon. For example, we can imagine a preparer $X$ who sends
packets of $n$ particles with probability $p(n)$. The internal
state of the preparer can be described by the density matrix
$\rho_X=\sum_n p(n) |n\ra\la n|$, with entropy
$S(\rho_X)=H[p]=-\sum_n p(n) \log p(n)$. After the particles
interact with the black hole, our preparer is now correlated with
it because the final state is now the density matrix \be
\rho_{\,{\rm I,II},X}=\sum_n p(n) |\psi_n\ra\la \psi_n|\otimes |n\ra_X\la n|\;, 
\ee 
where $|\psi\ra_n=e^{iH} |0,0,n\ra_{abc}$.
Tracing over the black hole interior (region II), we obtain
the joint density matrix of the radiation field in region I and
the preparer $X$: \be \rho_{\,{\rm I},X}&=& \sum_n p(n)
\rho_{k|n}\otimes |n\ra_X\la n| \label{block} \ee with entropy

\be S(\rho_{\,{\rm I},X})= H[p]+\sum_n p(n) S(\rho_{k|n}) \ee owing to
the block-diagonal form of Eq.~(\ref{block}). The mutual entropy
between radiation field and preparer is then simply given by \be
H(X;{\rm I}) &=& S(\rho_{\,\rm I}) +S(\rho_X)-S(\rho_{\,{\rm I},X})\nonumber \\
&=&S\left(\sum_n p(n)\rho_{k|n}\right)-\sum_n p(n) S(\rho_{k|n})\;,\ \  \ee which turns out to be
the Holevo bound~\cite{Holevo1973}. The latter constitutes the maximum amount of
classical information that can be extracted from a quantum measurement, and its maximum (over the probability distribution of signal states) is the capacity of a quantum channel to transmit classical information~\cite{Holevo1998}. In other words, the black hole channel capacity {\em is} the Holevo capacity, with Hawking radiation playing the role of channel noise.

As a simple example, consider a binary channel where the
preparer either sends no particle (with probability $1-p$) or one
particle (with probability $p$) into the black hole. As Hawking radiation 
does not depend on this decision, the information would be lost if stimulated and scattered particles were not present in the total radiation field outside the horizon. In order for information to be recovered, an outside observer must be able to make measurements on the radiation field that betray the preparer's decision. If we send in $n$ particles in mode $k$ (to signal a logical `1') then we obtain $\rho_{k|n}$  in region I, which is diagonal in the number basis (its general expression is given in the Appendix). 
For a single incident particle, 
\be
\rho_{k|1}&=&\frac{\alpha_k^2}{(1+\beta_k^2)^2}\sum_{n_k=0}^\infty 
\left(\frac{\beta_k^2}{1+\beta_k^2}\right)^{n_k}\!\!\!(1+n_k 
\frac{\gamma_k^2}{\alpha_k^2 \beta_k^2}) |n_k\ra\la n_k|\;.  \label{rho1}
 \ee
This density matrix is clearly non-thermal except in the unphysical limit $\gamma_k=0$, 
where the probability of absorbing a single quantum would exceed 1. 

Let us calculate the capacity $\chi=\max_p H(X;{\rm I})$ for the worst-case scenario: a perfectly black hole (no reflection, i.e., $\alpha_k^2=1$). This is the worst-case scenario because any radiation reflected at the horizon would allow us to recover the information sent in even if the vast majority of particles disappear behind the horizon. That observation illustrates the true magic of Shannon's theorem: information is only lost if absolutely {\em no} trace of the information is left for us to decode.

For the non-reflecting black hole, the mutual entropy $H(X;{\rm I})$ is maximized at $p=1/2$, and the capacity can be written in terms of the parameter $z=e^{-\omega_k/T_\bh}$ as~\cite{AdamiVersteeg2014}
\be
\chi=1-\frac1{2(1+z)^3}\sum_{m=0}^\infty\biggl(\frac z{1+z}\biggr)^m (m+1)(m-2z)\log(m+1)\;. \label{latecap}
\ee 
The capacity $\chi$ is positive for all values $0\leq z\leq1$, which implies that the channel's capacity never vanishes, and information can always be recovered with perfect accuracy (see Fig.~\ref{cap}). 
\begin{figure}[htbp] 
   \centering
   \includegraphics[width=3in]{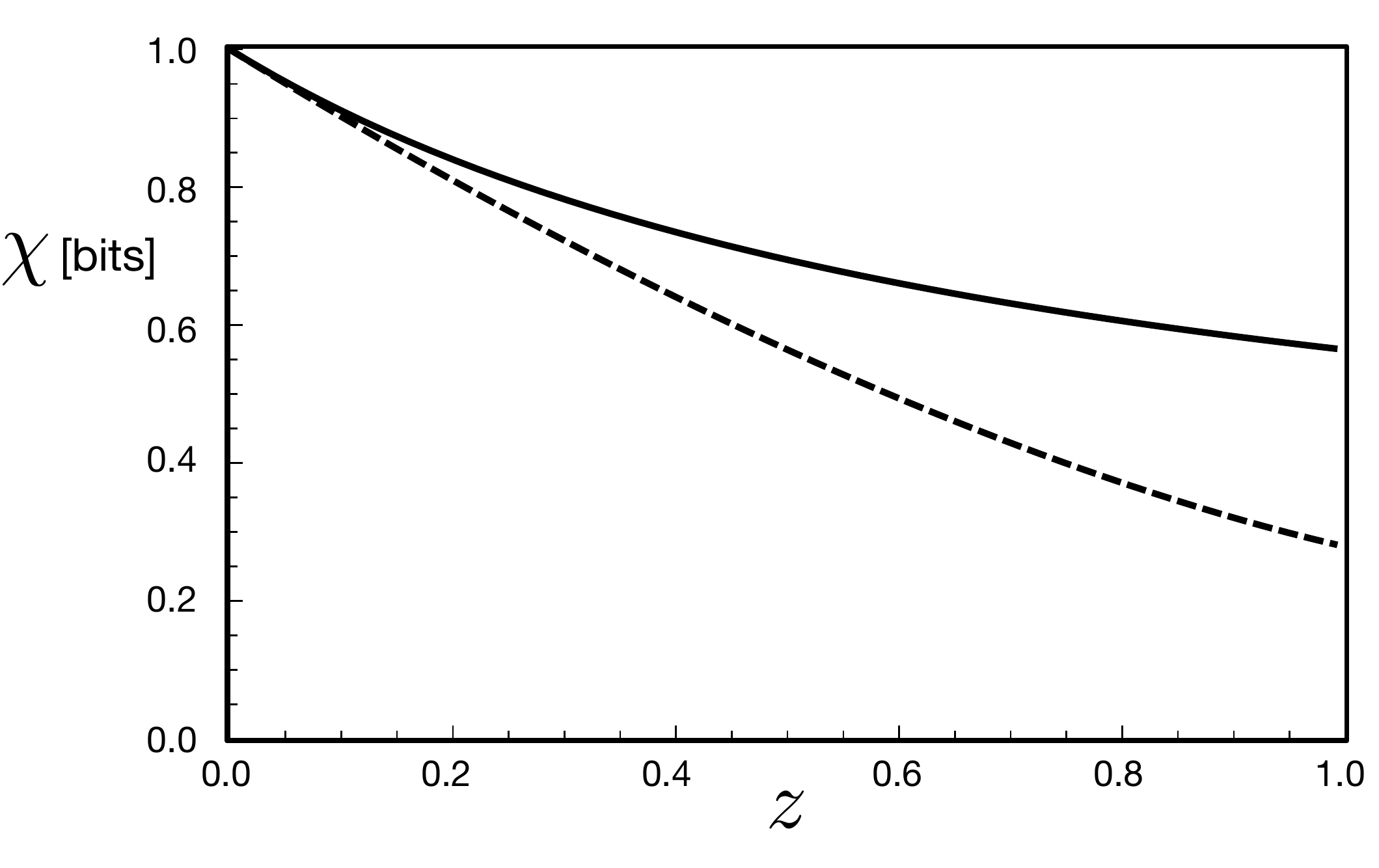} 
   \caption{Capacity $\chi$ for a binary non-reflecting black hole channel as a function of the parameter  $z=e^{-\omega_k/T_bh}$. $z=0$ corresponds to a black-hole with vanishing mass, while $z\to1$ as the mass of the black hole tends to infinity. The solid line represents the late-time capacity (\ref{latecap}). The dashed line is the capacity for early-time modes, which is obtained from the late-time capacity by rescaling $z\to\frac z{z-1}$~\cite{AdamiVersteeg2014}. Note that since $g_k\approx z$, we can see this plot also as depicting the dependence of the information transmission capacity on the mode coupling strength in the black-hole Hamiltonian (\ref{ham1}).}
   \label{cap}
\end{figure}
As the black hole shrinks (and the Hawking temperature increases), the capacity to transmit information (using particles with a given energy $\omega_k$) decreases. But a decreased capacity does not indicate information loss. Instead, it tells us what the maximum rate of {\em perfectly accurate} information transmission is. Thus, a capacity of 0.8 bits, for example, indicates that if we are sending in information at 1 bit per second, we can only retrieve perfectly accurate information at the rate of at most 0.8 bits per second, on account of the error correction necessary to protect the information. Non-optimal schemes of error correction will yield smaller rates of error-free information transmission.  

Incidentally, using the conditional probability for perfect absorption of {\em early-time} modes in Eq.~(\ref{condprob}) still leads to a non-vanishing capacity, though significantly reduced (see Fig.~\ref{cap}). It turns out that this early-time capacity was previously (and independently) derived for the Unruh channel (the case of accelerated observers, which is another case of quantum amplification). When replacing the black hole interaction strength $g_k$ with the acceleration $r$ (and keeping $g_k^\prime=g_k$ to ensure zero reflectivity), the black hole and the Unruh capacities are identical~\cite{Bradler2011}. In both cases, using the probability distribution $p(n|0)$ for spontaneous emission (pure Hawking radiation) yields a vanishing capacity. This quantifies what has been conjectured many times before: Hawking radiation carries no signal, no information. It is only noise.

The capacity of the binary channel using an encoding of $n$ particles as the logical `1', and either 0 particles or $n$ anti-particles as the logical '0' exceeds the one shown in Fig.~\ref{cap} (as the adventurous reader can confirm using the expressions in the Appendix) because these methods of encoding are significantly more robust to noise, that is, to the interference of the stimulated emission signal with the spontaneously emitted Hawking radiation. We will use such an encoding in the following section when discussing in which way a black hole can be understood in terms of quantum cloning machines.

\subsection{Black holes are quantum cloning machines} \label{cloning}
I wrote earlier that the stimulated emission process essentially "clones" the incoming particles so that these copies are available outside of the event horizon and information is not lost. Perfect cloning is, of course, forbidden in quantum mechanics: the "quantum no-cloning theorem" is a direct consequence of the linearity of quantum mechanics (see, for example, ~\cite{Dieks1982,WoottersZurek1982}). The proof of this theorem is simple, and worth repeating here. Suppose we define a cloning operator $U_C$ in quantum physics so that it will copy arbitrary quantum states $|\phi\ra$
\be
U_C|\phi\ra|0\ra=|\phi\ra|\phi\ra\;  \label{copy}
\ee
onto a prepared ancilla state $|0\ra$. After the copying operation, the original quantum state is in a product state with its copy, as desired. The action of this operator on a quantum superposition $\sigma |\phi\ra+\tau|\psi\ra$ (with complex $\sigma$ and $\tau$ that satisfy $|\sigma|^2+|\tau|^2=1$) does not produce a product state of that superposition, however, as
\be
U_C(\sigma|\phi\ra+\tau|\psi\ra)|0\ra=\sigma|\phi\ra |\phi\ra\ +\tau|\psi\ra|\psi\ra\neq(\sigma|\phi\ra+\tau|\psi\ra)(\sigma|\phi\ra+\tau|\psi\ra)
\ee
instead. In fact, this hypothetical "cloning operator" $U_C$ in (\ref{copy}) turns out to be an "entanglement operator", and is typically (in the space of qubits) the "controlled NOT" (CNOT) operator $U_C=P_0\otimes\one+P_1\otimes\sigma_x$, where $P_0$ and $P_1$ project on the respective states, and the Pauli matrix $\sigma_x$ flips a bit. In the following I discuss the cloning of binary states (quantum bits, or qubits), but the formalism can be extended to quantum states of arbitrary dimension~\cite{CerfFiurasek2005}.

While perfect cloning of arbitrary states is not possible, it is of course possible to clone "known" (that is, prepared) states, with the help of an operator as described above that uses projectors onto those known basis states. It is, however, also possible to make "approximate" copies of arbitrary quantum states, using unitary operators that are referred to as "quantum cloning machines".
A quantum cloning machine is designed to maximize the probability that a quantum state $|\psi\rangle$ (or, in general, $N$ identically prepared states $|\psi\ra^{\otimes N}$) is cloned into a state $|\psi_{\rm out}\ra=U|\psi_{\rm in}\rangle|0\ra |R\ra$ (or $M$ copies of that state, with $M>N$) with a unitary transformation $U$ acting on the input state $|\psi_{\rm in}\ra$, a "blank" state $|0\ra$ that will ultimately hold those multiple copies of the cloned state, and an ancillar state $|R\ra$. The formalism of quantum cloning machines was introduced by Buzek and Hillery~\cite{BuzekHillery1996}, and has given rise to a large body of work (see, e.g., the reviews~\cite{CerfFiurasek2005,Scaranietal2005}). 
\begin{figure}[htbp] 
   \centering
   \includegraphics[width=2in]{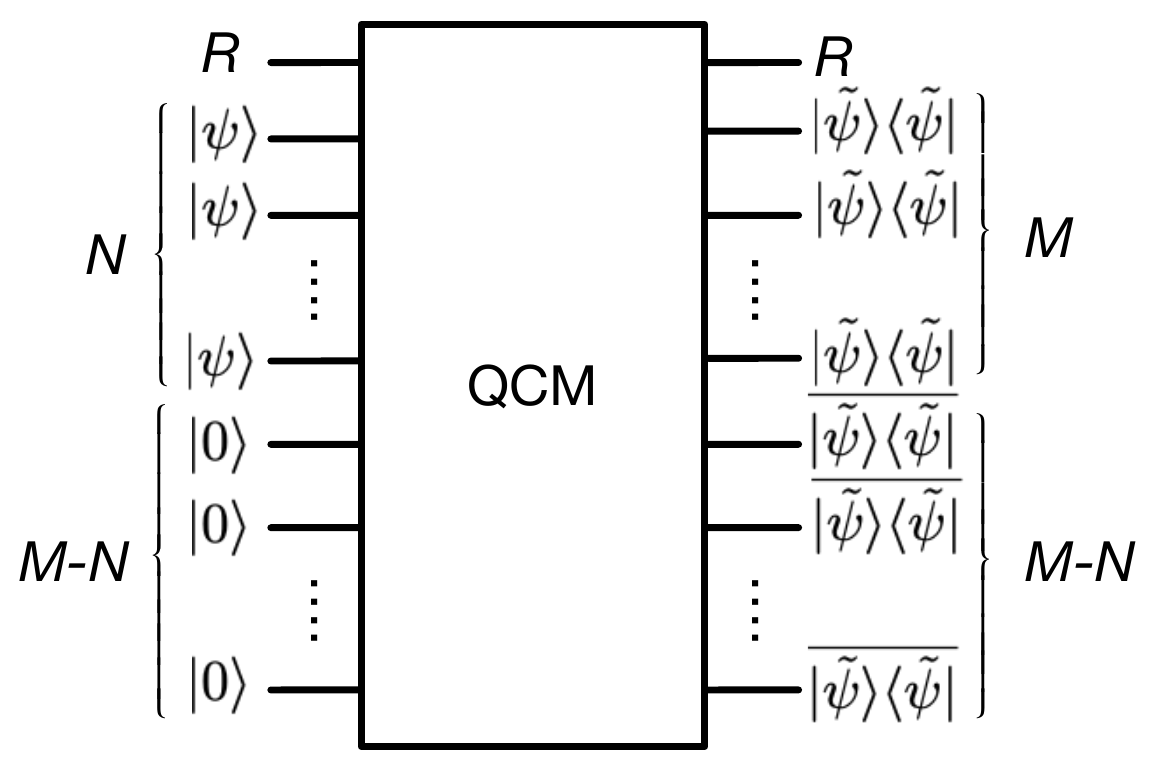} 
   \caption{Schematics of an $N\to M$ quantum cloning machine ($M>N$). The machine receives as input the product of $N$ identical copies of the quantum state, supplemented by an internal state $R$ and $M-N$ states in a prepared blank state $|0\rangle$. The $M$ approximate clones are mixed states $|\tilde \psi\ra\la\tilde \psi|$, while the $M-N$ anti-clones are the output of an approximate logical NOT transformation of the clones, indicated by  $\overline{|\tilde \psi\ra\la\tilde \psi|}$. }
   \label{fig:cloner}
\end{figure}
For the general case of an $N\to M$ cloning machine (see Fig.~\ref{fig:cloner}),  the accuracy of the cloning process is determined by a {\em fidelity} measure, defined as the expectation value 
of one copy of the $M$ cloned states $\rho_{\rm out}^j$ evaluated in the basis of the input state
\be
F_j=\mbox{}_{\rm in}\la\psi|\rho^j_{\rm out}|\psi\ra_{\rm in}\;.  \label{fidelity}
\ee
The largest possible  $N\to M$ cloning fidelity of universal quantum cloning machines (devices that copy any input state $|\psi\ra$ with the same fidelity) is~\cite{GisinMassar1997,Brussetal1998}
\be
F_{\rm opt}=\frac{M(N+1)+N}{M(N+2)}\;. \label{optfidel}
\ee
For example, the optimal (and therefore maximal) fidelity of a $1\to 2$ approximate cloning machine is $F_{1\to2}=5/6$. In the limit $M\to \infty$, the cloning fidelity approaches $\frac{N+1}{N+2}$, which happens to be the fidelity of the best possible state preparation via state estimation (using classical information only) from a finite quantum ensemble~\cite{MassarPopescu1995}.
Quantum cloning machines that reach the cloning limit are called "optimal cloners", and can be constructed using simple quantum optical elements (parametric amplifiers and beam splitters) for discrete states such as polarization~\cite{Simon2000}, but also for continuous variables~\cite{Fiurasek2001,Braunsteinetal2001}.

Because black holes stimulate the emission of copies of accreting particles, we can seek to apply the formalism of quantum cloning machines to black holes to answer the question: How well do black holes clone quantum states? We will see that the answer is "almost perfectly". 

Since the cloning transformation that takes $|\psi\ra_{\rm in}$ into $|\psi\ra_{\rm out}$ is a unitary transformation, our candidate for the cloning operator is $U=e^{-iH}$, which
transforms late-time incoming particles into outgoing radiation with Eq.~(\ref{ham2}) as the Hamiltonian. But before we do this, let us first study a much simpler case: the fully absorbing ($\Gamma=1$) black hole with only early-time particles, described by Eq.~(\ref{ham1}). Understanding this case will set important limits to allow us to better understand the general cloner based on (\ref{ham2}). 

Let us apply $U=e^{-iH}$ to an arbitrary incoming quantum state $|\psi_{\rm in}\ra$ (a state that includes the blank and ancillar states), where $|\psi\ra$ encodes information in the ``particle-antiparticle" basis using a so-called "dual rail" encoding. In a dual rail encoding, the logical one $|1\ra_L$ is encoded with $N$ particles and zero anti-particles in mode $a$ (in region I outside the horizon), while for the logical zero particles and anti-particles are interchanged, i.e.,
\be
|1\ra_L&=&|N,0\ra_a|0,0\ra_b  \label{log1} \\ 
|0\ra_L&=&|0,N\ra_a|0,0\ra_b \label{log2}\;.
\ee
Naturally,  sending information to future infinity using particles or anti-particles that are just outside of the black hole horizon after the formation of the black hole is not feasible as we remarked above before introducing late-time modes: this is purely an exercise to set some limits. We will even consider the limit where 
information is encoded in modes just inside the black hole horizon (the $b$ modes) for the same reason. 

We can construct an arbitrary quantum state from the logical states (\ref{log1}-\ref{log2}) by writing
\be
|\psi\ra_{\rm in}=\sigma|1\ra_L+\tau|0\ra_L\;.
\ee
However, because the quantum cloner defined by the unitary mapping $U=e^{-iH}$ using Hamiltonian~(\ref{ham1})  is  "rotationally invariant" (the action of the cloning machine does not depend on the state), we can simply choose $\sigma=1$ and $\tau=0$, that is, we will attempt to clone $N$ particles. 

Writing $|\psi_{\rm in}\ra = |N,0\ra_a|0,0\ra_b=|1\ra_L^{\otimes N}$ we obtain using (\ref{ham1})
\be
|\psi_{\rm out}\ra=e^{-iH}|\psi_{\rm in}\ra =\frac1{\alpha_k^{2+N}}\sum_{jj^\prime}^{\infty}e^{-(j+j^\prime)\frac{\omega}{2T}}\sqrt{\binom{j+N}{N}}|j+N,j^\prime\ra_a|j^\prime,j\ra_b\;. \label{clonerout}
\ee
In order to describe an $N\to M$ cloning machine and calculate its fidelity, we need to fix the number of particles and antiparticles in region I to $M$. In quantum optics, this "postselection" 
to a fixed number of clones is achieved by entangling the quantum state with a trigger signal and then measuring the trigger. If $M$ particles are detected in the trigger, then we know that only the component of (\ref{clonerout}) with $M$ clones survives. For black holes, we can perform the same operation, but must send the trigger towards future infinity (but away from the black hole horizon). Even though the quantum state reconstruction is conditioned on the trigger, it only uses the quantum states coming from the black hole.

Fixing the number of output clones to $M$ reduces the state (\ref{clonerout}) to 
\be
|\psi\ra_M\sim \sum_{j=0}^{M-N}\sqrt{\binom{M-j}{N}}|M-j,j\ra_a|j,M-N-j\ra_b\;,\ \ 
\ee 
which, it turns out, is (up to normalization) {\em identical} to the wavefunction emanating from the optical quantum cloner of Simon et al.~\cite{Simon2000} that achieves the optimal fidelity~(\ref{optfidel}). Thus, for quantum states sent into a black hole at early times so that they remain just outside the horizon, the black hole behaves as a universal, optimal, quantum cloning machine.

Let us now study what happens to quantum states {\em behind} the horizon (modes $b$). Just as in the optical realization of the optimal unversal cloner, the quantum state behind the horizon contains $M-N$ {\it anticlones} of the initial state $|N,0\ra_a$, that is, "inverted" states 
obtained from the initial states by the application of the optimal universal NOT gate~\cite{Buzeketal1999,GisinPopescu1999}. The fidelity of these anticlones is
\be
F_{\rm anti}=\frac{N+1}{N+2}
\ee
for each anticlone, which as we already saw is the fidelity of the best possible state preparation via state estimation~\cite{MassarPopescu1995}. This implies that the black hole has induced the {\it maximal} disturbance on the states behind the horizon.

Consider now instead the fidelity of cloning when $N$ {\em anti}-particles are sent into mode $b$. We are now following anti-particles that are traveling just {\em inside} the horizon: $|\psi\ra_{\rm in}=|0,0\ra_a|0,N\ra_b$. Thinking of these modes as impinging on the black hole horizon from {\em inside} the black hole (a horizon that looks to those modes as a mirror, that is, a {\em white} hole with $\Gamma=0$), we expect those modes to stimulate clones of these anti-particles inside the black hole, but also anticlones on the outside.

Let us calculate the fidelity of those anticlones generated on the outside region. The wavefunction is
\be
|\psi\ra_M\sim \sum_{j=0}^{M}\sqrt{\binom{j+N}{N}}|j,M-j\ra_a|M-j,j+N\ra_b\;,
\ee
which implies the probability to observe $j$ particles and $M-j$ antiparticles outside the horizon
\be
p(j,M-j|N)=\frac{\binom{j+N}{N}}{\sum_{j=0}^{M}\binom{j+N}{N}}\;.
\ee
As can easily be checked using (\ref{fidelity}), the fidelity of these anticlones is also $N+1/N+2$, that is, it is equal to the fidelity of the anticlones when sending in $|N,0\ra_a$. But the anticlones of the antiparticle states are, of course, clones of the particle states! 

In summary, sending in $N$ particles in mode $a$ creates particle clones in region I with {\em optimal fidelity}. These clones could in principle be entangled with particles that the $N$ incoming particles were entangled with before they were sent into the black hole, very much like in the quantum optical case, where the initial particles are entangled with a trigger~\cite{Simon2000} (we will exploit this later when calculating the quantum capacity of black holes in section \ref{quantum}). Sending in $N$ antiparticles in mode $b$ towards the horizon (which, to those modes, looks like a perfectly reflecting mirror) gives rise to {\em classical} particle clones in region I instead. They are classical because they cannot share entanglement with any particles that the $N$ antiparticles inside the black hole could have been entangled with\footnote{The classical roots of the probability $\frac{N+1}{N+2}$ are also apparent by noting that this is Laplace's {\em rule of succession}: the likelihood that an event takes place given that we have $N$ successive observations of it.}. We note that those anti-particles traveling towards a white hole horizon can be seen as particles moving away from the horizon (backward in time) after absorption on a non-reflecting horizon. Thus, it is plausible that white holes are just time-reversed black holes. Box 1 makes this analogy more clear, and shows how stimulated emission saves microscopic time-reversal invariance.
 \addtocounter{figure}{-1}
 \begin{center}
\vskip 0.25cm
\noindent\shabox{
\begin{minipage}{5.2in}
\centerline{\bf Box 1: Microscopic Time-Reversal Invariance in the Presence of Black Holes}
\mbox{}\vskip 0cm
\noindent 
Newtonian gravity is invariant under time-reversal at the micro-level: every particle trajectory in a gravitational field, when time-reversed, gives rise to another plausible particle trajectory. This is illustrated schematically in Fig.~\ref{TR}(a) (left panel), where for the sake of simplicity the effect of the gravitational field on a particle is shown as if a perfect mirror reflected the particle. Under time reversal, the trajectory simply reverses (middle panel). If we use charge-conjugation symmetry, we see that this trajectory is equivalent to that of an anti-particle moving backwards in time (anti-particles are depicted as dashed lines). Comparing the first and the last panel, we note that time reversal is equivalent to CP symmetry, thus embodying the celebrated CPT theorem. 
\begin{wrapfigure}{c}{9cm}
 \captionsetup{width=0.9\linewidth,font=small}
\includegraphics[width=\linewidth]{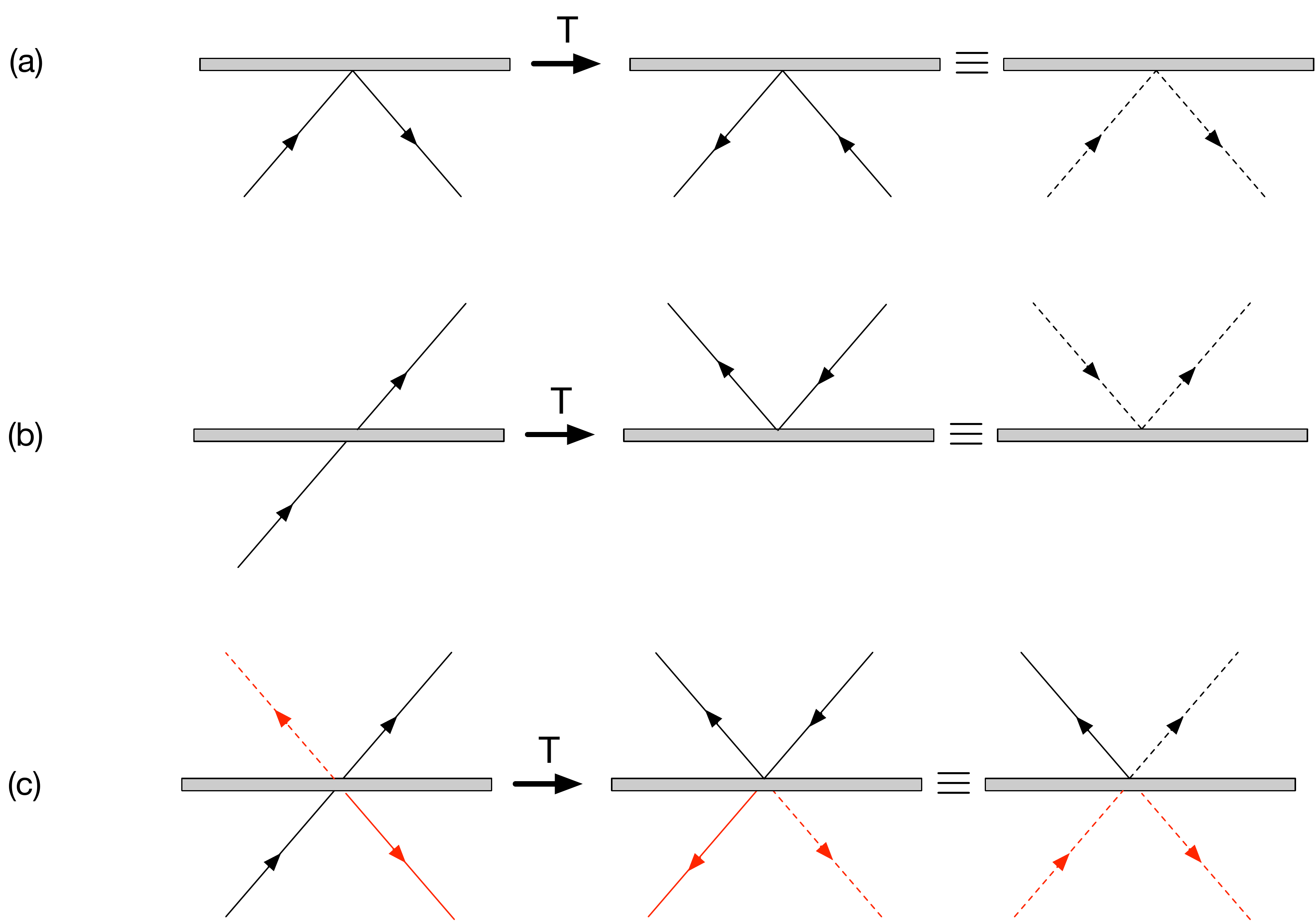}
\caption{Schematic representation of particle and anti-particle trajectories in the presence of white- and black hole horizons. Stimulated particles/anti-particles are rendered in red, solid lined denote particle trajectories, while dashed lines illustrate anti-particles. (a) Classical trajectory reflected at a mirror (or falling back under the influence of gravity), its time-reversed trajectory, and the equivalent trajectory in which particles moving forward in time are replaced by anti-particles moving backwards in time. (b) Classical trajectory of a particle absorved by a black hole horizon, its time-reversed trajectory, and the equivalent process. (c): Quantum trajectories including stimulated emission effects.\label{TR}}
\end{wrapfigure}
However, classical black holes violate this theorem. If instead of a perfect mirror the particle encounters a perfectly absorbing black hole horizon (Fig.~\ref{TR}(b), left panel) reversing the arrow of time (middle panel) does not produce a time-reversed picture of the left panel, since a particle from inside of the black hole cannot escape to the outside. Instead, it is "reflected" at the horizon, that is, from inside the black hole must act like the perfect mirror in Fig.~\ref{TR}(a). Replacing particles by anti-particles moving backwards in time produces the right panel, which clearly is not the anti-particle version of the left panel, thus breaking CPT invariance. Adding Hawking radiation (spontaneous emission of particles) to this picture does not restore CPT invariance. However, the stimulated emission of radiation does. Fig.~\ref{TR}(c) shows the stimulated particle/anti-particle pair in red (for illustrative purposes I disregard factors of $\beta^2$ here) that must accompany the absorption process (the case $\Gamma=1$ is shown here). Time-reversing this trajectory produces the middle panel, where the particle from inside of the black hole indeed reflects at the horizon, but it also stimulates a particle/anti-particle pair outside of the horizon (shown in red). In fact, this is the white-hole stimulated emission process described in the main text ($\Gamma=0$). Re-interpreting particles moving backwards in time as anti-particles moving forwards in time produces the right panel in Fig.~\ref{TR}(c), which indeed is the same process as in the left panel, only with particles replaced with anti-particles. Thus, stimulated emission of pairs restores CPT invariance. The spontaneous emission of pairs (Hawking radiation) is not shown in these diagrams as it has no influence on CPT invariance: it only serves to safeguard the no-cloning theorem.
\end{minipage}
}
\end{center}
\mbox{}\vskip 0.25cm

Let us now analyze the more realistic cloning scenario where we send quantum states into the already-formed black hole at late times, using modes $c$ (see Fig.~\ref{penrose1}) that are strongly blue-shifted with respect to the early-time modes $a$ and $b$. Because we also use the beam-splitter term with strength $g_k^\prime\geq g_k$, we are studying "gray holes" with arbitrary reflectivity.

We first discuss $1\rightarrow M$ cloning. Because the full Hamiltonian (\ref{ham2}) is also rotationally invariant, we can again restrict ourselves to study cloning of one particular state. To clone the state $|1\ra_L=|0,0\ra_a|0,0\ra_b|1,0\ra_c$, for example, we obtain (I omit the subscript $k$ in the particle numbers and coefficients in the following, as we send in modes of only one particular frequency): 
\be
\rho_a=\Tr_{bc}\left(U|1\ra_L\la1|U^\dagger\right) = \rho_{k|1}\otimes\rho_{-k|0}\;,
\ee
where $\rho_{k|1}$ is given by Eq.~(\ref{rho1}) derived earlier and $\rho_{k|0}$ given by the standard result (\ref{rho0}).

With (\ref{rho0}) and (\ref{rho1}), the $1\rightarrow M$ cloning fidelity can be calculated as before (with  $\xi=\frac{\gamma^2}{\alpha^2\beta^2}$)
\be
F_{1\rightarrow M}=\frac{\sum_{j=0}^{M}\frac{M-j}M p(M-j|1)p(j|0)}
{\sum_{j=0}^Mp(M-j|1)p(j|0)}=\frac{3+\xi+2\xi M}{3(2+\xi M)}\;.\ \ \label{bhfidel}
\ee
Let us investigate this result in a number of physical limits. As the black hole becomes more and more reflective, $\Gamma_0=\alpha^2\rightarrow0$, which implies $\xi=\frac{e^{\omega/T_\bh}(1-\Gamma)}{\Gamma_0^2}\rightarrow\infty$.
In this case, the fidelity (\ref{bhfidel}) approaches the optimal value
\be
\lim_{\Gamma_0\rightarrow0} F_{1\rightarrow M}=\frac23+\frac1{3M}\;,\label{optfidel1}
\ee
as is seen in Fig.~\ref{fig-fid2}. For arbitrary $N$, the limit is exactly equal to the Gisin-Massar optimal fidelity (\ref{optfidel}) of an $N\to M$ cloning machine. We can recognize this result as the special case we treated earlier: If the black hole perfectly reflects incoming states, the black hole behaves just as if early-time modes ($a$-modes) were traveling just outside the horizon (except for the redshift). Because those modes (by definition) never enter the black hole, this is akin to a perfectly reflecting black hole.

\begin{figure}[thbp]
\begin{center}
\includegraphics[width= 8cm,angle=0]{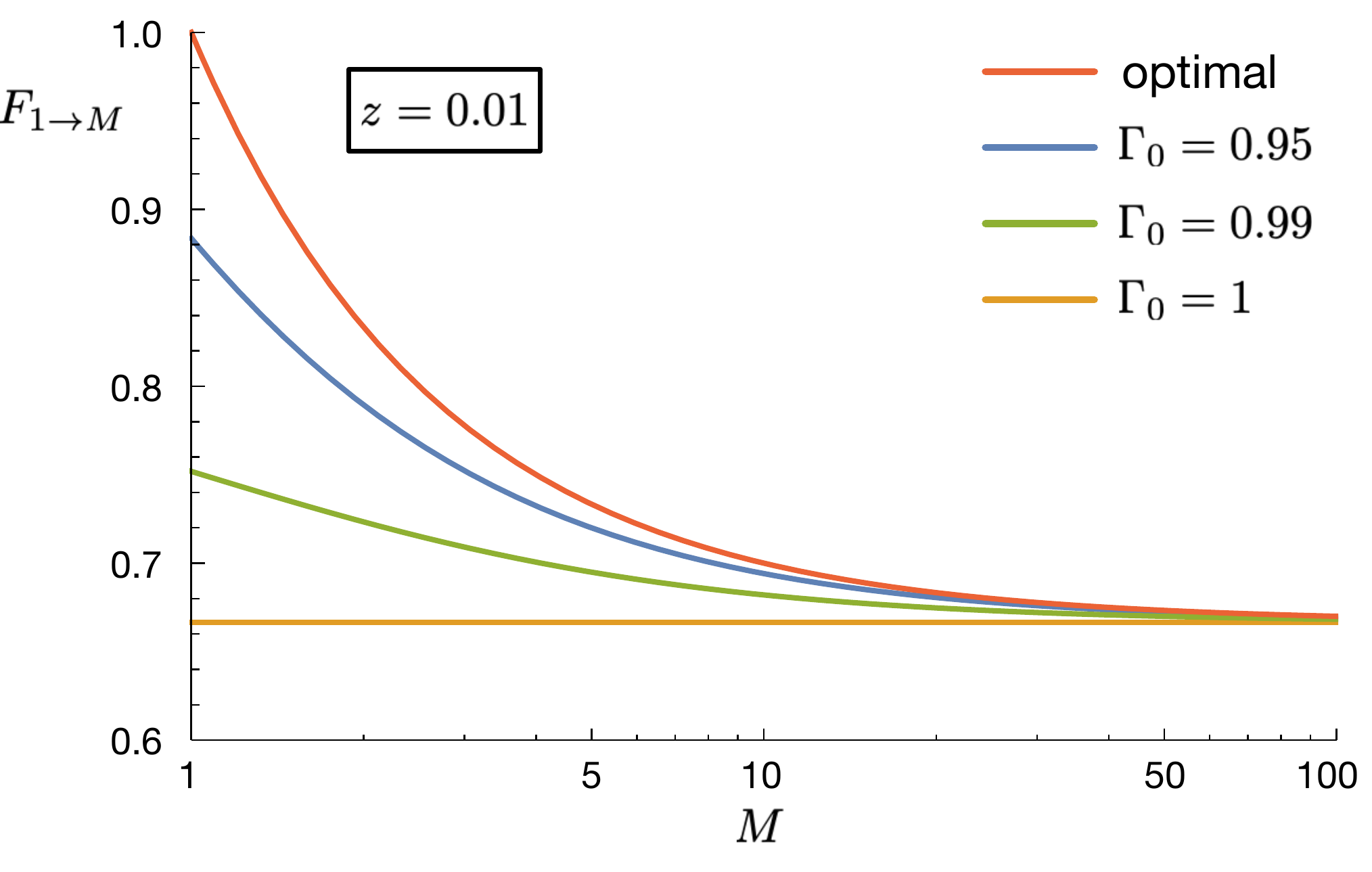}
\caption{Cloning fidelity $F_{1\to M}$ of the quantum black hole as a function of the number of copies $M$, for different values of the quantum absorption probability $\Gamma_0$  and a fixed $z=e^{-\omega/T_\bh}=0.01$.   Adapted from~\protect\cite{AdamiVersteeg2015}.
\label{fig-fid2}}
\end{center}
\end{figure}

Another limit of note is that of full absorption: $\Gamma_0\rightarrow1$. In that case $\xi\to1$ and $F_{1\to M}\to 2/3$, the fidelity of a classical cloning machine. It can be shown in general that for full absorption, the $N\rightarrow M$ cloning fidelity is equal to $N+1/N+2$ independently of $\omega/T_\bh$, which is the result we obtained earlier when sending $N$ antiparticles in mode $b$ directly behind the horizon. This is again not surprising, as the absorption of $c$-modes stimulates the emission of $b$ anti-modes behind the horizon, who in turn give rise to anticlones of the antiparticles, that is, clones. But they must be "classical" clones, so the fidelity is that of state estimation by classical means.

While in the limit $\Gamma_0\rightarrow1$  the best we can do to reconstruct the quantum state is to make classical measurements that allow us to optimally estimate the quantum state, note that in the limit $N\to\infty$ the probability to do this correctly tends to one, implying that the quantum state information can be reconstructed with arbitrary accuracy. In a sense, this result mirrors a result from calculating the classical capacity of the black hole channel, where you can show~\cite{AdamiVersteeg2014} that in the limit $N\to\infty$ the capacity of the quantum black hole channel to transmit classical information becomes equal to the noiseless channel capacity, even for full absorption. 
\begin{figure}[tbp]
\begin{center}
\includegraphics[width= 8cm,angle=0]{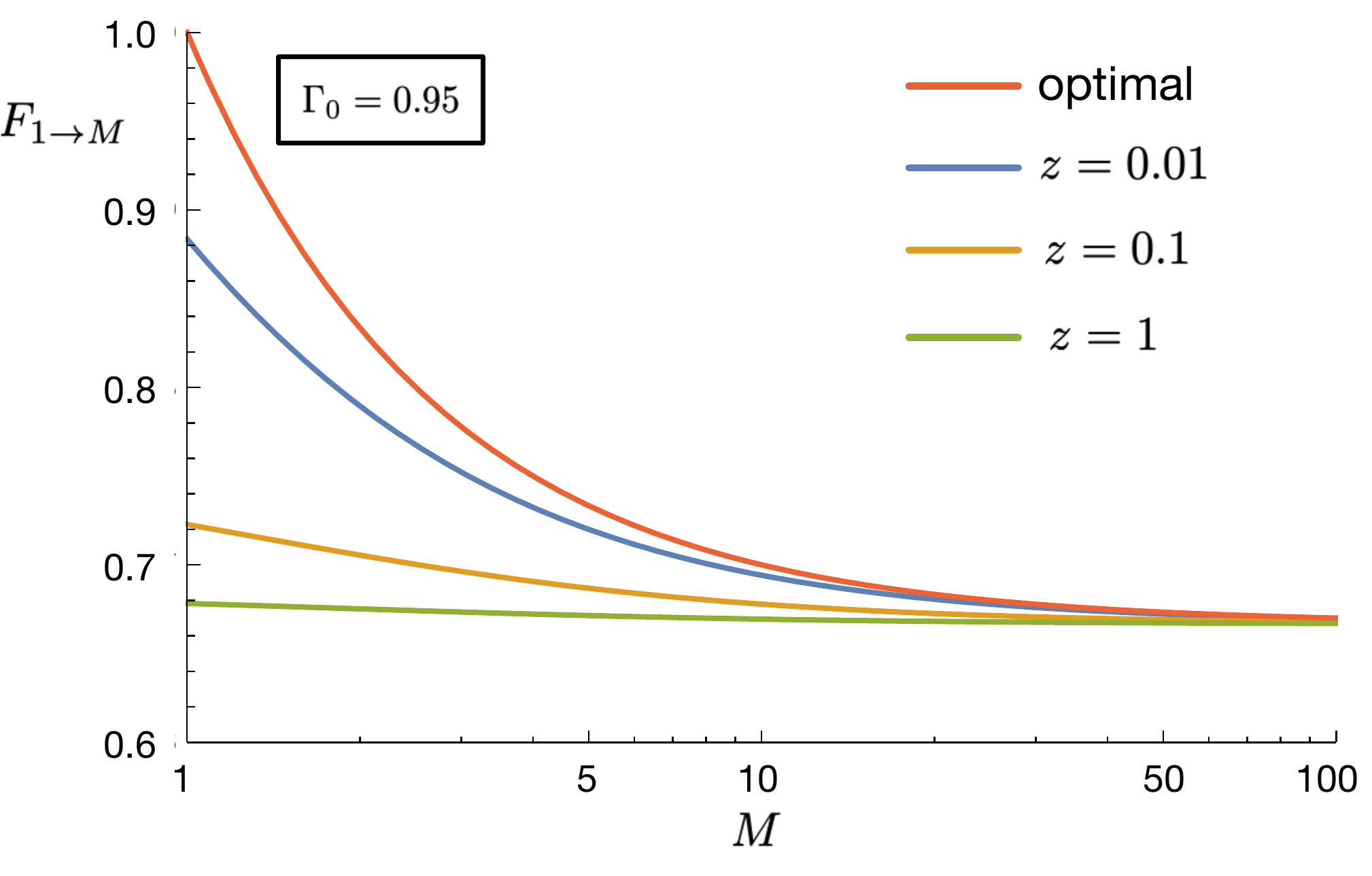}
\caption{Cloning fidelity $F_{1\to M}$ as a function of the number of copies $M$, for different $z=e^{-\omega/T_\bh}$ and a fixed quantum absorption probability $\Gamma_0=0.95$. Adapted from~\protect\cite{AdamiVersteeg2015}.
\label{fig-fid3}}
\end{center}
\end{figure}

In the previous discussion we had kept the ratio between the mode energy $\omega$ and the mass of the black hole constant (by keeping $z=e^{-\omega/T_\bh}$ constant). In the limit of 
 large black holes (where the Hawking temperature approaches zero), the limit $\omega/T_\bh\to\infty$ implies $\xi\to\infty$ (as long as $\Gamma_0<1$),  and we again recover the optimal universal quantum cloning fidelity (\ref{optfidel1}), as seen in Fig.~\ref{fig-fid3}.
Given that even modest-sized black holes have $\omega/T_\bh\geq10$, most black holes are therefore nearly-optimal universal quantum cloners, unless the absorption probability is exactly equal to 1. These results also hold for $N\to M$ cloning machines. Just as in the case $N=1$, the black hole cloner approaches the optimal cloner in the limit $T_\bh\to0$ or $\Gamma_0\to0$, and turns into a classical cloning machine in the limit $\Gamma_0\to1$ and for $M\to\infty$.

Now that we have seen in which way the black hole is a nearly optimal quantum cloning machine, we can turn our attention to the {\em quantum} channel aspects of the black hole. In the previous section we focused our attention on how classical information fares when sent into the black hole horizon. But in our discussion of quantum cloning, we clearly had the opportunity to study how entangled quantum states are affected by black holes. For example, in order to make sure that we measure exactly $M$ copies of the initial quantum state at future infinity, it was necessary to entangle a late-time particle with another whose state we would measure. It turns out that 
 the {\em quantum} channel defined by the mapping~(\ref{clonerout}) is an example of so-called "cloning channels"~\cite{Bradler2011}. More precisely, the black hole acts a weighted ensemble of cloning machines. In the next section, we will study the fate of {\em quantum information} (more precisely, quantum entanglement) that interacts with the black hole horizon. We should be mindful that there is no law of physics that prescribes that quantum entanglement must be preserved when interacting with a black hole. We will find that for the most part it is, but if the black hole is perfectly absorbing, it is not.

\section{Quantum Information Capacity} \label{quantum}
\setcounter{figure}{0}
Quantum information is a relatively new concept within the canon of theoretical physics. While classical information (defined in 1948) was a concept that was available to workers in the field of quantum gravity who worried about information conservation, quantum information theory rose to prominence in the mid 1990s spurred on by Peter Shor's discovery~\cite{Shor1994} that a quantum algorithm can factor numbers faster than any classical algorithm. One of the central results of classical information theory is the calculation of the capacity of various channels to transmit information, and the results in the previous section borrow heavily from that theory. Understanding the transmission of {\em quantum} information through a quantum channel requires an entirely different formalism, however, mainly because quantum information is something altogether different from classical information. As a consequence, we will see that for most channels it is not even possible to write down a closed expression for the capacity.

Classical information characterizes the relative state of two systems, specifically where one system can make predictions about the state of another. Quantum information is {\em entanglement}, that is, it is given by a quantum state relative to another. However, knowing this relative state does not make it possible for one system to make predictions about the other because entangled states are not separable: when two states are entangled they become one. For example, recall Eq.~(\ref{outwf}): the quantum state at future infinity (for the simplified situation without a beam splitter that would give rise to gray-body factors) with no initial particles present at past infinity, 
\be
|0\ra_{\rm out}=e^{-iH}|0,0\ra_a|0,0\ra_b=\prod_{k=-\infty}^\infty \frac1{\cosh^2 g_k} \sum_{n_k,n_k'}e^{-(n_k+n'_{-k})\omega_k/2T_\bh}|n_k,n'_{-k}\ra_a|n_k',n_{-k}\ra_b\;. \label{symmC}
\ee
This is a highly entangled state with both particles and anti-particles present behind and in front of the horizon, because the Bogoliubov transformation is, at heart, an entangling operation. Now let us study what happens to entangled states interacting with black holes.

When we say that we want to "send quantum entanglement through a quantum channel", what we mean is that the entanglement that one party has with a quantum system is {\em transferred} to another party. Say the two parties are called "Alice" and "Bob", and Alice is entangled with a reference system called "R". To simplify things further, suppose Alice's state is a qubit. In that case we can without loss of generality write the entangled state between Alice's qubit and R as
\be
|\psi\ra_{\rm in}=\sigma |0\ra_A|0\ra_R+\tau|1\ra_A|1\ra_R\;, \label{alice}
\ee
with complex coefficients $\sigma$ and $\tau$ that satisfy the condition $|\sigma|^2+|\tau|^2=1$.  
Now imagine that Bob is an inertial observer at future infinity, and Alice (for the present discussion) delivers her qubit into the forming black hole at past infinity. Is it possible for Bob to be entangled with R in the same way that Alice was? We will see in a moment that for the situation described here (namely a channel descibed by the unitary $U=e^{-iH}$ with the Hamiltonian given by Eq.~(\ref{ham1})), the answer is definitely "No": the quantum capacity of this channel vanishes. This, however, does not signal a breakdown of any law of physics: there are plenty of quantum channels with vanishing capacity. But understanding how black holes affect quantum entanglement is still an interesting question, which we now delve into.

Before we define and then calculate the quantum capacity of this channel, we need to discuss quantum channels more generally. It turns out that the mathematics of quantum channels is far more complex than that of classical channels, with many cases where it is not just impossible to calculate the capacity, we cannot even write down an expression for it! The reason for this is that the capacity of channels is defined asymptotically.  Classically, the capacity is the maximum rate at which information can be sent through the channel with vanishing error rate, in the limit where the number $n$ of uses of the channel goes to infinity. Because classically each use of the channel is independent of the previous use, such a limit is easily taken. In quantum physics, however, one message sent earlier can be entangled with a message sent later. Because these uses of the channel are now not independent anymore, the limit $n\to\infty$ becomes highly nontrivial. Let us first define the channel.

We begin with the entangled state between Alice and R written as in (\ref{alice}). This state becomes the input to the unitary dynamics of the Bogoliubov transformation $U=e^{-iH}$ with a suitable $H$ (see Fig.~\ref{qchannel}). 
\begin{figure}[htbp] 
   \centering
   \includegraphics[width=3in]{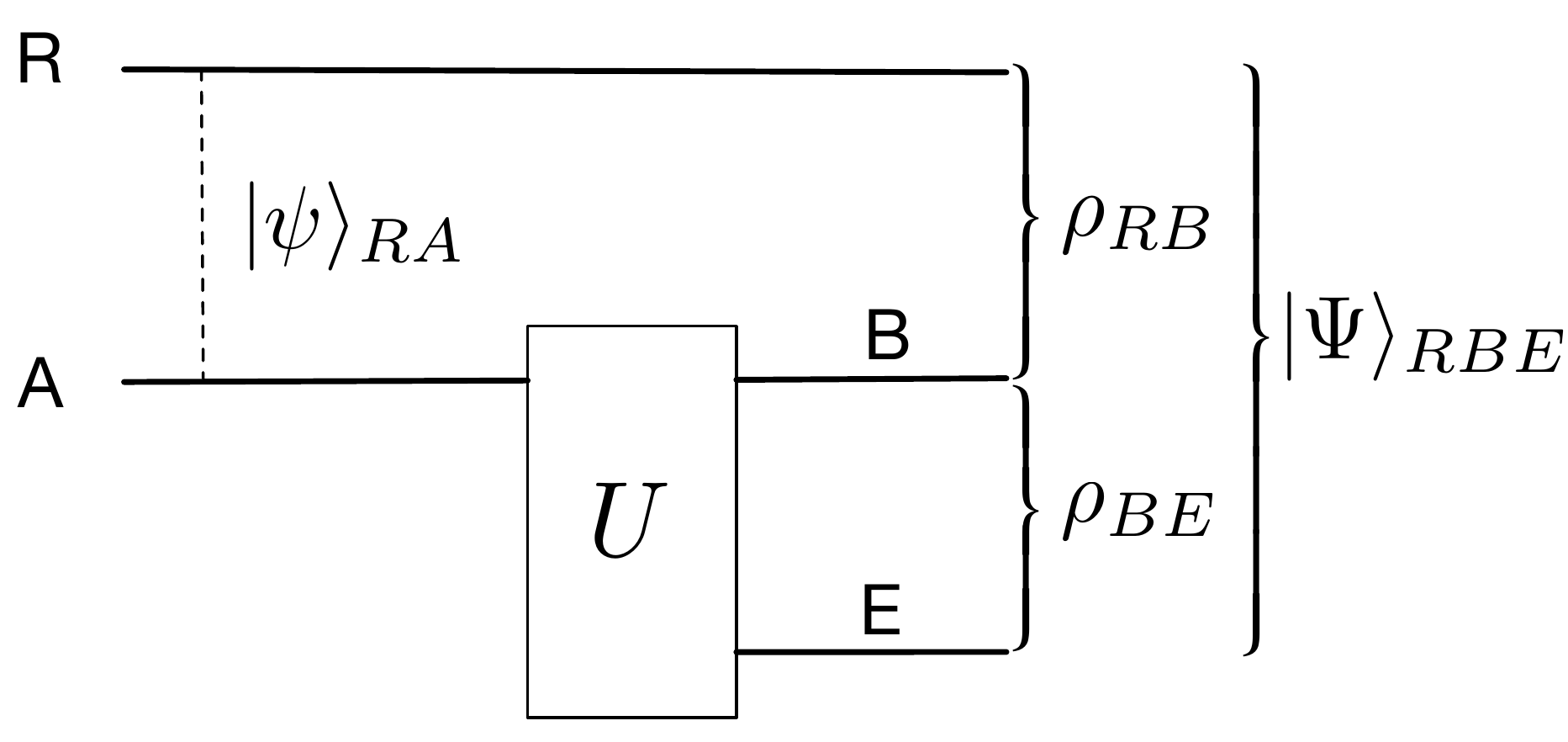} 
   \caption{Quantum channel with entangled input state $|\psi\ra_{RA}$ (the entanglement between R and A is indicated by the dashed vertical line), and outputs B and E. Generally, the "environment" E is an unobserved variable that provides the noise in the channel. For the black hole quantum channel, A is at past infinity (or when discussing late-time signaling, A sends in her quantum state at future infinity) while B is an inertial observer at future infinity. The complementary channel output E is the inside of the black hole, region II in Fig.~\ref{penrose}. The ouput of the joint channel is the pure state $|\Psi\ra_{RBE}$. The density matrices $\rho_{RB}$ and $\rho_{BE}$ are obtained from $|\Psi\ra_{RBE}$ by tracing over the unobserved output.}
   \label{qchannel}
\end{figure}
After the action of $U$, we ask whether the ouput of the channel, namely Bob's quantum state at future infinity, is now entangled with R in the same manner as Alice's qubit was. Note that the channel has two outputs: the recipient A, as well as a secondary ouput denoted by E. This secondary output is called the "complementary" channel: it is where everything that does not make its way to B must go. In classical information theory, what does not make its way to the receiver is lost to the environment, and in principle someone who has access to it could reconstitute the information from it, that is, such an agent could "eavesdrop" on the channel (hence sometimes this observer is called "Eve").  One of the most important differences between classical and quantum channels is that it is strictly impossible to both have quantum information perfectly reconstructed by B and by E: it is forbidden by the no-cloning theorem. 

Let us now return to the asymptotic property of channels. This is the second important distinction between classical and quantum channels. As mentioned earlier, the classical capacity is an asymptotic quantity: it is the rate at which information can be sent through the channel with arbitrary accuracy in the limit of $n\to \infty$ uses of the channel. The {\em quantum capacity} for Alice to send a single qubit (her share of the entangled state $|\psi_{RA}\ra$) to Bob is known as the "coherent information", and is defined as~\cite{Barnumetal1998}
\be
C_1(|\psi\ra_{RA})=\underset{|\psi\ra_{RA}}{\rm max}\bigl[S(B)-S(E)\bigr]\;, \label{singleshot}
\ee
where $S(B)$ is the von Neumann entropy of Bob's density matrix $\rho_B=\Tr_{RE}( |\Psi\ra_{RBE}\la\Psi| )$ given by~\cite{vonNeumann1927}
\be
S(B)=-\Tr_B\,\rho_B\log \rho_B\;,
\ee
and $S(E)$ is the entropy of the complementary channel, defined in a similar manner. Note that Shannon entropy [for example, Eq.~(\ref{shannon})] is simply a von Neumann entropy evaluated in the basis in which the density matrix is diagonal, defined by von Neumann 21 years before Shannon introduced his "uncertainty function"\footnote{It is interesting in this context to note that it was von Neumann who suggested to Shannon to call his measure  {\em entropy} because (as recounted in~\cite{TribusMcIrvine1971}) \protect{``}your uncertainty function has been used in statistical mechanics under that name\protect{"}.}.

That quantum channels are not "asymptotically" additive becomes quite obvious when you can show that two channels that each have zero $C_1$ could in fact transmit quantum information when used in parallel~\cite{SmithYard2008}, something that is completely impossible for classical channels. The true quantum capacity of a channel is
\be
C_Q=\underset{n\to\infty}{\lim}\frac1n C_1(|\psi\ra_{RA}^{\otimes n})\;,  \label{CQ}
\ee
where $|\psi\ra_{RA}^{\otimes n}$ represents $n$ copies of the state $|\psi\ra_{RA}$. However, the calculation of this limit $n\to\infty$ is in most cases intractable (see, for example~\cite{Wilde2013}, which should be consulted for a more comprehensive introduction to quantum channels). There are some exceptions: channels where the "regularization" [the limit $n\to\infty$ in (\ref{CQ})] is unnecessary. One such example is the "symmetric" quantum channel: a channel where the outputs B and E in the channel are interchangeable. This is a very peculiar channel, because if 
B and E are interchangeable, then we can say that both B and E receive the same amount of quantum information. But this is impossible according to the no-cloning theorem, and as a consequence the quantum channel capacity of a symmetric channel vanishes. This situation is, perhaps, analogous to the classical binary channel where a bit is equally likely to be flipped or not. That channel also has zero capacity. 

It turns out that the channel defined by the mapping (\ref{symmC}) is a symmetric channel, as can be seen by the symmetry between the $a$ modes in front of the horizon and the $b$ modes behind it. As a consequence we know this capacity, and it is zero. This, in hindsight, is not surprising. The input state to this channel is zero-dimensional: the vacuum. We would need at least an entangled qubit as input to have a nonzero capacity. 

We will now study a channel with input.  We use a dual-rail encoding of the logical bit like before, but rather than using particles and anti-particles, we instead encode the qubit in two particle modes that are available to Alice, for example the basis states $|10\ra_a$ and $|01\rangle_a$. This way, we can simplify the calculation by ignoring the anti-particle component in (\ref{ham1}). Doing this will also allows us to drop the subscript $k$, which we used mostly to indicate particle/anti-particle status. For simplicity, we will first show the calculation without the beam-splitter term [that is, Hamiltonian (\ref{ham1})], then later perform the full calculation with Hamiltonian (\ref{ham2}). 

Using $U=e^{-iH}$ with Hamiltonian (\ref{ham1}) on the input state $(\sigma|01\ra_a+\tau|10\ra_a)|0\ra_b$, it is not difficult to see that the channel we have constructed is in fact the same as the quantum cloner discussed in (\ref{clonerout}). We did not calculate the quantum capacity of the quantum cloning channel, so we will do this now. First, consider the $N\to M$ single use of the channel, which we term ${\cal N}_{N\to M}$. The $U$ in Fig.~\ref{qchannel} is then the quantum cloning machine QCM in Fig.~\ref{fig:cloner}. Using the single-shot formula for the quantum capacity~(\ref{singleshot}), Bradler et al.\ were able to show that~\cite{Bradleretal2010}
\be 
C_1({\cal N}_{N\to M})=\log_2(M+1)-\log_2(M-N+1)=\log_2\left(\frac{M+1}{M-N+1}\right)\;. \label{cloncap1}
\ee
While the calculation in~\cite{Bradleretal2010} was not performed in the context of black holes, once we realize that the quantum black hole channel is simply a cloning transformation, their results carry through. Of course, taking into account the redshift will modify these results, but using late-time particles as the signal (as we will do later) should recover this expression as late-time particles do not suffer a redshift. 

Now, the full quantum channel is not an $N\to M$ cloning machine, as the number of output particles is not fixed. Instead, the general channel is a superposition of cloning channels~\cite{Bradler2011}. Let us focus on the case with a single input qubit. The channel ${\cal N}$ is then the superposition
\be
{\cal N}=\sum_{M=1}^{\infty}p_M {\cal N}_{1\to M}\;, \label{cloningchannel}
\ee
where~\cite{Bradleretal2010,Bradler2011,BradlerAdami2014}
\be
p_M=\frac12(1-z)^3M(M+1) z^{M-1}\;. \label{clonprobs}
\ee
Here (as before), $z=e^{-\omega/T_\bh}$, and $\sum_{M=1}^{\infty}p_M=1$. Note that the complex phases $\sigma$ and $\tau$ do not appear anywhere, which is due to the earlier observation that cloning channels are invariant under unitary rotations of the input state, so we can simply set $\sigma=1$, for example. 

We would like to calculate the one-shot capacity of the channel ${\cal N}=\sum_{M=1}^{\infty}p_M {\cal N}_{1\to M}$, but up to this point we only know the capacity of a cloning channel with fixed $N$ and $M$, Eq.~(\ref{cloncap1}). Fortunately, it is possible to show that the capacity of a convex mixture of channels is equal to the mixture of capacities (see Appendix A in~\cite{BradlerAdami2014}):
\be
C_1({\cal N})=\sum_{M=1}^\infty p_M C_1({\cal N}_{1\to M})\;. \label{cloningcap1}
\ee
However, as I remarked before, in general the one-shot capacity is not equal to the quantum capacity of the channel because in quantum physics calculating a capacity requires the regularization (\ref{CQ}), which is intractable in the general case. On the other hand, we also saw that there are exceptions where regularization is not required. We will now see that the cloning channels are another such exception, but to understand this exception we have to first become familiar with another property of channels, termed {\em degradability}~\cite{DevetakShor2005}.

Roughly speaking, a channel is said to be degradable if the output of the channel can be "degraded" in such a manner that it looks like noise. Recall that in our construction Fig.~\ref{qchannel} the channel ${\cal N}$ maps the input state $\rho_A=\Tr_R |\psi\ra_{RA}\la \psi|$ to the output, that is $\rho_B={\cal N}(\rho_A)$. The noise of the channel is represented by the channel from Alice to the environment, which is called the "complementary channel" ${\cal N}_c$, so that $\rho_E={\cal N}_c(\rho_A)$.

A degradable channel is one where there exists a mapping ${\cal D}$ so that
\be
{\cal D}(\rho_B)=\rho_E\;,  \label{degradmap}
\ee
that is, where the map ${\cal N}$ followed by the map ${\cal D}$ equals the map ${\cal N}_c$: ${\cal N}_c={\cal D}\circ {\cal N}$. It is not difficult to prove (see, for example, Appendix A in~\cite{Cubittetal2008}) that all degradable channels have additive capacities, meaning that the regularization (\ref{CQ}) is unnecessary and that therefore $C_Q=C_1$.

Because a degradation always adds noise, a channel that is not degradable is one where Bob's output is noisier than Eve's. There are also channels where the environment "simulates" the channel itself (as opposed to the converse where the output simulates the noise). Such channels are called "anti-degradable", and naturally they require the existence of an anti-degrading map ${\cal D}_c$ so that ${\cal D}_c(\rho_E)=\rho_B$. The relationship between these maps is sketched in Fig.~\ref{degrad}.
\begin{figure}[htbp] 
   \centering
   \includegraphics[width=1.5in]{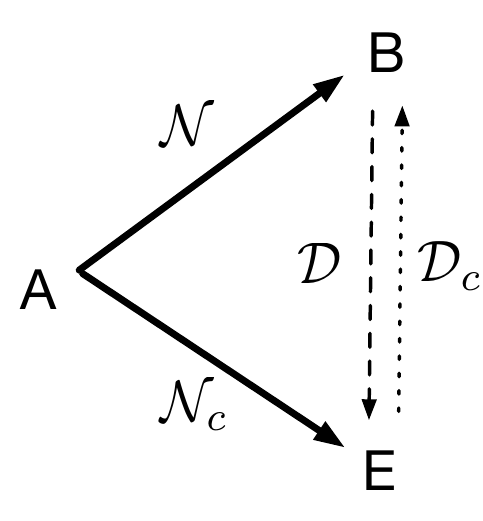} 
   \caption{Relationship between the channel ${\cal N}$, the complementary channel ${\cal N}_c$, and the degrading and anti-degrading maps ${\cal D}$ and ${\cal D}_c$.}
   \label{degrad}
\end{figure}
Let us check if we can find a degradable map for the cloning channel. To keep things simple, let us focus on the simplest $1\to 2$ cloning machine. One way to describe this channel is via its action on an arbitrary qubit, written in the Bloch-state representation as
\be
\rho_A=\frac12(\one+\hat n\cdot \vec \sigma)\;,
\ee
where $\hat n\in {\mathbbm R}^3$ is the unit vector in the Bloch sphere, and $\vec \sigma$ are the Pauli matrices. The vector $\hat n$ is determined by the complex coefficients $\sigma$ and $\tau$ of an arbitrary qubit, the exact form of which is not important for this argument. The optimal cloner returns~\cite{Bradler2011}
\be
\rho_{B_i}&=&\frac12(\one+ \frac23 \hat n\cdot \vec \sigma)\;,\\
\rho_E&=&\frac12\left(\one +\frac13(n_c\sigma_x-n_y\sigma_y+n_z\sigma_z)\right) \label{rhoe}\;,
\ee
for $i=1,2$. One thing we can do to turn $\rho_{B_i}$ into $\rho_E$ is to apply a {\em depolarizing} map that shrinks the Bloch vector by a factor of two. But this is not enough, as there is a minus sign multiplying $\sigma_y$ in (\ref{rhoe}).  However, applying a complex conjugation after the depolarization will indeed turn $\rho_B$ into $\rho_E$. Such combined maps are called "conjugate degrading" maps, and the cloning channel is therefore {\em conjugate degradable}. It is possible to prove that not only is the optimal $1\to 2$ cloner conjugate degradable, but so are all $N\to M$ cloners, and by extension the general cloning channel (\ref{cloningchannel})~\cite{Bradler2011}. Fortunately, conjugate degradable channels {\em also} are additive, so that the capacity of the quantum cloning channel, and therefore the quantum capacity of the black hole channel (as they are one and the same thing) is given by (\ref{cloningcap1}). Evaluating ({\ref{cloningcap1}) using (\ref{cloncap1}) (with $N=1$) and (\ref{clonprobs}) we arrive at the expression~\cite{BradlerAdami2014}
\be
C_Q=\frac12(1-z)^3\sum_{M=1}^\infty M(M+1) z^{M-1}\log\left(\frac{M+1}{M}\right)\;, \label{optcoh}
\ee
which is shown in Fig.~\ref{fig:cloncap} as a function of $z=e^{-\omega/T_\bh}$. Note that unlike the classical capacity of the black hole, which was still positive even as the temperature of the black hole diverges ($M_\bh\to 0$), the quantum capacity vanishes in this limit.
\begin{figure}[htbp] %
   \centering
   \includegraphics[width=3in]{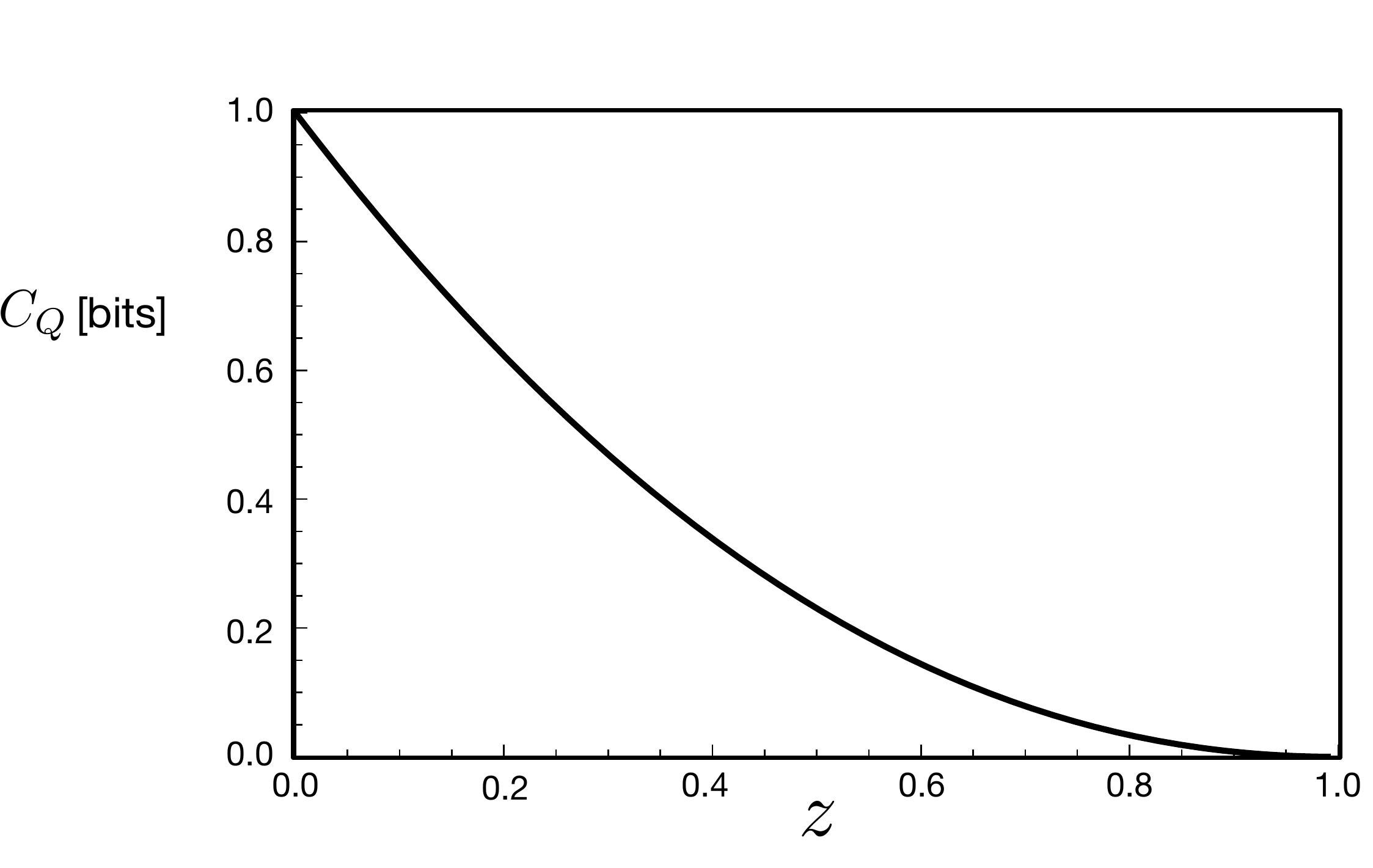} 
   \caption{Quantum capacity of the channel (\ref{cloningchannel}) with a single incoming qubit ($N=1$) as a function of  $z=e^{-\omega/T_\bh}$.}
   \label{fig:cloncap}
\end{figure}

The previous discussion of the capacity of black holes to transmit quantum information was intentionally naive (we sent in information early during the formation of the black hole but neglected the red shift, and we did not treat scattering off the black hole horizon). However, we will see that we can reuse some of that calculation in the full case that  we discuss now. 

To correctly treat scattering off of the black hole horizon with particles sent towards the horizon long after the formation of the black hole, we define (as before, in Eq.~(\ref{bogol})) the outgoing annihilation operator
\be 
A_k &=& e^{-iH}a_k e^{iH} = \alpha_k
a_k - \beta_k b^\dagger_{k}+ \gamma_k c_k\;,\label{bogol2} 
\ee
except for simplicity we will use a scalar as opposed to a complex field since we will not need to describe anti-particles. In the following, I will only treat the case $\alpha_k=0$ (a perfectly reflecting black hole, which we can call a "white hole"), and a perfectly absorbing black hole ($\alpha_k=1$). The reason I only treat these extreme cases is that, in this formalism, they are the only ones that are tractable. I will discuss the "gray hole channel" using the formalism of Gaussian states later.

\subsection{Quantum capacity of perfectly white holes}
\label{whitecap}
If we set $\alpha_k=0$, the Bogoliubov transformation (\ref{bogol2}) becomes
\be 
A_k = \gamma_k c_k - \beta_k b^\dagger_{k} \;,\label{bogol3} 
\ee
This transformation is formally identical to the transformation (\ref{bol}), which was the naive description of black holes without scattering. Using $\gamma_k^2=1+\beta_k^2$, the outgoing density matrix given $m$ incident late-time particles is precisely (\ref{denmat}). I previously pointed out (right after that equation) that it was not consistent to claim that the case treated there should be seen as a perfectly absorbing black hole ($\Gamma=1$), as detailed balance would be violated. In fact, we now see that the case described there (and therefore the case studied by Hawking in his initial publication~\cite{Hawking1975}) only consistently describes a perfectly {\em reflecting} black hole: a white hole. This makes sense in hindsight: if you take a look at the Penrose diagram Fig.~\ref{penrose} that describes the time evolution of the operators $a_k$ and $b_k$, it is clear that the particles that travel towards future infinity just outside of the horizon (the modes $a_k$) never enter the horizon\footnote{This is also consistent with how Hawking introduced his gray body factor, by following particles from region II {\em backwards in time} into region I. If particles can be transmitted with $\Gamma=1$ from the inside to the outside (a black hole from the inside), then its time reversal must be a white hole with $\Gamma=0$ from the outside).}. Thus, if viewed as the signaling particles, they are never absorbed and for them the black hole is a perfect mirror.

We can now be confident that the capacity shown in Fig.~\ref{fig:cloncap}, when treating late-time particles incident on the horizon, is in fact the quantum capacity of a white hole. It is finite for all but vanishing-mass black holes, which implies that (given suitable quantum error correction methods), Alice's quantum state can be perfectly reconstructed by Bob. Now let us take a closer look at the complementary channel (the one to Eve, that is, beyond the horizon). The quantum no-cloning theorem tells us that this capacity better vanish, as otherwise quantum information could be perfectly reconstructed by both Eve and Bob. 

It turns out that the complementary channel ${\cal N}_c$ gives rise to a quantum state straddling the horizon that is {\em separable}, meaning that the regions inside and outside of the horizon are not entangled~\cite{Bradler2011}: the horizon has "broken" the potential entanglement between Eve and Alice. Channels that do this are known as "entanglement-breaking channels"~\cite{Bradleretal2010}, and have zero capacity. In particular, it can be shown that all anti-degradable channels (see Fig.~\ref{degrad}) are entanglement-breaking~\cite{Bradleretal2010}. Since the cloning channel is degradable (along with being conjugate degradable), the no-cloning theorem ensures that the complementary channel is anti-degradable and therefore entanglement-breaking. 

\subsection{Quantum capacity of perfectly black holes}
\label{blackcap}
We now study the case $\alpha_k=1$, that is, black holes that do not reflect any of the incoming radiation. In this case, the Bogoliubov transformation reads
\be 
A_k =a_k - \beta_k b^\dagger_{k} + \gamma_k c_k \;,\label{bogol4} 
\ee
but because $\alpha_k^2-\beta_k^2+\gamma_k^2=1$, setting $\alpha_k=1$ implies $\beta_k=\gamma_k$ [we previously identified this case with $g_k=g_k^\prime$ in (\ref{ham1})]. 
Acting with $U=e^{-iH}$ on an initial state without any incoming particles $|000\ra_{abc}$ using Hamiltonian (\ref{ham1}) but with $\alpha_k=1$ and $\beta_k^2=\gamma_k^2\equiv g_k^2$ results in
\be
U|000\ra_{abc}=\frac1{1+\frac12\beta_k^2}\sum_{n=0}^\infty \sum_{m=0}^n\left(\frac{2\beta_k}{2+\beta_k}\right)^n (-\beta_k)^m\sqrt{n\choose m}|n-m\ra_a|m\ra_b|m\ra_c\;,
\ee
In order to encode a qubit in a dual-rail fashion, we also need to study how a single-late time particle (in mode $c_k$) fares under the transformation. We find~\cite{BradlerAdami2014}
\be
U|001\ra_{abc}&=&\left(\frac1{1+\frac12\beta_k^2}\right)^2\sum_{n=0}^\infty \sum_{m=0}^{n+1}\left(\frac{2\beta_k}{2+\beta_k}\right)^n (-\beta_k)^m\sqrt{n\choose m}\times\nonumber \\
&&\!\!\!\!\!\!\!\!\!\!\!\!\!\!\!\!\!\!\!\!\left( \sqrt{m+1}|n-m\ra_a|n\ra_b|m+1\ra_c +g_k\sqrt{n-m+1}|n-m+1\ra_a|n\ra_n|m\ra_c\right)\;.
\ee
If Alice's qubit is described by the density matrix $\rho_{\rm in}$, we can show that the output matrix is a superposition of channels\footnote{We should not confuse the depolarizing map ${\cal D}_M$ with the degrading map ${\cal D}$ introduced in Eq.~(\ref{degradmap}), as the depolarizing map is neither degrading nor anti-degrading.}
${\cal D}_M$ (where $M$ is again the number of clones)~\cite{BradlerAdami2014}
\be
\rho=\sum_{M=1}^\infty p_M {\cal D}_M(\rho_{\rm in}). \label{depol}
\ee
Let us look at the first term, $M=1$. The output under this channel can be calculated to  be
\be
{\cal D}_1(\rho_{\rm in})=\frac13\rho_{\rm in}+\frac13\one\;,  \label{depol1}
\ee
where $\one$ is the unit matrix. It turns out that this is the output of a {\em quantum depolarizing channel}~\cite{CalderbankShor1996,AdamiCerf1997,King2003}, whose action on an input $\rho_{\rm in}$ is given by $\rho_{\rm depol} = (1-p) \rho_{\rm in} + \frac p2\one$ ($p$ is the depolarizing parameter of the channel). Thus, the quantum channel for full absorption ($\alpha_k=1$) is a quantum depolarizing channel where the depolarizing parameter is given by the "classical" cloning fidelity $F=2/3$, which we recall from section~\ref{cloning} is the worst cloning fidelity we can achieve. More importantly, it was shown in~\cite{Bradler2011} that a depolarizing channel with $F=2/3$ is in fact entanglement-breaking, and therefore has zero capacity. Indeed, all channels ${\cal D}_M$ in (\ref{depol}) have this property, and therefore the quantum capacity for a perfectly black hole vanishes. 

In hindsight, we could have guessed this result. After all, the quantum channel with $\alpha_k=1$ is the {\em complementary} channel for the $\alpha_k=0$ channel, viewed from behind the horizon. And because that channel has positive capacity, the capacity of its complementary channel must vanish so that we cannot reconstruct quantum information in two different places. 

Clearly, the channels with $\alpha_k=0$  and those with $\alpha_k=1$ are extreme cases. Because the reflectivity of a black hole depends primarily on the impact parameter of scattering, it is important to understand the black hole quantum channel for all values between $\alpha_k=0$ and $\alpha_k=1$. To study the capacity for black holes with arbitrary $0\leq \alpha_k^2\leq1$, we are going to have to deploy more sophisticated artillery.


\subsection{Quantum capacity of black holes with arbitrary transmissivity}
Previously, I pointed out the relationship between the Bogoliubov transformation engendered by (\ref{ham1}) and those that we encounter in quantum optics in order to motivate a discussion of quantum dynamics in terms of optical elements: the two-mode squeezer and the beam splitter. In this section we are going to use this analogy in order to marshal the considerable quantum optics literature of so-called {\em Gaussian states}, and study the quantum capacity of Gaussian channels. This will allow us to make some statements about the quantum capacity of black holes with arbitrary transmissivity, but we will also see that, because of the problem of regularization of quantum capacities, we cannot as yet answer all questions about the capacity of those channels.

Earlier, we discussed an encoding of information using states with defined particle number. For example, a logical zero would be encoded using $m$ anti-particles, and a logical one would correspond to sending $m$ particles instead. However, creating quantum states with defined particle number is exceedingly difficult. In standard quantum optics applications, it is more convenient to construct states with a defined {\em mean} number of particles $\la m\ra$ instead. 
A typical Gaussian quantum state with fixed mean  number of particles is a {\em thermal state}, defined by the density matrix
\be
\rho_{\rm therm}=\frac1{\la m\ra+1}\sum_{m=0}^\infty \left(\frac{\la m\ra}{\la m\ra+1}\right)^m|m\ra\la m|\;.
\ee
This is a thermal state because the mean number of particles $\la m\ra =\Tr(\rho_{\rm therm} a^\dagger a)$ is 
\be
\la m\ra=\frac{e^{-\omega/T}}{1-e^{-\omega/T}}\;,  \label{mean}
\ee
and we immediately recognize that the output of a black hole channel without any incident particles (\ref{mat}) is, in fact, a thermal state with mean particle number $\la m\ra=\beta_k^2$ in each mode $k$.

Thermal states are a particular example of the more general Gaussian states, which are defined in terms of correlation matrices acting on {\em quadratures}, 
rather than the creation and annihilation operators that we have used throughout. 
The relation between quadrature operators $q,p$ and creation/annihilation operators $a^\dagger,a$ for a single mode $k$ is simply (I have reinstated $\hbar$ here)
\be
q_k=\sqrt{\frac{\hbar}2}(a_k+a_k^\dagger)\;,\; p_k=i\sqrt{\frac{\hbar}2}(a_k-a_k^\dagger)
\ee
so that $q_k$ and $p_k$ observe the standard uncertainty relation
\be
[q_k,p_{k'}]=i\hbar\delta_{kk'}\;.
\ee
To express an arbitrary density matrix $\rho$ in the $n$-mode quadrature basis~\cite{Weedbrooketal2012}, 
first we define the column vector of $2n$ operators
\be
{\bf x}=[q_1,p_1,\cdots,p_n,q_n]^\top
\ee
so that $x_1=q_1$, $x_2=p_1$, $x_3=q_2$ and so forth until $x_{2n}=p_n$. Then we can write the commutation relation for all $2n$ operators as
\be
[x_i,x_j]=i\hbar\Omega_{ij}\;,
\ee
where the matrix $\bm{\Omega}$ is the direct sum of matrices $\bm{\omega}$ for each mode:
\be
\bm{\Omega}=\bigoplus_{k=1}^n \bm{\omega}=   \begin{pmatrix} 
      \bm{\omega} & &  \\
       & \ddots & \\
       & &  \bm{\omega}
   \end{pmatrix}\;, \; \bm{\omega}= \begin{pmatrix} 0 & 1\\ -1& 0  \end{pmatrix}\; .
\ee
The first moment of a density matrix $\rho$ can then be expressed in this basis as 
\be
\bm{\bar x}=\Tr{(\rho \bm{x})}. 
\ee
The all-important {\em second} moment of $\rho$ is given by the matrix $\bm{V}$ with elements
\be
V_{ij}=\frac12\Tr(\rho\{\Delta x_i,\Delta x_j\})
\ee
where $\Delta x_i=x_i-\bar x_i$ and $\{\ ,\ \}$ is the anti-commutator. Gaussian quantum states are then defined as those states for which higher-order moments beyond the second moment vanish, and the covariance matrix $\bm{V}$ is a real symmetric matrix that satisfies the uncertainty principle~\cite{Weedbrooketal2012}
\be
\bm{V}+i\frac\hbar2\bm{\Omega}\geq 0 \label{positivity}\;.
\ee
The "positivity" requirement for a matrix such as that written in (\ref{positivity}) stipulates that all the eigenvalues of the matrix need to be positive, which puts constraints on the real-valued elements. 

With these preliminaries out of the way, we can study how Gaussian states behave under the transformations of the black hole channel. As before, we need to look at how the Bogoliubov transformation affects the two vacuum modes $a_k$ and $b_k$, as well as the signal mode $c_k$. We can write this transformation in matrix form [see Eqs.~(\ref{alpha}-\ref{gamma}) for the definitions of $\alpha_k$, $\beta_k$, and $\gamma_k$]
\be
   \begin{pmatrix} 
      A_k  \\
      B^\dagger_k\\
      C_k \\
   \end{pmatrix}=\begin{pmatrix} \alpha_k & -\beta_k & \gamma_k\\
   \beta_k & 1+\frac{\beta_k^2}{1+\alpha_k}& -\frac{\beta_k\gamma_k}{1+\alpha_k} \\
   -\gamma_k & \frac{\beta_k\gamma_k}{1+\alpha_k}  &  1+\frac{\gamma_k^2}{1+\alpha_k}
   \end{pmatrix} \begin{pmatrix}       a_k  \\
      b^\dagger_k\\
      c_k \\
   \end{pmatrix}\;.
\ee
We now have to write this transformation in terms of operators acting on quadratures instead. When we do this (see~\cite{BradlerAdami2015} for the details), we can write the action of the channel in terms of its effect on the covariance matrix of an input Gaussian state (the first moments of the Gaussian state can always be set to zero). In particular, an input Gaussian one-mode state (sent in at late time) $\bm {V}_{\rm in}$ transforms with a transmission matrix $\bm T$ and a noise matrix $\bm N$ as
\be
{\bm V}_{\rm out}={\bm T}{\bm V}_{\rm in}{\bm T}^\top +{\bm N}\;.
\ee
Here, $\bm T$ and $\bm N$ are $2\times2$ matrices (they act on the single late-time mode) that take the very simple (and diagonal) form
\be
{\bm T}=   \begin{pmatrix} 
      \sqrt{\gamma_k^2} & 0 \\
      0 & \sqrt{\gamma_k^2}
   \end{pmatrix}\; \ \ {\bm N}= \begin{pmatrix} 
      \alpha_k^2+\beta_k^2 & 0 \\
      0 & \alpha_k^2+\beta_k^2
   \end{pmatrix}\;. \label{matrices}
\ee
This form is particularly pleasing because it allows us to compare this channel directly to a complete characterization of all possible one-mode Gaussian (OMG) channels that previously appeared in the literature~\cite{Weedbrooketal2012,Schaferetal2013}. Of the eight channels listed there, three make an appearance in the black hole Gaussian channel. First, let us examine the parameter space of this channel, which is characterized by its transmission potential (parameterized by $\gamma_k^2$), and the noise level (described by $\alpha_k^2+\beta_k^2$). This much is not surprising: $\gamma_k$, after all, is the amplitude of the Bogoliubov transformation affecting our signal state, while the black hole's modes $a_k$and $b_k^\dagger$ are in a vacuum state and provide the noise to the channel. 

Fig.~\ref{OMG} depicts the OMG channel parameterized by the noise level and the transmissivity of the channel. 
\begin{figure}[htbp] %
   \centering
   \includegraphics[width=3in]{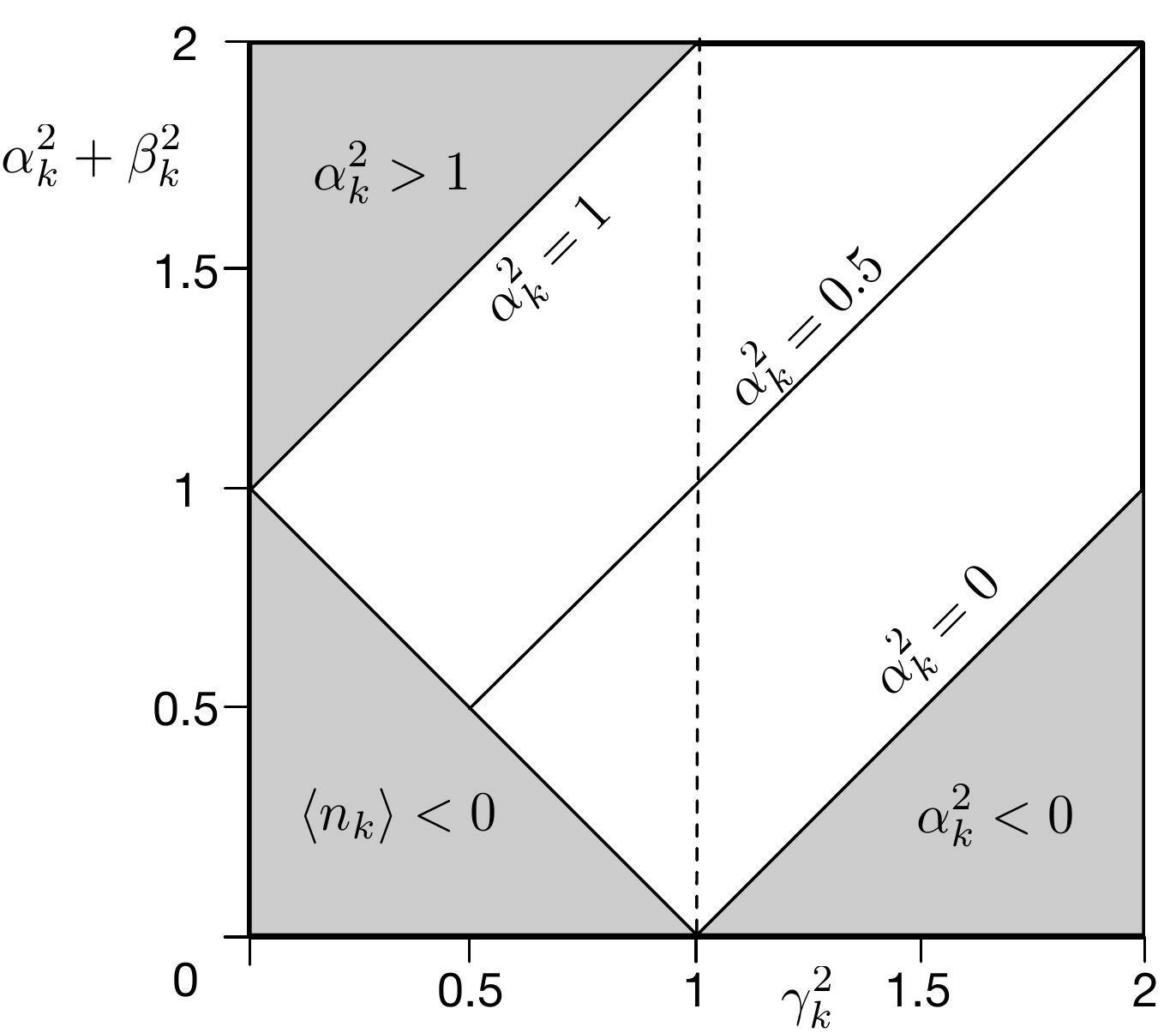} 
   \caption{Allowed regions of the parameter space (white) of the black hole OMG channel in terms of transmissivity $\gamma_k^2$ and noise level $\alpha_k^2+\beta_k^2$. Unphysical regions are in gray. }
   \label{OMG}
\end{figure}
While the regions in gray are not physical channels, we would like to calculate the capacity of the channel to transmit quantum information for the permissive region. Different conditions lead to the disallowed regions: two of them stem from the fact that the black hole's diffraction parameter $\alpha_k$ 
is bounded from below and above by zero and 1, respectively. In fact, these bounds also correspond to the positivity condition for the output Gaussian state, which is equivalent to the condition $\gp_k\geq g_k$. 
The second condition that rules out the lower left triangle in Fig.~\ref{OMG} is the condition that the mean particle number $\la n_k\ra\geq0$, as  $\alpha_k^2+\beta_k^2=(2\la n_k\ra+1)(1-\gamma_k^2)\geq 1-\gamma_k^2$, using $\la n_k\ra=\frac{\beta_k^2}{\alpha_k^2-\beta_k^2}$, consistent with (\ref{mean}).

Within the family of OMG channels, the channel with $\gamma_k^2<1$ is known as the ``lossy channel" ${\cal C}_{\rm loss}$ (area to the left of the vertical dashed line in Fig.~\ref{OMG}). For black holes, this corresponds to the cases where the effective transmissivity $\Gamma=1-\gamma_k^2$ lies between zero and one (recall that the effective transmissivity is strictly smaller than the black hole ``bare" transmissivity $\alpha_k^2$, the parameter that characterizes the beam-splitter, since $\Gamma=\alpha_k^2(1-e^{-\omega_k^2/T_\bh})$. OMG channels exist for which $\gamma_k^2>1$: these are the so-called ``amplifying channels", which, as the name implies, {\em amplify} the incoming signal. While it is unclear whether black holes exist with such a property, we'll discuss these hypothetical channels here for completeness.

The allowable region of the channel is bounded by the two lines $\alpha_k^2+\beta_k^2=1+\gamma_k^2$ and $\alpha_k^2+\beta_k^2=-1+\gamma_k^2$, which correspond to the two cases for which we have been able to calculate the quantum capacity of the black hole channel in section~\ref{whitecap} ($\alpha_k^2=0$) and section~\ref{blackcap} ($\alpha_k^2=1$).  For the latter case we determined that the quantum capacity vanishes, and indeed the analysis of~\cite{Carusoetal2006} revealed that the entire region between $\alpha_k^2=0.5$ (the balanced beam splitter) and $\alpha_k^2=1$ corresponds to a channel that can be written as the composition of an arbitrary channel and an anti-degradable channel, and therefore must have zero capacity\footnote{We noted earlier that anti-degradable channels are additive and must have zero capacity as they are entanglement-breaking.}. This area is shaded in yellow in Fig.~\ref{OMG-cap}.

\begin{figure}[htbp] %
   \centering
   \includegraphics[width=3in]{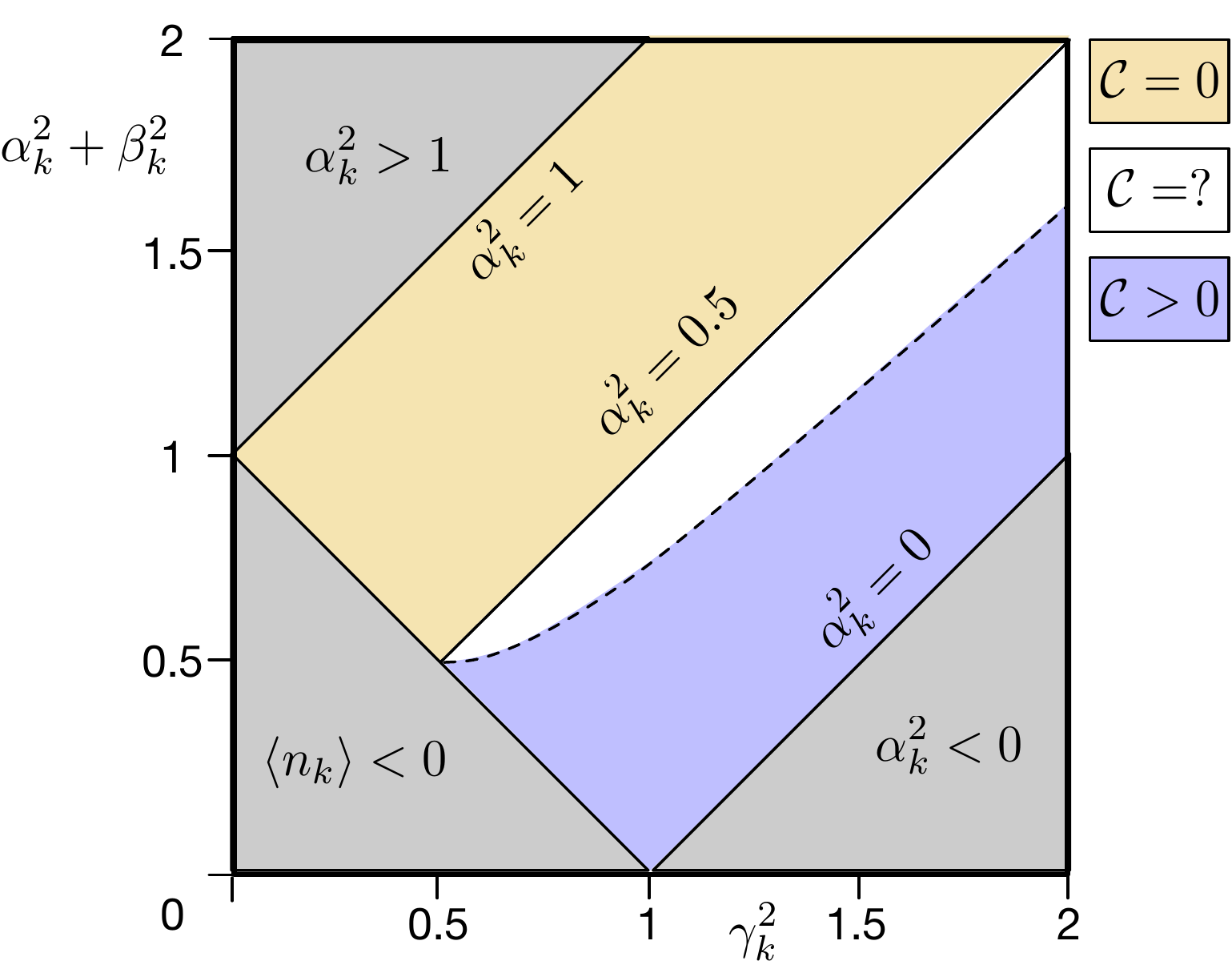} 
   \caption{Quantum capacity for the black hole OMG channel. The region in yellow has vanishing capacity, while for the purple region we have a non-zero lower bound for the capacity. For the region in white (bounded from below by ${\cal I}_{\rm coh}=0$), the capacity is currently not known because it could be super-additive.}
   \label{OMG-cap}
\end{figure}

We can now attempt to calculate the quantum capacity for the lossy channel. To do this, we must calculate the coherent information of the channel, and optimize it. As we discussed previously, this can only be achieved if the quantum capacity is {\em additive}, but for the channels with $\alpha_k^2<1/2$ we do not know this. However, if we can derive a {\em positive lower bound} using the additive capacity (\ref{optcoh}), then we can be assured that the capacity is positive. If the lower bound is zero, then we can only say that it is {\em possible} that the capacity is non-zero, but we simply do not know.

Let us then find out under what circumstances the limit $n\to \infty$ of the "$n$-shot" coherent information
$\lim_{n\to \infty} {\cal I}_{\rm coh}(n)$ 
is positive. The coherent information $ {\cal I}_{\rm coh}(n)$ has previously been calculated for both the lossy and the amplifying channel. In particular, it is possible to show that~\cite{BradlerAdami2015}
\be
 {\cal I}_{\rm coh}(n)=g(n)-2g(\xi)\;,
 \ee
 where $\xi=\sqrt{1+4n\alpha^2}$ and
 \be
 g(x)=(1+x)\log(1+x)-x\log x\;.
 \ee
 It turns out that in the limit $n\to\infty$ $g(\xi)$ vanishes~\cite{Bradler2015}, while
\be
{\cal I}_{\rm coh}=\lim_{n\to\infty} g(n)=\frac{\beta_k^2}{\alpha_k^2-\beta_k^2}\log\frac{\beta_k^2}{\alpha_k^2}+\log \frac{\gamma_k^2}{\alpha_k^2}\;. \label{coherent}
\ee
The dashed curve in Fig.~\ref{OMG-cap} corresponds to the boundary where ${\cal I}_{\rm coh}=0$. We thus see that for parameters where ${\cal I}_{\rm coh}>0$ (purple region in Fig.~\ref{OMG-cap}) the quantum capacity must be positive (as it is lower-bounded by~(\ref{coherent})), while the quantum capacity in the white region still cannot be determined.

For $\gamma_k^2>1$, the expression (\ref{coherent}) also turns out to be a lower bound~\cite{BradlerAdami2015}. The case $\gamma_k=1$, however, has to be treated separately. In this limit ${\cal I}_{\rm coh}$ diverges as $n\to \infty$, however, this channel is trivial: it represents the so-called "zero-added classical noise" channel~\cite{Holevo2007},  termed ${\cal B}_2$ in~\cite{Schaferetal2013}. For this channel, the transmission matrix ${\bm T}$ in (\ref{matrices}) is the identity, and the noise vanishes. Such channels have infinite capacity also in classical physics~\cite{CoverThomas1991}.

To summarize this section, we have seen that it is possible to make statements about the quantum capacity of black holes for values of the beam-splitter variable $\alpha_k^2$ other than the extreme cases $\alpha_k=0$ (for which we saw that the capacity is positive), and $\alpha_k=1$, where the capacity vanishes. We saw that when the beam-splitter absorbs more than it reflects ($\alpha_k^2\geq1/2$) then the quantum capacity must vanish (so as to conform to the no-cloning theorem). When, in turn, the beam-splitter reflects more than it transmits ($0\leq \alpha_k^2<1/2$) the capacity is positive for some parameters (those for which ${\cal I}_{\rm coh}>0$), but since for the remaining parameter region ${\cal I}_{\rm coh}\leq0$, we cannot establish whether the quantum capacity is positive since ${\cal I}_{\rm coh}$ is only a lower bound to the capacity. Needless to say at this point: a vanishing quantum capacity does not point to a flaw in the laws of physics. Rather, when it vanishes it does so because we must {\em conform} to the laws of physics, which stipulate that quanta cannot be cloned.

\section{Unitary Evaporation of Black Holes}
\setcounter{figure}{0}
In everything we have been discussing up to this point, the black hole was treated as a static quantity: it had already formed, and its mass was fixed at $M_\bh$. This approximation, which essentially treats the gravitational force as a {\em background} field, was necessitated by taking the static-path approximation to the time-dependent operator (\ref{unitary-sorkin}). In this approximation, the back-reaction of the radiation on the metric field is neglected: that is the essence of the semi-classical approach. 

However, this assumption also precludes us from studying the evaporation of the black hole microscopically. Hawking noticed early on that the energy of the outgoing Hawking radiation must be provided by the black hole, and that therefore the black hole must ultimately disappear. But this seemed to open up another fundamental problem: if we were to assume that a black hole can form from a quantum mechanical pure state (a state $\rho$ with vanishing von Neumann entropy) that in the future produces Hawking radiation with entropy $S_H$, then (since Hawking radiation is thermal) the final state after black hole evaporation would be a mixed state with positive entropy~\cite{Hawking1976b}. However, in a closed system such a transition from a pure state to a mixed state is forbidden: it is tantamount to the non-conservation of probability, a state of affairs I have previously referred to as an abomination. 

We contemplated this abomination when it appeared that classical information was lost inside of the black hole, but were able to recognize in the previous sections that all these problems arise simply from ignoring the stimulated emission process. However, there is no "incoming" signaling particle when discussing black hole evaporation, so stimulated emission will not help us understand how this process unfolds. To understand how black hole evaporation returns space-time to the pure state it started out as, we need a description of the interaction of black holes with radiation that goes beyond the semi-classical approach. The dynamics that such a treatment should reveal is that of the celebrated "Page curves". Page first discussed how the quantum entropy of one system might depend on the "size" of a subsystem that it is entangled with, while both together are in a pure state. 
Specifically, Page asks us to imagine a pure initial state formed using $n$ particles, $|\psi\ra_{\rm in}=|n\ra$. After this state is entangled with another system, the density matrix of the outgoing system becomes  $\rho_{\rm out} =\sum_{i=0}^n p_i |i\ra_{\rm out}\la i|$, with entanglement entropy 
\be
S_e(\rho_{\rm out})=-\sum_{i=0}^n p_i \log p_i\;.
\ee
The maximal entropy is reached when all the states $|i\ra \la i|$ are equiprobable, so that $S_{\rm max}=\log(n+1)$.  Page imagined that as the black hole pure state decoheres, the entanglement entropy of the outgoing radiation must also increase until it reaches its maximal value~\cite{Page1993}. As the black hole continues to evaporate, Page argued (using a toy quantum mechanical model) that the entanglement entropy must start to {\em decrease} (after a time now dubbed the "Page time"), as in Fig.~\ref{page}.

\begin{figure}[htbp] 
   \centering
   \includegraphics[width=3in]{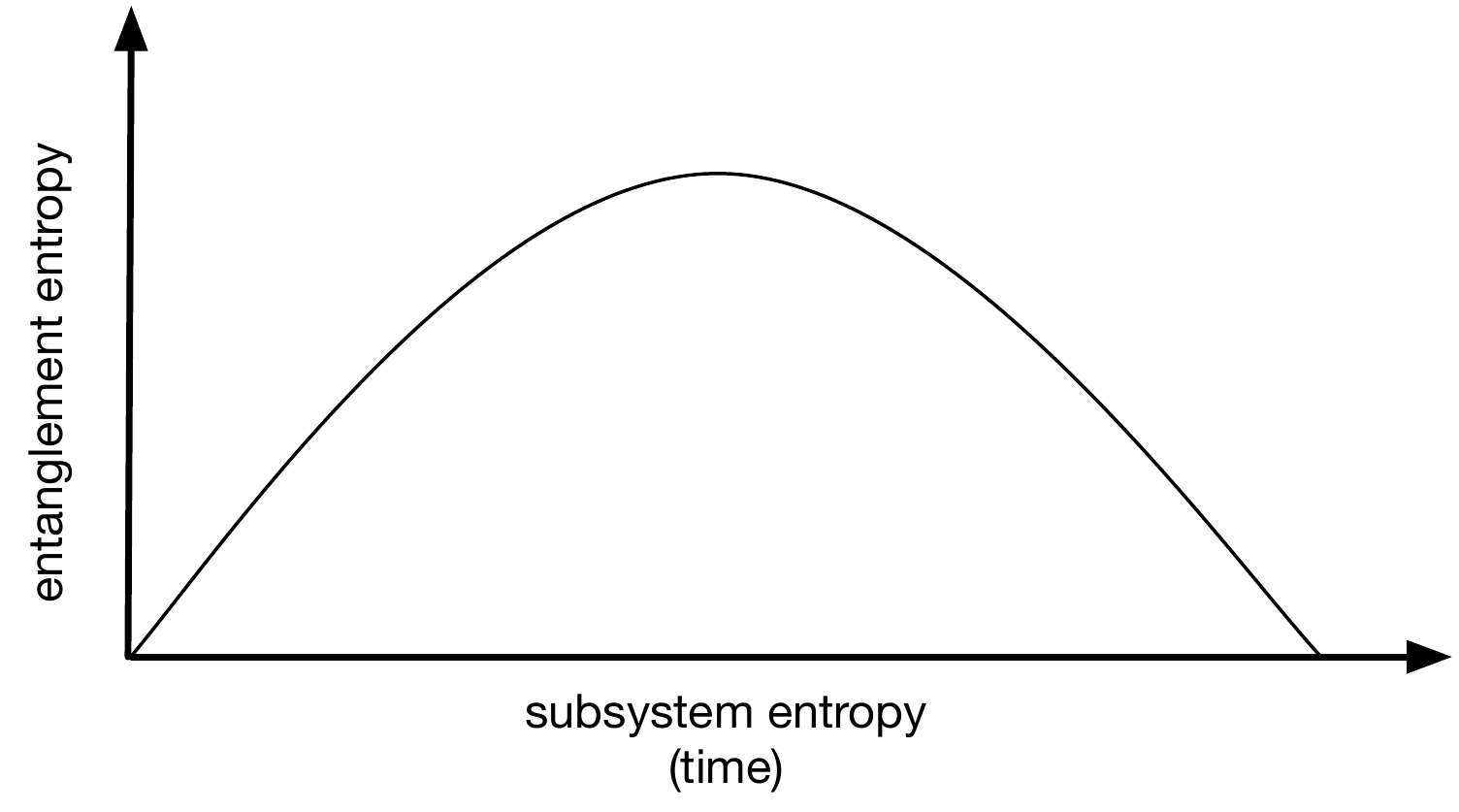} 
   \caption{A typical Page curve showing the entanglement entropy of Hawking radiation as a function of the "size" of the subsystem (determined by the number of particles in the subsystem). If we assume that the subsystem sizes increases as the black hole evaporates, we can take the subsystem size as a measure of elapsed time since formation of the black hole from a pure state.}
   \label{page}
\end{figure}

If a black hole evaporated via a unitary process (that is, if black hole evaporation can be described by an $S$-matrix), Page argued that ultimately the entanglement entropy of the black hole must disappear, leaving only a vacuum with zero entropy, in contradiction to Hawking's assessment that black holes must turn pure states into mixed states. Unfortunately, the semi-classical treatment of black holes prevents us from testing this prediction directly: we do not know what the interaction Hamiltonian $H_{\rm int}$ in (\ref{smatrix}) is (whose matrix elements would form the black hole $S$-matrix). Indeed, to proceed we had used the {\em free-field} Hamiltonian consisting of the two-mode squeezer and the beam-splitter analogues, which, along with using the single-time-slice approximation of the path integral, gave rise to a consistent picture of black hole dynamics when interacting with classical or quantum information. 

It is important to realize at this juncture that the two-mode squeezing (or "optical parametric amplifier", OPA) Hamiltonian (\ref{ham_opa}) is itself an approximation that assumes that the number of pump quanta is so large that the down conversion process does not change the "store" of pump quanta. In other words, in this approximation it is assumed that the down-conversion process does not "react back" on the pump, which is therefore "undepletable", much like the black hole mass is held constant in the semi-classical approximation. But unlike in quantum gravity where we do not know how to move beyond this approximation, in quantum optics it is possible to write down the interaction between the pump modes and the signal and idler modes that represents the canonical {\em extension} of the OPA to depletable pumps: it is a {\em tri-linear} Hamiltonian 
\be
H_{\rm tri}=ir(d_p a^\dagger_s b^\dagger_i-d_p^\dagger a_sb_i)\;. \label{tri} 
\ee
Here, the annihilation operators $a_s$ and $b_i$ refer to the signal and idler modes as before, but $d_p$ and $d^\dagger_p$ create and annihilate {\em pump} modes instead. The coupling constant $r$ is related to the gain $\eta$ of the OPA in Eq.~(\ref{ham_opa}) and the expected number of pump modes, and is in principle time-dependent.  

Given that the quantum optics analogy has been so successful when transferred to black hole dynamics, what if we used the interaction Hamiltonian (\ref{tri}) to calculate the black hole $S$-matrix, where the pump modes play the role of black hole modes, and the signal and idler modes are identified with the Hawking and partner modes (just as before)? This was in fact attempted by Nation and Blencowe~\cite{NationBlencowe2010}, and later by Alsing~\cite{Alsing2015}. Both found that 
the entanglement entropy of the Hawking modes decreases after reaching a maximum, 
but they could not reproduce Page curves because, using effectively a one-time-slice or "static path" approximation (SPA) of the path integral as in (\ref{smatrix}), the calculation quickly became unreliable as the time step $\Delta t$ is taken to be large. A good introduction to the quantum optics/black hole physics analogy using trilinear Hamiltonians can be found in~\cite{FlorezGutierrez2022}.

We will now see what happens if we use the tri-linear Hamiltonian (\ref{tri}) to calculate the $S$-matrix of black hole evaporation, by going beyond the SPA and approximating the black hole $S$ matrix using enough time slices that $\Delta t$ can be kept small. In this way, we can follow the evaporation of the black hole (or, in the words of quantum optics, the depletion of the pump) accurately as long as the number of initial quanta $n$ is not too high. While for black holes the number $n$ surely must be astronomical, we will have to keep this number comparatively small since the evaluation of the path integral can only be done numerically. 

We begin by writing the initial state at $t=0$ as
\be
|\Psi(0)\ra=|n\ra_d|0\ra_a|0\ra_b\equiv |n\ra_d|0\ra_{ab}\;.
\ee
Here, the Hawking modes (annihilated by operators $a_k$ in region I, as in Fig.~\ref{penrose}) and the partner modes (annihilated by $b_k$ in region II) are interacting with black hole modes created and annihilated by $d^\dagger_k$ and $d_k$, see Fig.~\ref{blackhole}. 
\begin{figure}[htbp] %
   \centering
   \includegraphics[width=4in]{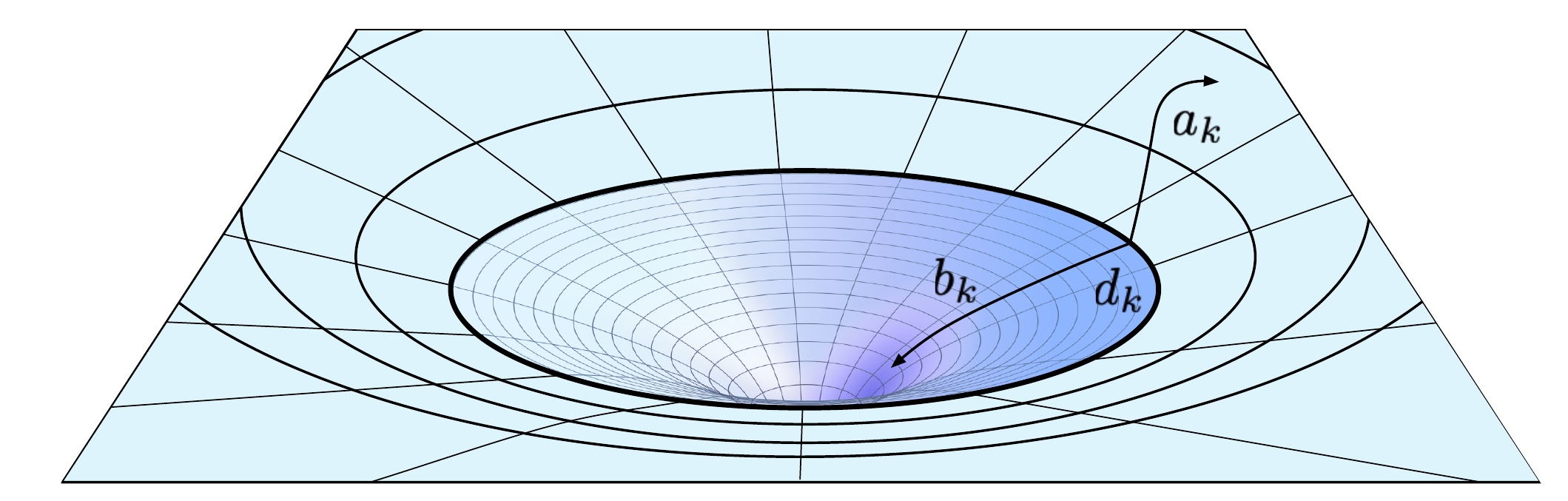}
   \caption{Schematics of a black hole (approximate Schwarzschild radius in bold) where black hole modes $d_k$ are transformed into Hawking modes $a_k$ and partner modes $b_k$ at the horizon.} 
   \label{blackhole}
\end{figure}

We write the time evolution of the joint state $\ket{\Psi(t)}$ in terms of black hole $S$-matrix acting on $|\Psi(0)\ra$
as 
\begin{equation}\label{eq:psi}
\ket{\Psi(t)}=S(t,0)\ket{\Psi(0)}=\mathsf{T}e^{-i\int_0^tH_{\mathrm{tri}}(t')dt'}\ket{n}_d\ket{0}_{ab}\;,
\end{equation}
using the trilinear Hamiltonian 
\begin{equation}
H_{\mathrm{tri}}=\sum_{k=-\infty}^\infty i r_{\om_k}(t)\big(d_ka_k^\dg b_k^\dg-d_k^\dg a_kb_k\big)\;. \label{hamtri}
\end{equation}
Here, $r_{\om_k}(t)$ is the time-dependent coupling strength that sets the Hawking temperature $T_\bh(t)$ and the black hole mass $M_\bh(t)=\frac1{8\pi T_\bh(t)}$, via the standard relation~\cite{NationBlencowe2010}
\be
T_\bh(t)=\frac{\omega_k}{2\ln {\rm cotanh}\ r_{\omega_k}(t)}\;.
\ee
In the following, I will again focus on a single mode $k$ with energy $\omega_k$, and omit the index $k$ for convenience.

In order to evaluate (\ref{eq:psi}), we need to introduce small time slices $\Delta t$ that allow us to discretize the path integral so that with $t=N\Delta t$
\begin{equation}\label{eq:prod}
U(t)=\mathsf{T}e^{-i\int_0^tH_{\mathrm{tri}}(t')}dt'\approx \prod_{i=1}^N e^{-i\Delta t H_i}\;. 
\end{equation}
In (\ref{eq:prod}), the $i$-th time-slice Hamiltonian $H_i=ir_0\big(d a_i^\dg b_i^\dg-d^\dg a_ib_i\big)$ acts on the black hole state and the $i$-th slice of the $ab$ Hilbert space $\ket{0}_{a_ib_i}$. The initial value of the coupling strength $r_0$ simply sets the energy scale, and we can set $r_0=1$ in the following without loss of generality.

Let us now apply the discretized (\ref{eq:prod}) to the initial state, so that~\cite{HilleryZubairy1982}
\begin{equation}  
\ket{\Psi(t)}_{dab}=W\ket{n}_d\ket{0}_{ab}=\prod_{i=1}^N e^{-i\Delta t H_i} \ket{n}_d\ket{0}_{ab}\;,  \label{product}
\end{equation}
where I defined he time-sliced basis
\be
\ket{0}_{ab}\df\ket{0}_{a_Nb_N}\otimes\hdots\otimes\ket{0}_{a_1b_1}\; \label{prodbasis}
\ee
as well as the unitary operator acting on time slice $i$
\be
W^{(i)}=e^{-i\Delta t H_i}   \label{slice-ham}
\ee
so that
\be
W=W^{(N)}\otimes\hdots\otimes W^{(1)}\;.
\ee 
Assuming that the basis states for each time slice appear as product states in Eq.~(\ref{product}) implies that after a black hole mode has been converted to Hawking and partner modes, those modes will never interact with the black hole again, as depicted in Fig.~\ref{sliced}(a). 

While this is certainly reasonable for the outgoing Hawking modes, this is questionable for the partner modes $b$ behind the black hole horizon. In fact, this is an approximation that is also often made in quantum optics, where the non-linear crystal is assumed to be so thin that the two modes (the signal and idler modes) do not interact with the crystal degrees of freedom after they have been produced. We will test this assumption later by allowing the $b$ modes to interact with the black hole again, as depicted in Fig.~\ref{sliced}(b).
\begin{figure}[htbp] 
   \centering
   \includegraphics[width=3.5in]{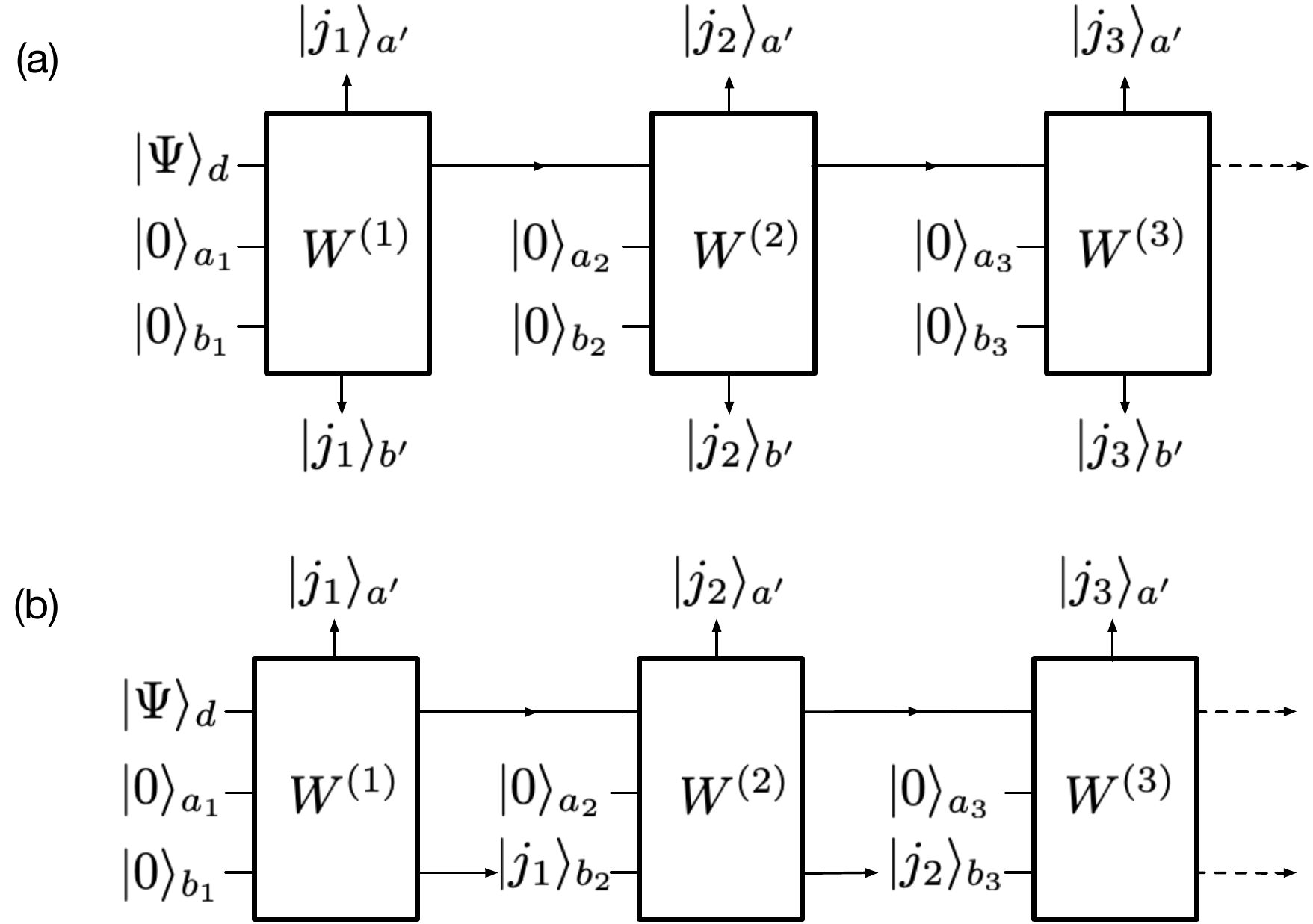} 
   \caption{Action of the discretized $S$-matrix $W^{(i)}$ on the discretized basis states. (a) By assuming that once a black hole mode is converted, the Hawking and partner mode never interact with the black hole again allows us to write the sliced basis as a product (\protect\ref{product}). (b) Relaxing this assumption for the $b$ modes creates a far more complex set of interactions.}
   \label{sliced}
\end{figure}

Let's evaluate the first time slice:
\begin{equation}\label{eq:W1action}
\ket{\Psi(1)}=W^{(1)}|n\ra_d\ket{0}_{a_1b_1}=\sum_{j=0}^nU_{nj}^{(1)}\ket{j\ra_d|n-j}_{a_1b_1}\;,
\end{equation}
with amplitudes $U_{nj}^{(1)}$ determined below. The probability $p(j|n)=|U_{nj}^{(1)}|^2$ reflects the probability to convert $n$ black hole modes into $j$ Hawking and partner modes in one interaction, and I will outline its calculation (and others like it) below. The full quantum state after time $t$ in this approximation becomes 
\be\label{eq:WNaction}
 \ket{\Psi(t)}_{dab}  = \sum_{j_1...j_N}U_{nj}^{(1)}\hdots U_{j_{N-1}j_N}^{(N)} 
  |j_N\rangle_d|j_{N-1}-j_N\rangle_{a_Nb_N}\hdots|n-j_1\rangle_{a_1b_1}\;.
\ee
As I pointed out earlier, it is possible to approximate the path integral using a single time-slice in the static path approximation (SPA), see e.g.~\cite{Arveetal1988,Langetal1993}. Such an approximation can yield good results at very low temperatures, when self-consistent temporal fluctuations can safely be ignored. However, SPA calculations of the black hole entropy using the trilinear Hamiltonian lead to an oscillating behavior of the black hole entropy~\cite{NationBlencowe2010,Alsing2015}, suggesting that self-consistency of fluctuations are an important element of Page curves. 

Using the time-dependent out-density matrix
\be
\rho_{\rm out}(t)=\ket{\Psi(t)}_{dab}\la \Psi(t)|
\ee
we can define the black hole density matrix $\rho_\bh$ by tracing over the Hawking and partner modes:
\be
\rho_\bh(t)=\Tr_{ab}\, \rho_{\rm out}(t)\;,
\ee
so that the black hole entropy is
\be
S_\bh(t)=-\Tr_d\,\rho_\bh(t) \log \rho_\bh(t)\;.
\ee
The density matrix $\rho_\bh$ can be written entirely in terms of the probabilities $p(j|i)=|U_{ij}|^2$ introduced earlier, which stand for the probability to turn $i$ black hole modes into $j$ Hawking/partner modes (there are always as many partner modes as there are Hawking modes since they are always created in pairs). We find
\be
\rho_\bh(t)=\sum_{j_N=0}^n\limits p_{j_N}|j_N\ra\la j_N|
\ee
where
\begin{equation}\label{eq:pi}
p_{j_N}=\sum_{j_1>j_2>\hdots>j_{N-1}}^n |U_{nj_1}^{(1)}|^2\hdots |U^{(N-1)}_{j_{N-2}j_{N-1}}|^2 |U_{j_{N-1}j_N}^{(N)}|^2.
\end{equation}
The probabilities $p(j|i)=|U_{ij}|^2$ are difficult to evaluate.  Unlike in the case when we were dealing with the "free-field" Hamiltonian (\ref{ham1}) that allowed the associated $U=e^{iH}$ to be factorized using the SU(2) and SU(1,1) disentangling theorems, the unitary operator $W^{(i)}$ does not appear to be factorizable in a simple way. The usual formal factorization formulas~\cite{Magnus1954,Suzuki1976,Trotter1959} are not suitable for practical calculations. 

In the absence of a disentanglement decomposition of $W^{(i)}$, we might entertain the idea to simply perform a Taylor expansion of the exponential in (\ref{slice-ham}) in terms of $r_0t$. However, even for moderate $r_0t$, the Taylor expansion is prohibitively inefficient, requiring of the order of about $2^{500}$ terms for $n=50$ and $\Delta t=1/15$. Fortunately, a method developed by Br\'adler~\cite{Bradler2015} makes it possible to evaluate matrix elements of $W^{(i)}$ in terms of an integer lattice known as a generalized Dyck path~\cite{Stanley1999} as long as $W^{(i)}$ acts on any state generated by the repeated action of $da^\dg b^\dg$ on a ground state $|0\ra$, defined by $d^\dagger ab|0\ra=0$. It so happens that the basis elements $\{|j\ra_d|00\ra_{ab}\}_{j=0}^n$ spanning the input Hilbert space of $W^{(i)}$ are all ground states of $H_{i}$.

Using Br\'adler's nearly miraculous Dyck-path representation of $W^{(i)}$ (which generates a polynomial rather than exponential number of terms) we can evaluate $\rho_\bh$ for black holes with initial quanta up to $n=50$, using the discretized path integral (\ref{eq:WNaction}). Fig.~\ref{fig:entropy} shows the black hole entropy $S_\bh$ as a function of the number of time slices used, for a small $\Delta t=0.15$, for black holes with $n=5$, $n=20$, and $n=50$. As the maximal entropy of a black hole with $n$ initial modes is $\log(n+1)$ (counting the $n$ states plus the vacuum state), we show in Fig.~\ref{fig:entropy} the {\em normalized} entropy in order to be able to compare the shape of the curves as the size of the black hole is changed. The resulting entropy curve turns out to be strikingly similar to the one predicted by Page~\cite{Page1993} as long as we observe evaporation for long enough (several thousand time slices). Most importantly, the entropy that started out as a pure state reaches a maximum (at about the time when half the black hole quanta have been converted, see~\cite{FlorezGutierrez2022}) and appears to vanish as $t\to\infty$, where the final black hole density matrix $\rho_\bh$ approaches a pure vacuum state $|0\ra_d$ in the limit $N\to\infty$, for all the input basis states $\ket{n}$ tried. It thus appears that, from the formation to the decay of black holes, pure states are turned into pure states, and the laws of physics remain inviolate.

\begin{figure}[htbp] %
   \centering
   \includegraphics[width=3in]{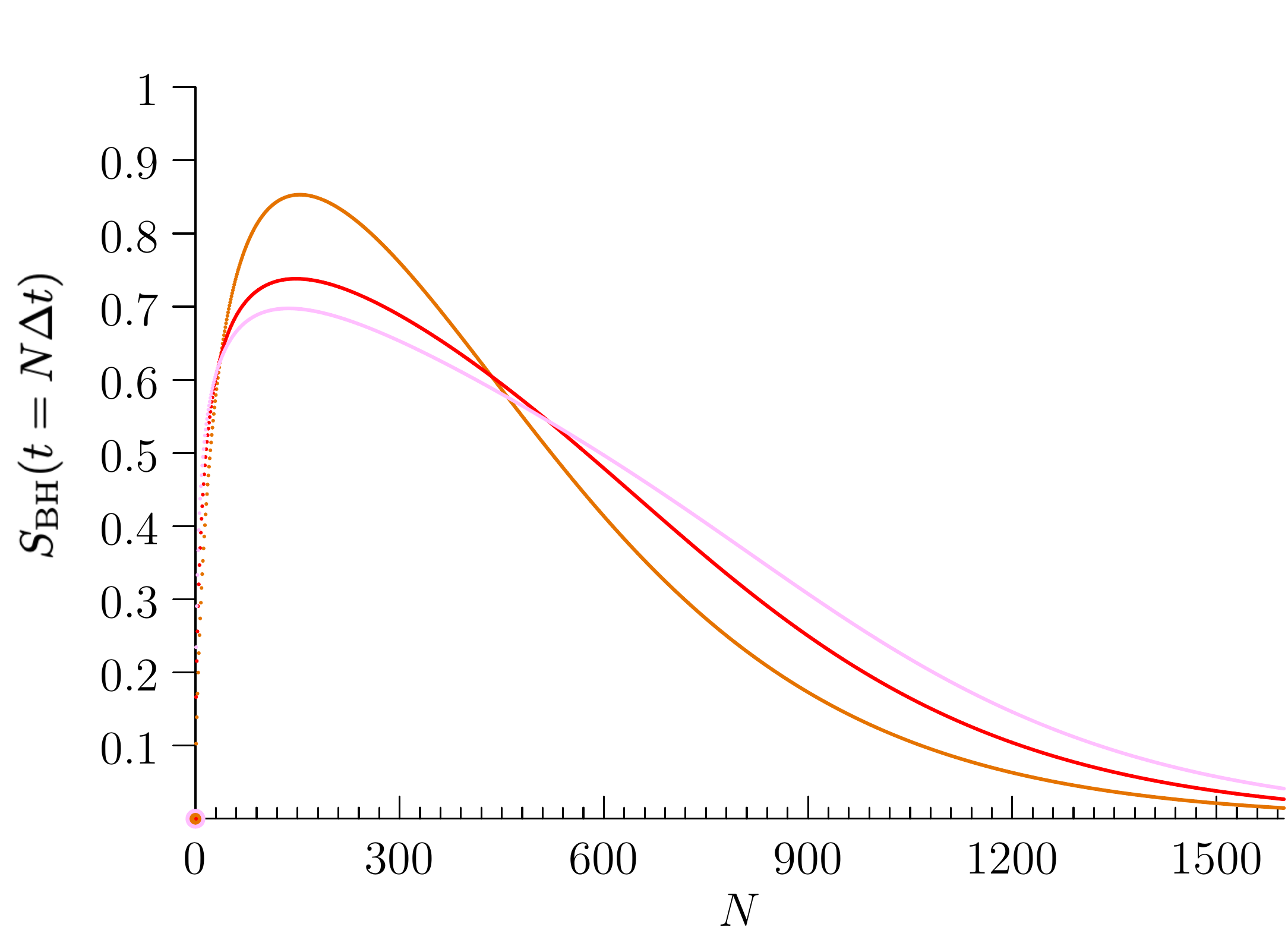} 
  \caption{Black hole entropy $S_\bh(t)$ as a function of the number of time slices $N$ for initial states 
  $\ket{n}_d\ket{0}_{ab}$ with $n=5$ (orange), $n=20$ (red) and $n=50$ (pink), with $t=\Delta t N$. To normalize entropies, logarithms are taken to the base $n+1$ so that the entropy $S_\bh(t)\leq1$ for any $n$.  The $S$-matrix was evaluated using $r_0=1$ and $\Delta t=1/15$ (modified from~\protect\cite{BradlerAdami2016}).}
   \label{fig:entropy}
\end{figure}

Note that using just $n=50$ initial modes already gives rise to an extremely large Hilbert space. Using a Taylor expansion of $W^{(i)}$ with $\Delta t=1/15$ to order up to 500 would require $2^{500}$ terms\footnote{A Taylor expansion of $W^{(i)}$ entails the expansion of operators of the type $e^{A+A^\dagger}=\sum_{i=0}^n \frac1{n!}(A+A^\dagger)^i$, where $A=d^\dagger a b$. Because $A$ and $A^\dagger$ in general do not commute, the number of summands in $(A+A^\dagger)^i$ is not $i+1$ but instead $2^i$.}, which is of course intractable. Br\'adler's Dyck path representation~\cite{Bradler2015} renders the calculation tractable, but it does require High-Performance Compute Clusters. Using a smaller $\Delta t=1/25$ with commensurately fewer time slices (to keep overall compute time constant) does not change the curves visibly, which suggests that $\Delta t=1/15$ is sufficiently small to allow for equilibration. 

While the curves shown in Fig.~\ref{fig:entropy} used the approximation that black hole modes converted to Hawking and partner modes only interact once (by virtue of the initial state (\ref{prodbasis})), it is possible to relax this assumption and allow the partner modes behind the horizon to interact with black hole modes again, as in Fig.~\ref{sliced}(b). Doing so complicates the calculation enormously so that only small systems can be evaluated, for fewer time slices. The overall shape of the entropy curves does not change appreciably when allowing partner modes to interact with the black hole repeatedly. The curves become somewhat more symmetric due to a slower rise of the entropy, and in so doing become more similar to the symmetric curves that Page had imagined. Further, while the calculation assumed that the black hole was initially in a pure state $|n\ra_d$, the results are unchanged if the initial state is instead in a rotated basis $|\psi\ra_d=\sum_{j=1}^n\epsilon_j |j\ra_d$. 

Technically speaking, the mapping from the black hole initial state to the final state is an example of an {\em erasure} map
\be
|n\ra_d|0\ra_{ab}\stackrel{W}\longrightarrow |\Psi(t)\ra_{dab}\stackrel{t\to\infty}{\approx} |0\ra_d \otimes \rho_{ab}(|n\ra)
\ee
that ultimately decouples the black hole from the Hawking and partner modes. It turns out that this map is an explicit realization of the fully-quantum Slepian-Wolf (FQSW) protocol~\cite{Abeyesingheetal2009}, which is the fundamental protocol in quantum information science that quantifies how well quantum entanglement can be transferred, stored, and distilled. The unitary interaction (\ref{eq:prod}) first creates the entanglement between the black hole (which plays the role of the reference in the FQSW protocol) and the outside and the inside of the black hole (re-enacting the parts of Alice and Bob). After the Page time, further dynamics erases the entanglement between the black hole and Hawking radiation, just as in the FQSW protocol the entanglement between Alice and the reference is erased. That the further dynamics reverses the prior entanglement is ensured by the continued unitary dynamics of the black hole's interaction with the radiation field, in the same manner as the erasure map in the FQSW protocol forces the transfer of entanglement from Alice to Bob. It is this same unitarity that enforces the existence of stimulated emission, which in turn preserves information in black hole dynamics. 

It is useful to consider the dynamics of black hole formation, evaporation, up to ultimate disappearance, in terms of entropy Venn diagrams. Those diagrams summarize how classical or quantum entropies are distributed among the subsystems of a closed system. In particular, if we begin with a system in a pure state (with zero entropy) those diagrams can reveal how the purity of the system is maintained as long as the sum of all entropies remains at zero. In Box 2 I show schematically how this is indeed achieved in the scenario I have outlined here (see also~\cite{AdamiCerf1999b}), and in particular suggest that the information about the black hole's formation is in fact {\em encrypted} in the Hawking and partner modes (a process that is otherwise called "scrambling", see~\cite{HaydenPreskill2007}). It is also clear that as a consequence, Hawking modes can be purely thermal and yet convey information about the black hole modes.
\noindent \rule{\textwidth}{1pt}
\centerline{\bf Box 2: Quantum Entropy Venn Diagrams in Unitary Black Hole 
Dynamics}
\mbox{}\vskip 0cm
\noindent Entropy Venn diagrams are a useful tool to study how classical or quantum information is distributed among the subsystems of a (larger) composite system. Such diagrams have been used extensively to study quantum information processing and communication~\cite{CerfAdami1997,CerfAdami1997a,AdamiCerf1997,AdamiCerf1999} as well as quantum experiments~\cite{GlickAdami2017,GlickAdami2020}. In those diagrams, a circle represents a subsystem, and the intersection of this circle with another circle (subsystem) refers to the shared entropy between the two subsystems. Fig.~\ref{Venn1} shows a simple bi-partite Venn diagram between systems $X$ and $Y$, labeling the conditional and shared entropies. 
\begin{SCfigure}[1.8][h]
\includegraphics[width=1.75in]{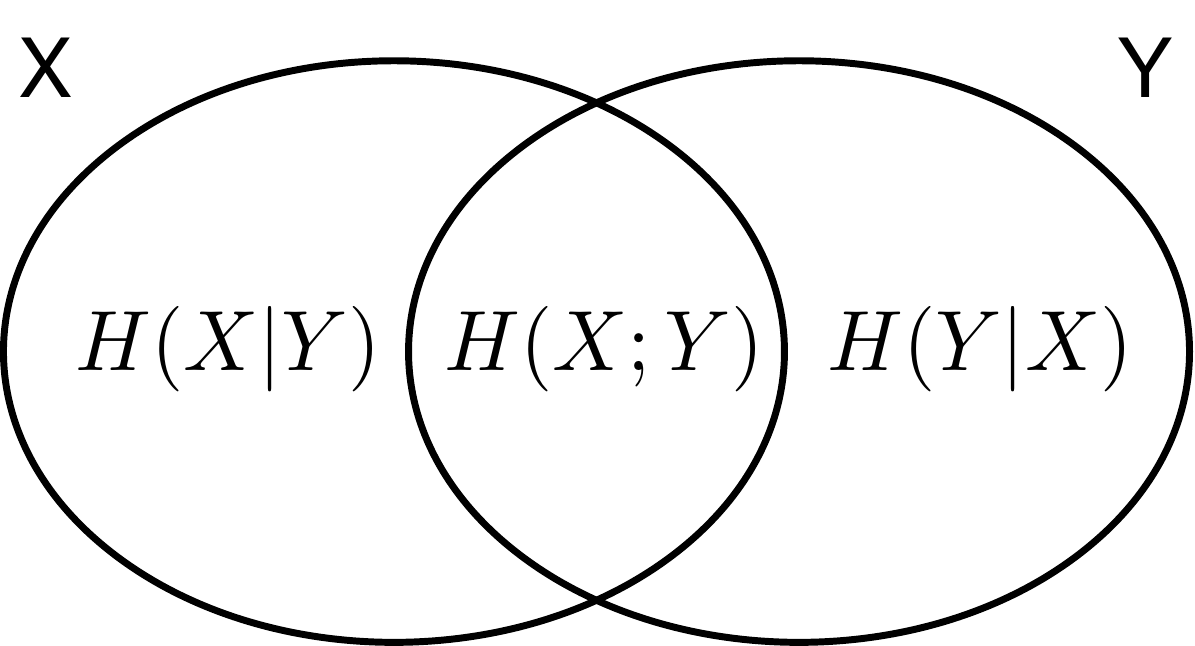}
\caption{\small Entropy Venn diagram showing how the entropy of $X$, $H(X)$, is composed of the conditional entropy $H(X|Y)$ and the shared entropy (information) $H(X;Y)$ (similarly for $Y$). \label{Venn1}}
\end{SCfigure}
Shared pairwise entropies (such as $H(X;Y)$ in Fig.~\ref{Venn1}) can never be negative (either in classical or quantum Venn diagrams), but they can exceeed the entropy of $X$ or $Y$ in the quantum case, and conditional entropies can be negative in quantum physics~\cite{CerfAdami1997}, something that is impossible for classical (Shannon) entropies. Shared entropies between three or more systems can be negative both in classical and quantum physics.

The quantum entropy diagram for the black hole pure state (Fig.~\ref{Venn2}) shows the black hole entangled with a reference state $R$ (as we did in the construction of the quantum channel, when we "purified" Alice's density matrix in Fig.~\ref{qchannel}).
\begin{SCfigure}[2.5][h]
\includegraphics[width=1.75in]{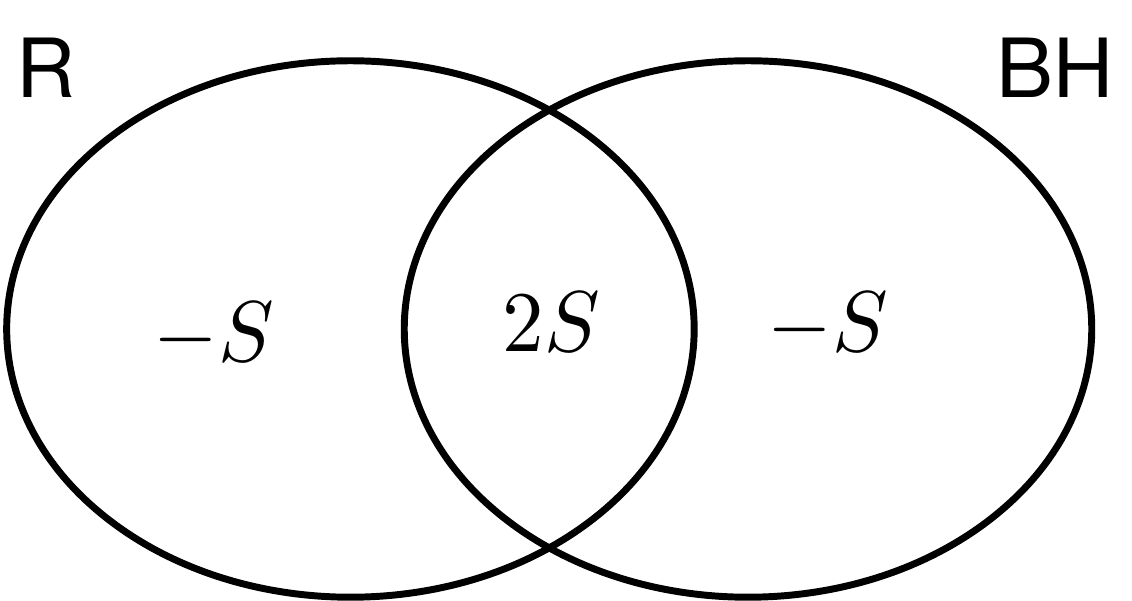}
\caption{Entropy Venn diagram showing the initial joint state of the black hole $BH$ purified by the reference $R$. The joint state has zero entropy, while the inital von Neumann entropy of the black hole is $S$. As the black hole evolves, the entropy of the reference remains at $S$, while the entropy of the black hole (minus the entropy of the Hawking radiation) will decrease.\label{Venn2}}
\end{SCfigure}
As the black hole evolves, it loses entropy due to pair formation of Hawking and partner modes at the horizon (modes annihilated by $a_k$ and $b_k$). I will use the letter $A$ to denote the Hawking modes (outside the black hole) and $B$ to denote the partner modes (inside the black hole). They should not be confused with the "Alice" and "Bob" systems defined earlier. Assuming that the entropy of Hawking modes $H(A)=\Delta S$ (because Hawking and partner modes are entangled, their entropy must always be the same), the entropy Venn diagram between the black hole and radiation must be that shown in Fig.~\ref{Venn3-4}(a). Note that in this diagram the reference state R with entropy $S$ is traced out, so that the joint state of black hole BH and the AB system must also have entropy $S$. We can also see that the information that the Hawking and partner modes have extracted from the black hole is {\em encrypted}: the negativity of the shared "triplet" information $H(BH;A;B)=-\Delta S$ is the tell-tale sign of a symmetric Vernam cipher~\cite{Vernam1926,Shannon1949}: each of the three systems is the cryptographic key to unlock the information between the other two (it is easily implemented via the "controlled NOT" (CNOT) operation, for example, A=B.CNOT.BH). Incidentally, the CNOT operation is precisely the one implementing the cloning operation (\ref{copy}). This relationship between black hole and Hawking/partner modes has previously been described as "scrambling"~\cite{HaydenPreskill2007}, except that in the Hayden-Preskill protocol the decoder has access to the partner modes, which is not possible here.
\begin{figure}[h]
\includegraphics[width=3.5in]{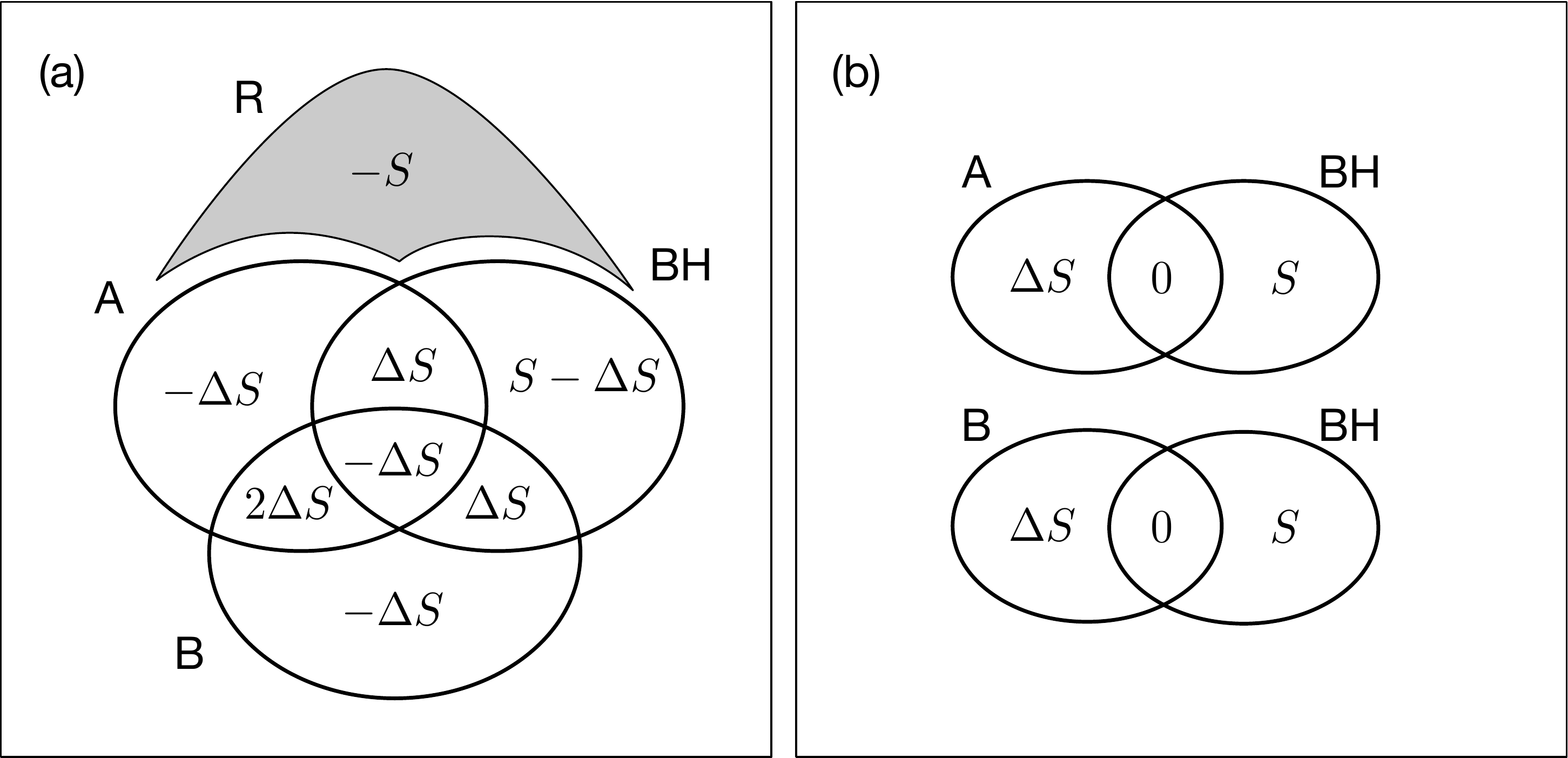}
\caption{(a): Entropy Venn diagram of the black hole BH and the Hawking and partner modes, A and B, respectively. As the reference R is traced over, the joint entropy of BH and radiation is also $S$. The reference state's existence is indicated by the gray area, with conditional entropy $-S$, reminding us that the joint system is still pure.
Each of the radiation modes have marginal entropy $\Delta S$. The entropy of the black hole {\em given} the radiation modes (that is, not counting their contribution to the total entropy) is $H(BH|A,B)=S-\Delta S$.
(b): Entropy Venn diagram showing that the black hole BH and the Hawking modes A are uncorrelated when tracing over reference and partner modes. This holds true also for BH and partner modes B (when tracing over Hawking modes). 
\label{Venn3-4}}
\end{figure}
Note further that when tracing over reference as well as partner modes, the entropy Venn diagram between Hawking radiation and the black hole indicates that they share zero information (Fig.~\ref{Venn3-4}b): one cannot be used to predict the state of the other. Yet, they are still entangled via their entanglement with the partner modes and the reference.  
Once $\Delta S$ has become as large as $S$, the entire entropy of the black hole modes has been converted, so that $H(BH|A,B)=0$ (see Fig.~\ref{Venn5}). The decoupling via "entanglement erasure" has been achieved, and the remaining state is that of a fully entangled pure GHZ (Greene-Horne-Zeilinger) state between reference, Hawking, and partner modes.

\begin{figure}[h]
\centering
\includegraphics[width=3.5in]{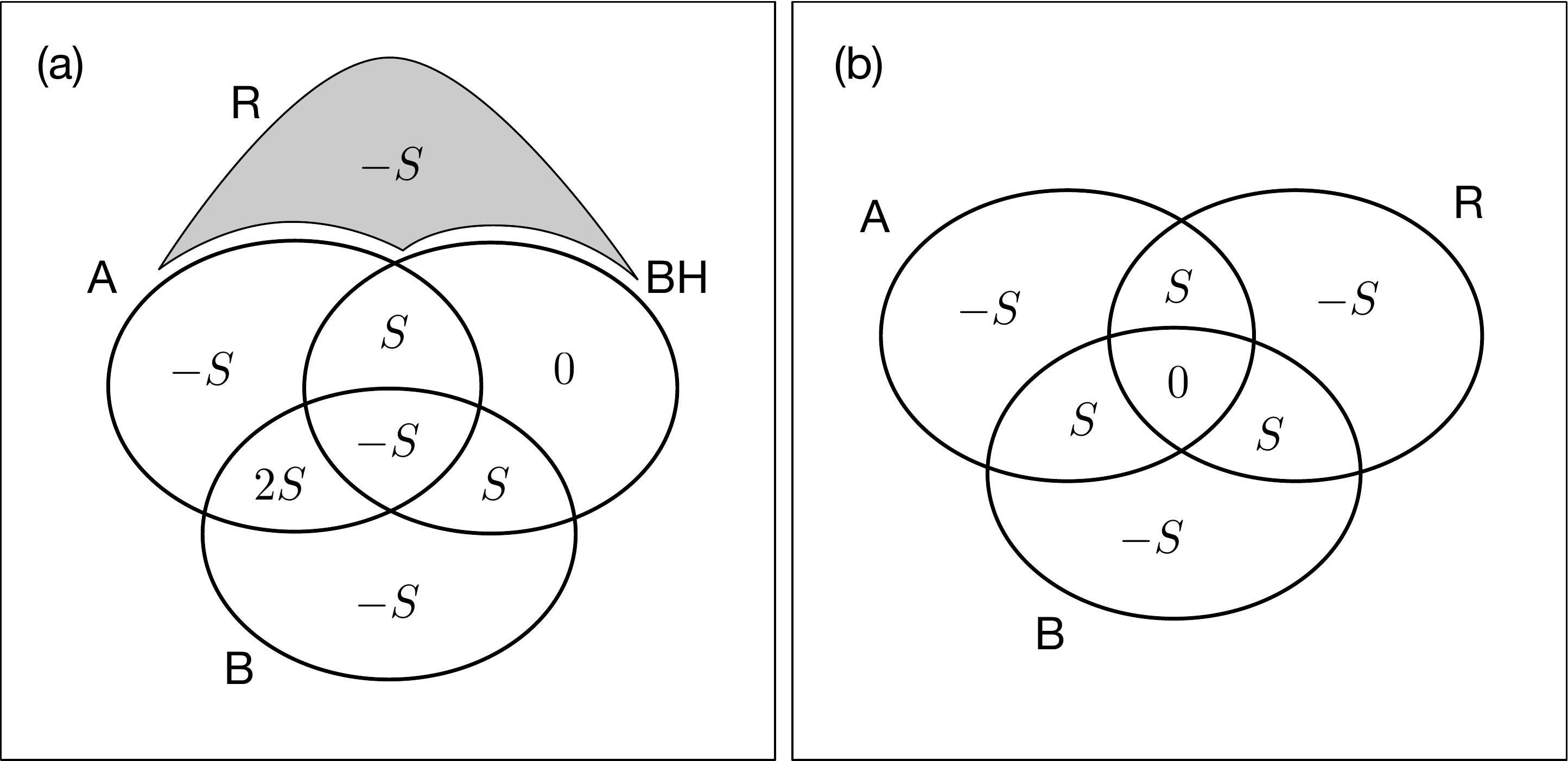}
\caption{(a): Entropy Venn diagram after complete evaporation of the black hole. While the diagram shows the outline of the black hole BH, it actually ceases to be a physical entity as the horizon has disappeared. (b): The entropy Venn diagram between radiation modes and the reference after full evaporation of the black hole. \label{Venn5}}
\end{figure}
\noindent \rule{\textwidth}{1pt}

Finally, a few words about the use of the tri-linear Hamiltonian to simulate interactions between black hole modes and radiation. Obviously, this interaction Hamiltonian does not follow from a fundamental theory of quantum gravity, but instead emerges from taking the quantum optics analogy suggested by the Bogoliubov transformation (\ref{bol}) seriously. That transformation suggested a free-field Hamiltonian (the squeezing Hamiltonian) which, in parametric down-conversion, can be extended so as to deal with a depletable pump using the term (\ref{tri}). Alternatives to using such a simple interaction require coupling the black hole to extraneous degrees of freedom,  such as for example the dilaton (as in the CGHS model~\cite{Callanetal1992}). But generally speaking (and as argued for by Strominger~\cite{Strominger1996}), {\em any} consistent unitary theory that obeys energy conservation rules must have a tri-linear intreraction term in the low-energy limit.

\section{Discussion}
In this review I have tried to marshal a set of arguments---some of which are nearly as old as Hawking's original derivation of the spontaneous emission of radiation by black holes---to support the point of view that black hole dynamics is unitary from formation to complete evaporation, and that information is not lost in, swallowed by, or destroyed in, black holes. Because of the effect of stimulated emission of radiation, which must accompany the spontaneous emission of radiation for any black body (as Einstein's original derivation showed), information is always copied at the horizon with an accuracy close to what the laws of physics allow. Stimulated emission not only ensures information preservation, it also ensures CPT invariance, as Box 1 suggests. The copying of information at the event horizon does not, however, violate the quantum no-cloning theorem, which after all is a consequence of the linearity of quantum mechanics. In fact, it is precisely the Hawking radiation that saves the no-cloning theorem, as it is Hawking radiation that prevents perfect cloning of quantum states (just as perfect cloning is impossible in quantum optics due to the "open port" in the nonlinear crystal that gives rise to vacuum fluctuations). 

How is it possible that Hawking missed this effect in his initial (and also subsequent) publications on the quantum properties of black holes? It is clear that he was aware of the possibility of stimulated emission, but in~\cite{Hawking1975} he describes stimulated emission in terms of the phenomenon of "superradiance", which he refers to as a "classical phenomenon" (citing Refs.~\cite{Misner1972,PressTeukolsky1972,Starobinski1973}). Superradiance is a term used both in quantum optics and in astrophysics, and generally refers to the emission of radiation due to excitations that owe their energy either to inertial motion at superluminal speeds (e.g., \v{C}erenkov radiation) or from rotational motion. In both cases, the energy powering the radiation must come from the kinetic energy of bulk matter. Soviet physicist Yakov Zeldovich had calculated that a rotating cylinder would emit radiation via stimulation~\cite{Zeldovitch1972}, while Alexei Starobinskii argued that a rotating black hole would emit radiation due to that same effect~\cite{Starobinski1973}. During Hawking's well-documented visit to Moscow in 1973 he talked to Zeldovitch as well as to Starobinskii, who suggested to Hawking that stimulated (superradiant) emission must be accompanied by {\em spontaneous} emission in the same modes but {\em without} input from bulk kinetic energy. After the visit, Hawking engaged in the calculation we now know. His calculation, however, treated non-rotating (Schwarzschild, rather than Kerr) black holes, and for this reason it seems he discarded the possibility of stimulated emission out of hand.

In hindsight, the importance of the stimulated process should have been apparent if only Hawking had investigated the effect of particles present in the initial vacuum (at past infinity ${\mathscr I}^-$). But Hawking explicitly disregarded those. His reasons for doing so are not immediately clear. As discussed earlier, to understand the effect of gray-body factors, Hawking relied on an argument following particle trajectories backwards in time from inside the black hole to the outside. In both Refs.~\cite{Hawking1974} and~\cite{Hawking1976b}, Hawking claims that following particles inside the horizon at future infinity back to ${\mathscr I}^-$ outside the horizon (that is, with $v>v_0$ in his notation) has zero amplitude [Eq. (4.22) in~\cite{Hawking1976b} concerns precisely such particles]. It is not clear what reasons he mustered for this assertion that there cannot be any particles at past infinity, but we must keep in mind that Hawking also did not seem to realize that a perfectly black hole will, when viewed from the {\em inside}, look like a perfectly reflecting mirror (see Box 1). For such a mirror (a white hole), those moving-backwards-in-time particles hitting the horizon would not be transmitted to past infinity, but stimulated emission of particles from those reflected at the white hole horizon, as discussed in section~\ref{cloning}, would. 

My best guess of why Hawking dismissed stimulated emission as an important element in black hole dynamics is that he thought that stimulated emission needs an energy source so as to lift the system out of its ground state. For Kerr black holes, this energy is supplied by the rotational motion. For atomic gases, inversion is created via pumping. But black holes are different: they are "forever in a pumped state": their heat capacity, after all, is negative. As a consequence, they can be stimulated to emit copies even if they lack a charge or angular momentum.

What empirical evidence do we have today that stimulated emission must play an important role in black hole dynamics? We have yet to detect Hawking radiation from black holes, and are unlikely to do so because the temperature (for average-sized black holes) is extremely low (about $6\times 10^{-8} \frac{M_\odot}{M}$[K] for a black hole of mass $M$, where $M_\odot$ is the solar mass), and the radiation is expected to be extremely faint. However, several teams have created "analogue black holes" by artificially creating the causal split between space-time regions that defines black holes, as initially suggested by Unruh~\cite{Unruh1981}. One way to achieve this is to use gravity waves in flowing fluids. As discussed for example by Leonhardt (see in particular Fig.~8.7 in Ref.~\cite{Leonhardt2010}) it is possible to create horizons for gravity waves by creating a fluid flow that is faster than the maximum speed for those gravity waves. Weinfurtner et al.~\cite{Weinfurtneretal2011} tested whether a water wave directed towards a (simulated) white hole horizon stimulated pairs of waves with the correct amplitude ratio (\ref{ratio}) predicted by Hawking. They found this to be the case, and argued that because stimulated emission must always be accompanied by spontaneous emission, it should be possible in principle to observe analogue Hawking radiation in this system also. This was achieved by another group, using atomic Bose-Einstein condensates as the medium~\cite{MunozdeNovaretal2019,Kolobovetal2021}. In this experiment, two regions were created where the flow velocity is larger (smaller) than the speed of sound in this medium, simulating the inside (outside) of the black hole. The group observed the formation of the black hole horizon, and the subsequent emergence of Hawking and partner modes. Because the experiment produces not only a black hole horizon but also a white hole horizon (an area inside of the simulated black hole that reflects waves) the dynamics creates \v{C}erenkov radiation that ultimately stimulates the emission of Hawking/partner pairs. 

Earlier speculations~\cite{CorleyJacobsen1999} and experiments~\cite{Steinhauer2014} with Bose-Einstein condensates had suggested that perhaps the reflection of waves at the white-hole "inner" horizon could stimulate the emission of more modes at the black-hole horizon, which in turn after reflection at the white-hole horizon could lead to ever-increasing amplification of these waves: a black hole laser. In fact, this idea was introduced rather speculatively already by Press and Teukolsky~\cite{PressTeukolsky1972} before the discovery of Hawking radiation. Those authors imagined a black hole encased in a spherical mirror, and radiation amplified superradiantly (in a Kerr black hole, of course) could reflect back onto the mirror, creating an instability they termed the "black hole bomb". However, a close analysis~\cite{Wangetal2017} of Steinhauer's experiment~\cite{Steinhauer2014} revealed that lasing at most played a minor role in the Bose-Einstein cavity, most likely because the location of the white hole horizon with respect to the black hole horizon is constantly changing, destroying the coherence required for the lasing phenomenon.

However, these discussions open up the possibility of a black hole laser that is not due to an "inner" horizon, but rather emerges for the brief period of time when a black-hole binary is inspiraling, just before the merger event. While the period of time where the binaries are close enough so that a significant amount of stimulated radiation could hit the partner is brief (on the order of a fraction of a second for a typical binary), this is sufficient for on the order of ten reflections (in the rotating frame), leading to significant amplification\footnote{An even more speculative notion is the idea that a wormhole that connects two black holes could create a cavity that would coherently amplify radiation trapped within it, giving rise to a {\em wormhole laser}.}. Could such a "flash" of coherent black-hole laser light be detected at the same time that we are recording the gravitational signature of such an event? This is a difficult experimental question, since as of yet we do not have a simulation of such a phenomenon that would allow us to tune experiments to detect this signature. Such a simulation would allow us to distinguish the tell-tale coherent signature from the light emitted by accretion disks (if any). To-date, no electromagnetic signature has been observed coming from regions where a binary black-hole merger was pinpointed to. This suggests that possibly those mergers do not have accretion disks, and consequently there would also be no source of radiation/matter that could initiate the black hole laser. Should we observe an electromagnetic counterpart to a binary black-hole merger in the future, it may be worth while to develop techniques that can test whether that light was due to the black-hole induced simulated emission of radiation, giving us the first direct experimental verification of (stimulated) Hawking radiation.
\mbox{}\vskip 0.25cm
\noindent{\bf Funding statement} This research was unsupported.

\begin{acknowledgments}I am indebted to my collaborators in black hole information theory: Greg Ver Steeg and Kamil Br\'adler. I also thank numerous friends and colleagues who patiently listened to my arguments that black holes do not violate any laws: Paul Davies, Nigel Goldenfeld, Nicolas Cerf, Arend Hintze, Claus Wilke, and Richard Lenski. I dedicate this contribution to the late Jonathan P. Dowling, who led the quantum computing group at the Jet Propulsion Laboratory where the ideas presented in this article were first hatched.
\end{acknowledgments}

\appendix
\section{Density matrix of outgoing radiation for arbitrary absorptivity}

\renewcommand{\theequation}{A.\arabic{equation}}
The density matrix of outgoing radiation in region I when $m$
particles are incident (notation $k|m$) can be calculated from the outgoing demsity matrix
$|\psi\ra_\out\la \psi|$ (constructed from the out-state Eq.~(\ref{state})) by repeated application of the disentangling theorems for SU(2) and SU(1,1), and tracing over the degrees of
freedom of region II (modes $b_k$ and $c_k$). If no antiparticles
are accreting on the horizon, the antiparticle part of the density matrix factorizes again and we can
write $\rho_{\rm I}=\rho_{k|m}\otimes\rho_{-k|0}$ where (I omit the index $k$ in the particle numbers and the coefficients $\alpha$, $\beta$ and $\gamma$ for succinctness in the following)
\be
\rho_{k|m}=\sum_{n=0}^{\infty}p(n|m)|n\ra\la n|\;,
\label{mpartmat} 
\ee 
with
\be
p(n|m)&=&Z_m^2\sum_{l=0}^\infty\frac{n!\,l!}{m!(l-m+n)!} 
 \left[\sum_{i=0}^{\min(n,m)}(-1)^i {m \choose i}{{l-m+n}\choose{n-i}}\Bigl(\frac{\gamma^2}{\alpha(1+\alpha)}\Bigr)^i\right]^2\nonumber \\
&\times&  \Bigl[\frac{\beta^2(1+\alpha)^2}{(1+\alpha+\beta^2)^2}\Bigr]^n
  \Bigl[\frac{\beta^2\gamma^2}{(1+\alpha+\beta^2)^2}\Bigr]^l  \label{myexp}\nonumber \\
\ee
where
\be
Z_m^2=\left(\frac{1+\alpha}{1+\alpha+\beta^2}\right)^{2}\left(\frac{\alpha^2(1+\alpha)^2}{\gamma^2}\right)^m\;.
\ee
It is possible to rewrite this expression\footnote{Converting the double sums in (\ref{myexp}) into the single sum in (\ref{BM}) is highly non-trivial and took two years to discover after the identity was confirmed numerically by G. Ver Steeg.}
for $p(n|m)$, the probability
to detect $n$ outgoing particles at ${\mathscr I}^+$ if $m$ were
incident on the static black hole, by using a resummation technique described by Panangaden and Wald~\cite{PanangadenWald1977} to read
\be
p(n|m)=R_{nm}\sum_{i=0}^{\min(n,m)}(-1)^i{{m}\choose {i}}{{m+n-i}\choose {n-i}}\left(1-\frac{\gamma^2}{\alpha^2\beta^2}\right)^i \label{BM}
\ee
with 
\be
R_{nm}=\frac1{1+\beta^2}\left(\frac{\beta^2}{1+\beta^2}\right)^{m+n}\left(\frac{\alpha^2}{\beta^2}\right)^m\;,
\ee
Expression (\ref{BM}) agrees precisely with the conditional probability derived by Bekenstein and Meisels~\cite{BekensteinMeisels1977} using maximum entropy methods, and by Panangaden and Wald~\cite{PanangadenWald1977} in quantum field theory (but using a very different method than the one described here). While the expression given by Bekenstein and Meisels looks quite different from (\ref{BM}) as they write (with $x=\omega/T_\bh$ as before)
\be
p(n|m)&=&\frac{(e^x-1)e^{mx}\,\Gamma^{n+m}}{(e^x-1+\Gamma)^{n+m+1}}\sum_{i=0}^{\min(n,m)}
\frac{(-1)^i(m+n-i)!}{i!(m-i)!(n-i)!}\nonumber \\
&\times&\biggl[1-2\frac{1-\Gamma}{\Gamma^2}(\cosh(x)-1)\biggr]^i \label{BM1}\;,
\ee
we can nevertheless see it agrees with (\ref{BM}) (and therefore also with (\ref{myexp})) by noting, for example,  that
\be
2\frac{1-\Gamma}{\Gamma^2}(\cosh(x)-1)=\frac{\gamma^2}{\alpha^2\beta^2}\;.
\ee
and
\be
\frac{(e^x-1)e^{mx}\,\Gamma^{n+m}}{(e^x-1+\Gamma)^{n+m+1}}=R_{nm}\;.
\ee
The expression (\ref{BM1}) has the advantage that it manifestly observes the detailed balance condititions, something that is not immediately apparent from Eqs.~(\ref{myexp}) and (\ref{BM}).


\bibliography{bh-RMP}

\end{document}